\documentclass[11pt]{article}

\usepackage{enumitem}
\usepackage[paperheight=297mm,paperwidth=210mm,width=150mm,top=30mm,bottom=30mm,bindingoffset=10mm, inner=30mm, outer=25mm]{geometry}
\usepackage[utf8]{inputenc}
\usepackage[hyphens]{url}
\usepackage{graphicx}
\usepackage{caption}
\usepackage{subcaption}
\usepackage[nottoc]{tocbibind}
\usepackage{datetime}
\usepackage[colorlinks=true, citecolor=blue, filecolor=black, linkcolor=blue, urlcolor=blue]{hyperref}
\usepackage{fancyhdr}
\usepackage{setspace}
\usepackage{kotex}
\usepackage{cite}
\usepackage{amsmath}
\usepackage{booktabs}
\usepackage{multirow}
\usepackage{tabularx}
\usepackage{rotating}
\usepackage{titlesec}
\usepackage{tocloft}
\usepackage{comment}
\usepackage[table,dvipsnames]{xcolor}
\usepackage[toc,page]{appendix}
\usepackage{pdfpages}
\usepackage{svg}
\usepackage{longtable}
\usepackage[T2A,T1]{fontenc}
\usepackage[russian,english]{babel}
\usepackage{blindtext}
\usepackage{footnote}
\usepackage{amsmath,amssymb,amsfonts}
\usepackage[noend]{algpseudocode}
\usepackage{algorithm}
\usepackage{rotating}
\usepackage{xcolor}
\usepackage{fancyhdr}
\usepackage{array,booktabs, makecell}
\usepackage{pgfplots}
\usepackage{longtable}
\usepackage{tabu}
\usepackage{lscape}
\usepackage{adjustbox}
\usepackage{pgf-pie}
\usepackage{tikz}
\usepackage{lipsum,lmodern}
\usepackage[most]{tcolorbox}

\usepackage[acronym,nomain,toc=true,nonumberlist]{glossaries}

\makeglossaries

\newacronym{apwg}{APWG}{Anti-Phishing Working Group}
\newacronym{asn}{ASN}{Autonomous System Number}
\newacronym{as}{AS}{Autonomous System}
\newacronym{BEC}{BEC}{Business Email Compromise}
\newacronym{bgp}{BGP}{Border Gateway Protocol}

\newacronym{c&c}{C\&C}{Command and Control}
\newacronym{cctld}{ccTLD}{Country-Code Top-Level Domain}
\newacronym{csam}{CSAM}{Child Sexual Abuse Material}
\newacronym{czds}{CZDS}{Centralized Zone Data Service}

\newacronym{dbl}{DBL}{Domain Block List}

\newacronym{DDoS}{DDoS}{Distributed Denial-of-Service}

\newacronym{DKIM}{DKIM}{DomainKeys Identified Mail}

\newacronym{dmarc}{DMARC}{Domain-based Message Authentication, Reporting and Conformance}
\newacronym{dns}{DNS}{Domain Name System}
\newacronym{dnssec}{DNSSEC}{Domain Name System Security Extension}

\newacronym{DRDoS}{DRDoS}{Distributed Reflective Denial-of-Service}

\newacronym{esp}{ESP}{Email Service Provider}

\newacronym{fqdn}{FQDN}{Fully Qualified Domain Name}

\newacronym{gtld}{gTLD}{Generic Top-Level Domain}

\newacronym{iana}{IANA}{Internet Assigned Numbers Authority}
\newacronym{icann}{ICANN}{Internet Corporation for Assigned Names and Numbers}
\newacronym{idn}{IDN}{Internationalized Domain Name}
\newacronym{ioc}{IOC}{Indicators of Compromise}
\newacronym{ip}{IP}{Internet Protocol}

\newacronym{ipv4}{IPv4}{Internet Protocol Version 4}
\newacronym{ipv6}{IPv6}{Internet Protocol Version 6}

\newacronym{isp}{ISP}{Internet Service Provider}

\newacronym{issp}{ISSP}{Information Society Service Providers}

\newacronym{kybc}{KYBC}{Know Your Business Customer}

\newacronym{ntp}{NTP}{Network Time Protocol}

\newacronym{rdap}{RDAP}{Remote Data Access and Protocol}

\newacronym{SAV}{SAV}{Source Address Validation}

\newacronym{SMTP}{SMTP}{Simple Mail Transfer Protocol}

\newacronym{SNMP}{SNMP}{Simple Network Management Protocol}

\newacronym{spf}{SPF}{Sender Policy Framework}

\newacronym{SSDP}{SSDP}{Simple Service Discovery Protocol}

\newacronym{tld}{TLD}{Top-Level Domain}

\newacronym{url}{URL}{Uniform Resource Locator}

\newacronym{whois}{WHOIS}{protocol for querying databases that store the registered users or assignees of an Internet resource}

\newcommand{\pagelabel}[1]{\phantomsection\label{#1}}
\newcommand\eupageref[2]{(Section~\ref{#1} p.~\pageref{#2})}
\newcommand\eupagerefrange[3]{(Section~\ref{#1} pp.~\pageref{#2}--\pageref{#3})}

\newdateformat{monthyeardate}{\monthname[\THEMONTH] \THEYEAR}
\renewcommand{\cftfigpresnum}{Figure\ }

\newlength{\mylenf}
\newcolumntype{C}[1]{>{\centering\let\newline\\\arraybackslash\hspace{0pt}}m{#1}}

\makeatletter
\let\sv@endpart\@endpart
\def\@endpart{\thispagestyle{empty}\sv@endpart}
\makeatother

\begin{document}

\newcommand{\gray}[1]{\textcolor{Gray}{#1}}
\newcommand{\blue}[1]{\textcolor{NavyBlue}{#1}}
\newcommand{\red}[1]{\textcolor{Red}{#1}}

\newcommand*\circled[1]{\tikz[baseline=(char.base)]{
            \node[shape=circle,draw,inner sep=2pt] (char) {#1};}}

\pagenumbering{gobble}
\begin{titlepage}
    \begin{center}
        \vspace*{0cm}

        \vspace{1cm}
                
        \Huge
        \sffamily\textbf{Study on\\ Domain Name System (DNS) Abuse
        \\ Technical Report}
               
        \vspace{1.5cm}
        \huge
                \vspace{1.5cm}

        \LARGE
        Jan Bayer$^1$, Yevheniya Nosyk$^1$, Olivier Hureau$^1$, \\ Simon Fernandez$^1$, Ivett Paulovics$^2$, Andrzej Duda$^1$, \\ Maciej Korczy\'nski$^1$ 
                        \vspace{1.5cm}
        
        \Large
                January 31, 2022
        \vspace{2cm}

        {\Large $^1$Grenoble INP-UGA Institute of Engineering \\ $^2$Fasano Paulovics Società tra Avvocati}
            
        \vspace{2cm}

        \vspace*{0cm}
        
    \end{center}
\end{titlepage}

\clearpage
\newpage
\tableofcontents
\newpage
\listoffigures
\newpage
\listoftables
\newpage
\printglossary[type=\acronymtype,title={List of Acronyms}]

\newpage
\setstretch{1.0}
\pagenumbering{arabic} 
\setstretch{1.5}
\addcontentsline{toc}{part}{Executive Summary}
\part*{Executive Summary}
\label{chapter:introduction}
\pagestyle{plain}

The European Commission contracted Fasano Paulovics Società tra Avvocati and Institut Polytechnique de Grenoble for carrying out a Study on Domain Name System (DNS) abuse, EC reference VIGIE 2020/0653.
This document is Appendix 1 - Technical Report and provides the analysis of the scope and magnitude of DNS abuse. 

\section*{Definition of DNS Abuse}
A safe and secure Domain Name System (DNS) is of paramount importance for the digital economy and society. Malicious activities on the DNS, generally referred to as ``DNS abuse'' are frequent and severe problems affecting online security and undermining users’ trust in the Internet. The proposed definition of DNS abuse is as follows:

\begin{quote}
\textbf{\emph{Domain Name System (DNS) abuse is any activity that makes use of domain names or the DNS protocol to carry out harmful or illegal activity.}}
\end{quote}

DNS abuse exploits the domain name registration process, the domain name resolution process, or other services associated with the domain name (e.g., shared web hosting service). Notably, we distinguish between:
\begin{itemize}
    \item \textbf{maliciously registered domain names}: domain name registered with the malicious intent to carry out harmful or illegal activity
\item \textbf{compromised domain names}: domain name registered by bona fide third-party for legitimate purposes, compromised by malicious actors to carry out harmful and illegal activity.
\end{itemize}

DNS abuse disrupts, damages, or otherwise adversely impacts the DNS and the Internet infrastructure, their users or other persons. 

DNS abuse can be categorized into three main types that can also appear combined:

\begin{itemize}[leftmargin=+1.8cm]
    \item[\textbf{Type 1:}] 
    Abuse related to \textbf{maliciously registered domain names}.

  \item[\textbf{Type 2:}] 	Abuse related to the \textbf{operation of the DNS and other infrastructures}.
  
  \item[\textbf{Type 3:}] Abuse related to \textbf{domain names distributing malicious content}.\footnote{This type of abuse may take advantage of maliciously registered or compromised domain names.}

\end{itemize}

The three types also differ in terms of which relevant entities are responsible and/or best positioned to put in place mitigation measures:

\begin{enumerate}

    \item 	\textbf{Abuse related to maliciously registered names (Type 1)} is usually best addressed by resellers (if any), registrars, and registries with the following proper remediation path:

\textbf{\texttt{Domain Reseller (if any) $\rightarrow$ Registrar $\rightarrow$ TLD Registry (at the DNS level)}}

\item 	\textbf{Malicious content} can be distributed using a \textbf{maliciously registered domain name} (Type 1 and 3) or it can be distributed using a compromised website (Type 3), where the domain under which the malicious content is made available is registered by an unaware third-party, which uses it legitimately.

\begin{enumerate}

    \item In case of \textbf{illegal/harmful content distributed using a maliciously registered domain name (Type 1 and 3)} (e.g., typosquatting domain name serving phishing content), the following remediation path is to be followed to effectively mitigate this abuse: 

\textbf{\texttt{Hosting Reseller $\rightarrow$ Hosting Provider (at the hosting level)}}

and

\textbf{\texttt{Domain Reseller (if any) $\rightarrow$ Registrar $\rightarrow$ TLD Registry (at the DNS level)}}

Mitigating the abuse only at the hosting or DNS level will prevent access to malicious content but will not block all elements of the malicious infrastructure. Therefore, both levels have to be involved in the mitigation of this kind of abuse.

\item While it is also possible for the reseller (if any) / registrar / TLD registry to take action in case of \textbf{malicious content hosted on compromised websites} (Type 3), addressing abuse at the DNS level can be counterproductive, as it can cause collateral damage to legitimate registrants and website visitors. In this case, the site operator, hosting provider (and where it exists, its reseller) are well positioned to take action to curb the abuse. Remediation path is as follows:

\textbf{\texttt{Site operator $\rightarrow$ Registrant (if different from site operator) $\rightarrow$ Hosting reseller 
$\rightarrow$  Hosting provider (at the hosting level)}}

Mitigating abuse at the hosting level includes removing malicious content from the hacked website and patching the vulnerability. Site operators are best positioned to mitigate  abuse in case of so-called unmanaged dedicated servers that they are in complete control and are responsible for their hosting servers and software. Hosting companies are best positioned to mitigate  abuse in case of so-called managed shared hosting as they maintain the operating system and application infrastructure.
\end{enumerate}

\item 	All entities related to the DNS infrastructure (registrars, registries, resellers, operators of authoritative name servers, and DNS resolvers) are concerned with \textbf{abuse related to DNS operations} (Type 2).

\end{enumerate}

To illustrate the types of DNS abuse and responsible entities, we give below the examples of DNS abuse cases:

\begin{itemize}

    \item
    maliciously registered domain names serving phishing content: 
    \begin{itemize}
    \item Type 1 and 3 $\rightarrow$ mitigation action at the DNS level and the hosting level
    \end{itemize}

 \item	compromised websites serving phishing content:     \begin{itemize}
    \item Type 3 $\rightarrow$ mitigation action at the hosting level
\end{itemize}
 \item	compromised websites used to distribute (deliver) malware: 
     \begin{itemize}
    \item Type 3 $\rightarrow$ mitigation action at the hosting level
\end{itemize}
 \item	maliciously registered domain name used to distribute (i.e., deliver) spam---emails containing malicious content: 
      \begin{itemize}
    \item Type 1 $\rightarrow$ mitigation action at the DNS level
    \end{itemize}
\item	maliciously registered domain name (e.g., algorithmically generated domain name - AGD) used for the malicious command-and-control (C\&C) communication (between compromised hosts and a malicious actor):
      \begin{itemize}
    \item Type 1 $\rightarrow$ mitigation action at the DNS level
  \end{itemize}
 \item	file sharing system abused to distribute child sexual abuse material (CSAM):     \begin{itemize}
    \item Type 3 $\rightarrow$ mitigation action at the hosting level
\end{itemize}
 \item	maliciously registered domain name used to distribute CSAM: 
     \begin{itemize}
    \item Type 1 and 3 $\rightarrow$ mitigation action at the DNS level and the hosting level
\end{itemize}
 \item	DDoS attack against a DNS server (e.g., NXNSAttack): 
     \begin{itemize}
    \item Type 2 $\rightarrow$ mitigation action at the DNS level
\end{itemize}
 \item	DDoS attack against a web server using DNS open resolvers as amplifiers/reflectors: 
     \begin{itemize}
    \item Type 2 $\rightarrow$ mitigation action at the DNS level
\end{itemize}
 \item	hijacked domain name (e.g., using zone or cache poisoning) 
 \begin{itemize}
    \item Type 2 $\rightarrow$ mitigation action at the DNS level
    \end{itemize}
\end{itemize}

\section*{Report Structure}

In this Technical Report, we first present the datasets as well as passive and 
active measurement methods used in the analysis of the scope and magnitude of DNS 
abuse. The analysis involves the following blacklist feeds: Spamhaus Domain Block Lists, the lists of SURBL, APWG, PhishTank, OpenPhish, URLHaus, and ThreatFox. 
We have also gathered the list of domain names for each TLD from zone files whenever available and with active web content crawling. We have identified  over 251 million active domain names for generic TLDs, new gTLDs, European Union country-code TLDs, and non-EU ccTLDs,
and  collected the following additional
data: 
\begin{itemize}
    \item ‘A’ resource records to calculate the reputation metrics for hosting providers,
\item registration information using RDAP/WHOIS protocols to calculate the reputation metrics for domain registrars,
\item TLD sizes to express the ``overall health'' of TLD ecosystems,
\item information about the deployment of DNSSEC (‘DS’, ‘DNSKEY’, and ‘RRSIG’ resource records), 
\item open DNS resolvers,  
\item SPF and DMARC entries in ‘TXT’,
records to measure the deployment of email security extensions.
\end{itemize}

Finally, we measured uptimes, i.e., how long a malicious URL (or a domain name) has been active since it appeared on one of the blacklists.

In the second part, we show the results of measurements and data analysis related to DNS abuse. 
We start with high-level statistics describing the distributions of the
malicious resources and abuse rates per TLDs, hosting providers, and countries for
different abuse types. 
We distinguish between compromised and maliciously registered domain names. While some domains are registered exclusively for malicious purposes, others are benign but get compromised and abused to, for example, serve malicious content.
This part also provides key statistics on maliciously registered and compromised sites concerning TLDs and types of abuse.
Since registrars cannot prevent abuse of vulnerable hosting software (unless they provide registration and hosting services), we calculate registrar reputation metrics based on domain names categorized as maliciously registered.
Finally, we discuss targeted brands and special domain abuse.

The last part concerns the abuse of DNS infrastructure---how technical DNS infrastructure can be abused to perform different types of illegal activities.
We first discuss the deployment of DNSSEC, followed by DNS resolver vulnerabilities, the study on SPF and DMARC adoption, the analysis of RFC-compliant email aliases, and inbound Source Address Validation that enables a great variety of attacks on DNS infrastructure.
The study covers the second quarter of 2021.

\section*{Key findings of the measurements}

The key findings of the measurements are as follows:

 \begin{enumerate}
 \item 	\textbf{``Overall health'' of TLDs}
\begin{enumerate}[label=(\alph*)]
    \item	In relative terms, \textbf{new generic TLDs (gTLDs), with an estimated market share of 6.6\%, are the most abused group of TLDs}. In the second quarter of 2021, 20.5\% of all abused domain names representing phishing, spam, botnet C\&C (Command and Control), and malware distribution combined were registered in new gTLDs \eupageref{sec:general_dataset}{fig:tld_distribution}.
\item However, not all new gTLDs suffer from DNS abuse to the same extent. The \textbf{two most abused new gTLDs combined account for 41\% of all abused new gTLD names} \eupageref{sec:tld_rep_results}{text:abused_new_gtlds}).
\item	\textbf{European Union country-code TLDs (EU ccTLDs) are by far the least abused} in absolute terms, relative to their overall market share. Only 0.8 percent of all abused (compromised and maliciously registered) domain names were registered under EU ccTLDs \eupageref{sec:general_dataset}{fig:tld_distribution}.
\end{enumerate}

\item \textbf{Malicious vs. compromised domains: where does the abuse occur?}
\begin{enumerate}[label=(\alph*)]
    \item	The vast \textbf{majority of spam and botnet C\&C domain names are maliciously registered}, which is expected given the nature of the abuse \eupageref{sec:tld_statistics_mal_vs_com}{fig:malcompabuse}.
\item	In the analysed data, \textbf{about 25\% of phishing domain names and 41\% of malware distribution domain names are presumably registered by legitimate users, but compromised at the hosting level}. In these cases, trying to address abuse at the DNS level can be counterproductive, as it can cause collateral damage to their legitimate users \eupageref{sec:tld_statistics_mal_vs_com}{fig:malcompabuse}.
\item	When looking at compromised domain names, it emerged that \textbf{for highly used TLDs such as European ccTLDs, there is a higher incidence (42\%) of hacked websites}. In TLDs with lower usage rates such as new gTLDs, attackers have a much stronger tendency to register directly the domains they intend to use for their malicious activities \eupageref{sec:tld_statistics_mal_vs_com}{fig:malcomptld}.
\item	TLD registries and registrars can \textbf{prevent malicious registrations (proactive measures) and mitigate maliciously registered domains (reactive measures) at the DNS level}. However, they have no control over the hosting infrastructure (unless they also provide a hosting service). Therefore, the authors have computed reputation metrics for domain names that are found to be maliciously registered and exclude domains that are likely compromised at the hosting level \eupageref{sec:regrep}{sec:regrep}.
\item	The \textbf{top five most abused registrars account for 48\% of all maliciously registered domain names} \eupageref{sec:regrep_results}{text:registrar_mal_reg}.
\item	The study reveals \textbf{hosting providers with a disproportionate concentrations of spam domains} reaching 3,000 abused domains per 10,000 registered domain names \eupagerefrange{sec:reghos_results}{sec:reghos_results}{text:tld_rep_max_end}.
\item	\textbf{Phishers make heavy use of free subdomain and hosting providers} because they incur no cost, which makes them practical for serving malicious content. These services are less suitable for distributing spam and botnet C\&C communication \eupagerefrange{sec:special}{sec:special}{tab:meanmedianspecial}.
\end{enumerate}
\item 	\textbf{Adoption of DNS security extensions and mail protection protocols}
\begin{enumerate}[label=(\alph*)]
    \item	The overall level of \textbf{DNSSEC adoption remains low}. In a large sample of 227 million domain names, only 9.4 million domains have all the required DNSSEC resource records (DNSSKEY, RRSIG and DS). 98.1\% of these are correctly signed and have been correctly validated \eupageref{sec:dnssec_measurements}{text:dnssec_correctly_signed}.
\item	As for EU ccTLDs, \textbf{.cz (59\%), .se (55\%), .nl (51\%), and .sk (48\%) have the highest percentage of domain names signed with DNSSEC}. The registry operators of these domains provide price incentives and technical support for DNSSEC adoption \eupagerefrange{sec:dnssec_measurements}{text:best_dnssec_eucctld}{text:best_dnssec_eucctld_end}.
\item	The measurements revealed \textbf{2.5 million open DNS resolvers} worldwide that can be effectively used as amplifiers in Distributed Denial-of-Service attacks \eupageref{sec:dns_resolvers_discussion}{sec:dns_resolvers_discussion}.
\item	In large sample of 247 million domain names, more than \textbf{60\% of domain names are without SPF and 97\% of domains are without DMARC records} that prevent email spoofing, one of the techniques used in Business Email Compromise scams \eupagerefrange{sec:spf_dmarc_results}{sec:spf_dmarc_results}{text:spf_dmarc_results_end}.
\end{enumerate}
 \end{enumerate}

\section*{Recommendations}

Based on the analysis and measurements, the authors propose the following set of recommendations. 

\noindent
\begin{enumerate}
\item{\textbf{Better DNS Metadata (for identifying resources and their attribution to intermediaries)}} 

\begin{enumerate}[label=(\alph*)]
\item	Likewise gTLDs, ccTLDs  registries should provide a \textbf{scalable and unified way of accessing complete registration (WHOIS) information} (in compliance with data protection laws), using the Registration Data Access Protocol (RDAP), necessary to attribute abused and vulnerable domain names to their respective registrars and obtain their contact information \eupageref{sec:whois}{text:whois_rec}. 
\item	In the same manner as gTLDs, ccTLDs  registries should consider \textbf{publishing DNS zone file data} through DNS zone transfer or a system similar to the Centralized Zone Data Service maintained by ICANN \eupageref{sec:general_dataset}{text:czds}.
\end{enumerate}

\noindent
\item{\textbf{Contact Information and Abuse Reporting}}

\begin{enumerate}[label=(\alph*)]
\item	The \textbf{email addresses of registrants and domain name administrators} that are not visible in the public WHOIS could be displayed as \textbf{anonymized email addresses}   to ensure privacy and the ability to contact domain owners and administrators directly to notify security vulnerabilities and abuses \eupageref{sec:eee_disc}{text:whois_anonym_rec}.
\item	With no direct contact with domain name registrants and administrators via the public WHOIS database, domain name administrators should \textbf{maintain RFC 2142 specific email aliases} for given domain names (e.g., abuse, hostmaster, webmaster) and an email in the DNS SOA record so that they can be contacted directly in the event of vulnerabilities and domain name abuse \eupageref{sec:eee_disc}{text:whois_rfc_email}.
\item	A \textbf{standardized (and potentially centralized) system for access to registration data (WHOIS data)} should be set up, identifying the minimum information necessary to process disclosure requests. The reaction time to such requests shall be clearly defined.

\item	The study also recommends to set up a \textbf{standardized (and potentially centralized) system for abuse reporting}, identifying the minimum information necessary to process such report. The receipt of abuse reports is to be acknowledged. The reaction time to such reports shall be clearly defined and the abuse reporter should be provided with information on the actions taken. The DNS service providers shall provide for an appeal proceeding against their decisions to a third neutral party. 

\item We encourage the exchange of information on threats between parties involved (e.g., CERTs, security organizations) using \textbf{collaborative platforms such as Malware Information Sharing Platform (MISP)} to report and mitigate abuse in a more effective and timely way.
\end{enumerate}

\noindent
\item{\textbf{Improved Prevention, Detection and Mitigation of DNS Abuse Type~1}}

\begin{enumerate}[label=(\alph*)]
\item TLD registries, registrars, and resellers should \textbf{verify the accuracy of the domain registration (WHOIS) data}. The identification of the registrants could be implemented through possibly harmonised Know Your Business Customer (KYBC) procedures. In case of registrants from the EU, KYBC could be carried out through eID authentication in accordance with the eIDAS Regulation, as amended by the forthcoming Regulation on the European Digital Identity. KYBC procedure shall use cross-checks in other publicly available and reputed databases \eupageref{sec:tld_rep_results}{text:kybc_rec}.

\item TLD registries are encouraged to \textbf{develop or improve existing similarity search tools or surveillance services} to enable third-parties to identify names that could potentially infringe their rights \eupageref{sec:regrep_results}{text:tld_ipr_rec}.

\item	TLD registries are encouraged to offer, directly or through the registrars/resellers, \textbf{services allowing intellectual property rights (IPR) holders to preventively block infringing domain name registrations} (similar to services already existing on the gTLD market).

\item	The use of \textbf{predictive algorithms to prevent abusive registrations} by TLD registries and registrars is also encouraged.
\item	The study recommends that the \textbf{abuse rates of TLD registries or registrars be monitored} on an ongoing basis by independent researchers in cooperation with institutions and regulatory bodies (e.g., ICANN, European Commission, European Union Agency for Cybersecurity – ENISA or national authorities). Abuse rates should not exceed predetermined thresholds. If thresholds are exceeded and the abuse rates do not improve within a given time period, accreditation may be revoked \eupageref{sec:tld_rep_results}{text:reg_monitor_rec}.
\item \textbf{TLD registries and registrars with lower abuse rates may be financially rewarded}, e.g., through a reduction in domain registration fees, to align economic incentives and raise barriers to abuse \eupageref{sec:tld_rep_results}{text:reg_reward_rec}.
\item	TLD registry operators are encouraged to:
\begin{itemize}
\item	\textbf{maintain access to existing domain/URL blacklists},
\item	identify the \textbf{registrars with the highest and lowest concentrations and rates of DNS abuse} in their ecosystems,
\item	propose incentive structures to \textbf{encourage their registrars to develop methods to prevent and mitigate malicious registrations} effectively \eupageref{sec:regrep_results}{text:tld_blacklist_rec}.
\end{itemize}
\end{enumerate}
\noindent
\item{\textbf{Improved Detection and Mitigation of DNS Abuse Type 3}}

\begin{enumerate}[label=(\alph*)]
\item In a similar way with respect to the TLD registries and the registrars, the \textbf{abuse rates of hosting providers should be monitored} on an ongoing basis by independent researchers in cooperation with institutions and regulatory bodies (e.g., European Commission, European Union Agency for Cybersecurity – ENISA or national authorities). Abuse rates should not exceed predetermined thresholds. Incentive structures should be studied to induce hosting providers to develop technical solutions that effectively curb hosting and content abuse \eupageref{sec:reghos_results}{text:monitor_abuse_rate_rec}.
\item	Since free services (e.g., free hosting and subdomains) are commonly exploited in phishing attacks, their operators should \textbf{employ advanced prevention and remediation solutions to quickly curb abuses of subdomain names and hosting infrastructure}. They should proactively detect suspicious domain names containing keywords of the most frequently targeted brands and names and work closely with the most heavily attacked companies and develop trusted notifier programs \eupageref{sec:special}{text:free_service_rec}.
\end{enumerate}

\noindent
\item{\textbf{Better Protection of the DNS Operation and Preventing DNS Abuse Type 2}}

\begin{enumerate}[label=(\alph*)]
\item	Similarly to gTLD registries, registry operators of ccTLDs should be required to \textbf{sign TLD zone files with DNSSEC} and facilitate its deployment according to good practices \eupageref{sec:dnssec_measurements}{text:cctld_signed}.
\item	To facilitate the implementation of DNSSEC, domain administrators (registrants) should have \textbf{easy access to DNSSEC signing of domain names within the TLD}. TLD registries  should require all registrars that offer domain names in the TLD to support DNSSEC signing for registrants \eupageref{sec:dnssec_measurements}{text:dnssec_rec}.
\item	As an incentive to the deployment of DNSSEC, TLD registries might \textbf{offer discounts for DNSSEC-signed domain names} \eupageref{sec:dnssec_measurements}{text:dnssec_discounts}.
\item	Internet Service Providers (ISP) that operate DNS resolvers should \textbf{configure DNSSEC validation} to protect end users from cache poisoning attacks and ensure the integrity and authenticity of domain name resolutions \eupageref{sec:dns_resolvers}{text:isp_dnssec_rec}.
\item	National CERT teams should subscribe to data sources that identify open DNS resolvers. National governments and Computer Emergency Response Team (CERT) teams should intensify \textbf{notification efforts to reduce the number of open DNS resolvers} (and other open services), which are among the root causes of distributed reflective denial-of-service (DRDoS) attacks \eupageref{sec:dns_resolvers_discussion}{text:cert_open_res_rec}.
\item	Security community should intensify efforts to continuously \textbf{measure the adoption of SPF and DMARC} protocols, especially for high risk domain names and raise awareness of the domain spoofing problem among domain owners and email service providers. Correct and strict SPF and DMARC rules can mitigate email spoofing and provide the first line of defence against Business Email Compromise scams \eupageref{sec:spf_dmarc_discussion}{text:sec_spf_dmarc_rec}.
\item	Network operators should deploy \textbf{IP Source Address Validation} not only for outgoing but also for incoming traffic at the edge of a network. It provides an effective way of protecting closed DNS resolvers from different external attacks against DNS infrastructure, including possible zero-day vulnerabilities within the DNS server software \eupageref{sec:isav}{text:sav_rec}.
\end{enumerate}

\noindent
\item{\textbf{DNS Abuse Awareness, Knowledge Building, and Mitigation Collaboration}}

\begin{enumerate}[label=(\alph*)]
\item At the EU level, the study recommends the harmonisation/approximation of the practices of EU ccTLDs by the \textbf{adoption of the good practices} available at the European and international levels.
\item	The study recommends to require the \textbf{DNS service providers to collaborate with EU and Member States’ institutions, law enforcement authorities (LEA) and so-called trusted notifiers or trusted flaggers}. Where informal collaborations exist, they are to be further strengthened and formal processes are to be set up for the parties to interact.

\item	The study encourages \textbf{awareness-raising and knowledge-building activities} to make the consumers, IPR holders, or other affected parties aware of existing measures tackling DNS abuse.
\item	The study encourages \textbf{knowledge-sharing and capacity-building} activities between all intermediaries and stakeholders involved in the fight against DNS abuse.
\end{enumerate}

\end{enumerate}

\newpage
\part{Background}
\label{chapter:background}
\rhead{Background}
\pagestyle{fancy}

The Domain Name System (DNS), along with the IP protocol, is a vital service of the Internet, mapping applications, hosts, and services from names to IP addresses.

ICANN delegates responsibility to registry operators to maintain an authoritative source for registered domains within the TLD (e.g., Donuts is the registry for the .lawyer TLD). TLDs are divided in gTLDs and ccTLDs, the first group is generic, the second group is reserved to countries, territories and geographical locations identified in the ISO 3166-1 country codes list.
The governance of the gTLD namespace by ICANN is contractual.
The management of the ccTLD namespace varies from informal to formal contracts between some countries or territories and ICANN. 

Domain registrars manage the registration of domain names. They are generally accredited by TLD registries and can be accredited by ICANN. Domain resellers are third-party organizations that offer domain name registration services through a registrar but may not be accredited by a TLD registry or ICANN. Registrants are individuals or organizations that register domain names through domain registrars or resellers for a specific time period.

DNS providers operate zone files and authoritative DNS servers that map domain and hostnames to the corresponding IP addresses. While registrars usually maintain zone files, registrants may choose to delegate the responsibility to third-party authoritative DNS services such as Cloudflare or their own servers.

Web hosting providers maintain the server infrastructure used to host content for a given domain. Hosting providers may sell their services to individuals or other Web hosting providers---hosting resellers. Web hosting providers may offer so-called managed hosting. A hosting company handles the setup, management, and support of a server and/or applications (such as content management systems). Often, multiple (sometimes thousands) domain names might be hosted on the same physical server sharing the same IP address. Hosting providers may also offer unmanaged hosting, for example, a dedicated server with, for example, only an operating system installed. A user (webmaster) needs to install all the necessary software and keep the software up-to-date. 

Note that the registrant/webmaster may choose to buy a reverse proxy service that can hide the characteristics of an origin (i.e., backend) server  such as its IP address. Finally, ISPs (access providers) typically maintain recursive DNS resolvers that resolve domain names on behalf of the end user's computer willing to access an application, a host, or a service associated with the domain name. Note that the same entity can provide different services (e.g., it is common that registrars offer authoritative DNS services and Web hosting plans).

 Because the DNS encompasses a large ecosystem of different types of intermediaries that maintain the technical DNS infrastructure and hosting, the role of intermediaries in addressing abuse depends on both the type of abuse and the services they provide. 
 In this sense, we should first formulate a comprehensive definition of DNS abuse and then detail the types of abuse to understand better the role of each intermediary involved in the DNS abuse mitigation process.

\section{Definition of DNS Abuse}

To start with, we propose the following definition of DNS Abuse:

\begin{quote}
\textbf{\emph{Domain Name System (DNS) abuse is any activity that makes use of domain names or the DNS protocol to carry out harmful or illegal activity.}}
\end{quote}

DNS abuse consists of three possibly overlapping types:

\begin{itemize}[leftmargin=+1.8cm]
    \item[\textbf{Type 1:}] 
    Abuse related to \textbf{maliciously registered domain names}.

  \item[\textbf{Type 2:}] 	Abuse related to the \textbf{operation of the DNS and other infrastructures}.
  
  \item[\textbf{Type 3:}] Abuse related to \textbf{domain names distributing malicious content}.\footnote{This type of abuse may take advantage of maliciously registered or compromised domain names.}
\end{itemize}

The three types also differ in terms of which relevant entities are responsible and/or best positioned to put in place mitigation measures:

\begin{enumerate}

    \item 	\textbf{Abuse related to maliciously registered names (Type 1)} is usually best addressed by domain resellers, registrars, and TLD registries with the following proper remediation path:

\textbf{\texttt{Domain Reseller $\rightarrow$ Registrar $\rightarrow$ TLD Registry (at the DNS level)}}

\item 	\textbf{Malicious content} can be distributed using a \textbf{maliciously registered domain name} (Type 1 and 3) or it can be distributed using a compromised website (Type 3) where the domain under which the malicious content is made available is registered by an unaware third-party, which uses it legitimately.

\begin{enumerate}

    \item In case of \textbf{illegal/harmful content distributed using a maliciously registered domain name (Type 1 and 3)} (e.g., typosquatting domain name serving phishing content), the following remediation path is to be followed to effectively mitigate this abuse: 

\textbf{\texttt{Hosting Reseller $\rightarrow$ Hosting Provider (at the hosting level)}}

and

\textbf{\texttt{Domain Reseller $\rightarrow$ Registrar $\rightarrow$ TLD Registry (at the DNS level)}}

Mitigating the abuse only at the hosting or DNS level will prevent access to malicious content but will not block all elements of the malicious infrastructure. Therefore, actors at both levels have to be involved in the mitigation of this kind of abuse.

\item While it is also possible for the reseller (if any) / registrar / TLD registry to take action in case of \textbf{malicious content hosted on compromised websites} (Type 3), addressing abuse at the DNS level can be counterproductive, as it can cause collateral damage to legitimate registrants and website visitors. In this case, the site operator, hosting provider (and where it exists, its reseller) are well positioned to take action to curb the abuse. Remediation path is as follows:

\textbf{\texttt{Site operator $\rightarrow$ Registrant (if different from site operator) $\rightarrow$ Hosting reseller (if any)
$\rightarrow$  Hosting provider (at the hosting level)}}

Mitigating abuse at the hosting level includes removing malicious content from the hacked website and patching the vulnerability. Site operators are best positioned to mitigate  abuse in case of so-called unmanaged dedicated servers that they are in complete control and are responsible for their hosting servers and software. Hosting companies are best positioned to mitigate  abuse in case of so-called managed shared hosting as they maintain the operating system and application infrastructure.
\end{enumerate}

\item 	All entities related to the DNS infrastructure (registrars, registries, resellers, operators of authoritative name servers, and DNS resolvers) are concerned with \textbf{abuse related to DNS operations} (Type 2). This type of abuse is to be addressed at the DNS level.

\end{enumerate}

To illustrate the types of DNS abuse and responsible entities, we give below the examples of DNS abuse cases:

\begin{itemize}

    \item
    maliciously registered domain names serving phishing content: 
    \begin{itemize}
    \item Type 1 and 3 $\rightarrow$ mitigation action at the DNS level and hosting level
    \end{itemize}

 \item	compromised websites serving phishing content:     \begin{itemize}
    \item Type 3 $\rightarrow$ mitigation action at the hosting level
\end{itemize}
 \item	compromised websites used to distribute (deliver) malware: 
     \begin{itemize}
    \item Type 3 $\rightarrow$ mitigation action at the hosting level
\end{itemize}
 \item	maliciously registered domain name used to distribute (i.e., deliver) spam---emails containing malicious content: 
      \begin{itemize}
    \item Type 1 $\rightarrow$ mitigation action at the DNS level
    \end{itemize}
\item	maliciously registered domain name (e.g., algorithmically generated domain name - AGD) used for the malicious command-and-control (C\&C) communication (between compromised hosts and a malicious actor):
      \begin{itemize}
    \item Type 1 $\rightarrow$ mitigation action at the DNS level
  \end{itemize}
 \item	file sharing system abused to distribute child sexual abuse material (CSAM):     \begin{itemize}
    \item Type 3 $\rightarrow$ mitigation action at the hosting level
\end{itemize}
 \item	maliciously registered domain name used to distribute CSAM: 
     \begin{itemize}
    \item Type 1 and 3 $\rightarrow$ mitigation action at the DNS level and hosting level
\end{itemize}
 \item	DDoS attack against a DNS server (e.g., NXNSAttack): 
     \begin{itemize}
    \item Type 2 $\rightarrow$ mitigation action at the DNS level
\end{itemize}
 \item	DDoS attack against a web server using DNS open resolvers as amplifiers/reflectors: 
     \begin{itemize}
    \item Type 2 $\rightarrow$ mitigation action at the DNS level
\end{itemize}
 \item	hijacked domain name (e.g., using zone or cache poisoning) 
 \begin{itemize}
    \item Type 2 $\rightarrow$ mitigation action at the DNS level
    \end{itemize}
\end{itemize}

The rationale of this approach resides in taking into account the perspectives of different actors along the entire abuse chain. Indeed, the following actors should be taken into consideration. 
Concerning abuse involving domain names, the following actors should be considered: 
\begin{itemize}
    \item the \textbf{abuser/attacker  } -- the registrant of the maliciously registered domain name or the actor compromising legitimately registered domain names (for example, by exploiting vulnerable websites),
    \item the \textbf{abused party} -- the Internet users and/or third parties affected by the abuse causing physical, psychological, or economic harms, such as minors in case of child sexual abuse material (CSAM), consumers that are the victims of online frauds, the owners of intellectual property rights, etc. 
    \item the \textbf{intermediaries} -- the DNS operators (notably registries and registrars) and information society service providers
    (ISSPs)\footnote{\url{https://eur-lex.europa.eu/legal-content/EN/TXT/PDF/?uri=CELEX:52016DC0872&from=DE }}, including providers of hosting, access, and online platforms operators, as well as regular Internet users of the misused infrastructures that facilitate the distribution of illegal content. They should also be considered as victims (unless they are willingly facilitating cybercrime) because DNS infrastructure and content abuse affect their reputation and impose economic costs related to abuse handling. At the same time, this third group of actors plays a key role in effective abuse prevention and mitigation.
\end{itemize}

Recent definitions of DNS abuse\footnote{\url{https://dnsabuseframework.org/media/files/2020-05-29_DNSAbuseFramework.pdf}}$^,$\footnote{\url{https://www.internetjurisdiction.net/uploads/pdfs/Papers/Domains-Jurisdiction-Program-Operational-Approaches.pdf}} generally categorize phishing, botnet C\&C, malware, pharming, and spam (when used as a delivery mechanism to other forms of abuse) as DNS ``technical'' abuse and therefore the registries and registrars ``must'' act upon this category. On the other hand, child abuse material, controlled substances, regulated goods, or intellectual property are considered website content abuses.

In practice (and depending on the type of abuse), there is a great deal of overlap between DNS technical and content-related abuse.
For example, an attacker may register a domain name to launch a phishing campaign to deceive potential victims into disclosing passwords to their bank accounts.
The fake website may use the official logo of the bank to look more trustworthy.
Therefore, we deal with content abuse (phishing of credentials, trademark, and copyright infringement of the bank) and DNS technical abuse (domain name is maliciously registered).
Both a hosting provider and a DNS service operator (i.e., registrar) have to react in such a case.
As long as the hosting provider removes the malicious content, but the malicious domain is not suspended, the attacker can purchase another hosting service from another provider and reuse the maliciously registered domain name. 
Let us assume that only the registrar (or a TLD registry operator) removes the domain name from the zone file. In that case, the attacker may reuse the hosting and register a new domain name with another operator to continue malicious operations.

In the example given, mitigating the abuse at the hosting or domain name level interrupts the malicious actions of the attacker. However, mitigating the problem at both hosting and DNS levels is required because the attacker abused hosting and DNS technical infrastructures.
Such an approach also leads to an increased cost for the attacker and thus creates higher barriers to DNS abuse.

Consider another example of an attacker that compromises a legitimate website using a vulnerable content management system and uploads malware to distribute it and infect end users.
This example represents content abuse (distributing malicious software), but it is not the abuse of the DNS infrastructure. The domain name is registered by a benign user and abused (the site is hacked). In this case, the malicious content should be removed and the hosting infrastructure (vulnerability) patched by the webmaster or hosting provider (depending on whether the hosting is managed or unmanaged).

Generally, in such cases, DNS service providers should not intervene at the DNS level, since suspending a benign domain name can cause collateral damage and disrupt legitimate activities of the domain owner and its users.
However, certain types of abuse, such as online distribution of CSAM materials or human trafficking, can cause serious psychological and physical threats.
Even if the DNS operator determines that the malicious user is not abusing the DNS infrastructure, 
the DNS operator must inform about the abuse the intermediaries involved in hosting (website operator, domain owner, and hosting provider). The DNS operator should also have the legal authority to temporarily suspend the domain if the content is not removed promptly.
If, on the other hand, the DNS service operator concludes that, for example, CSAM material is being distributed using a maliciously registered domain name (DNS infrastructure abuse), the domain must be suspended as well as the content removed by the hosting provider operating the infrastructure.

From the perspective of the domain name holder who registered a domain with malicious intent, all kind of abuses are prohibited and no distinction is made between ``technical'' and ``content-related'' abuses (see ICANN New gTLD Registry Agreement Specification 11 3a ``\textit{prohibiting Registered Name Holders from distributing malware, abusively operating botnets, phishing, piracy, trademark or copyright infringement, fraudulent or deceptive practices, counterfeiting or otherwise engaging in activity contrary to applicable law}''; ICANN Registrar Accreditation Agreement 1.13 \textit{``Illegal Activity'' means a conduct involving use of a Registered Name sponsored by Registrar that is prohibited by applicable law and/or exploitation of Registrar's domain name resolution or registration services in furtherance of conduct involving the use of a Registered Name sponsored by Registrar that is prohibited by applicable law"}).

From the perspective of the abused or affected party, the focus is on the harm suffered and not on the categorisation whether the origin of the harm is of the ``technical'' and/or ``content-related'' nature.

Moreover, in many cases such a clear-cut distinction (technical vs. content-related)  is not possible at high level and the borderline is blurred due to the great deal of overlap between different types of abuse. Recent definitions of DNS abuse generally categorise, for example, phishing as ``technical'' abuse and therefore the registries and registrars ``must'' act upon this category.\footnote{\url{https://dnsabuseframework.org/media/files/2020-05-29_DNSAbuseFramework.pdf }} However, previous research \cite{comar} showed that in the sample of manually labelled phishing domains (gathered from blacklisted URLs), 58\% of domaines 
were registered by seemingly malicious actors indicating DNS technical abuse, whereas all of the URLs served abusive content affecting, in the first place, Internet users (by tricking them into revealing sensitive personal or financial information), but also third parties (by incorporating well-known trademarks in the phishing websites), and finally, ISSPs whose infrastructure was misused to host malicious content. 

The remaining 42\% of domains were seemingly compromised, meaning that benign registrants registered the underlying domains, but an abuser most probably exploited vulnerable web hosting. Therefore, the registries and registrars should respond (by forwarding the abuse complaint to another intermediary such as the hosting provider) rather than being obliged to react at the DNS level because these domains do not directly abuse DNS infrastructure (unless the benign domain name was compromised at the DNS level, e.g., it was hijacked using zone poisoning\cite{zone}). 

URLs used to distribute malware is another example indicating that the clear-cut distinction between ``technical'' and ``content-related'' is not appropriate. Malware distribution URLs serve harmful content (malicious software) to infect end users. 
In the sample of manually labeled malware URLs \cite{comar}, as many as 57\% of domains were compromised \cite{comar} by exploiting, for example, web vulnerabilities, and therefore again, in most cases they do not directly abuse the DNS technical infrastructure. Therefore, registries and registrars should respond (following the principle of victims first) by forwarding the abuse complaint to the hosting provider rather than must react at the DNS level. Moreover, the recent definitions of DNS abuse generally categorise domains used to host websites offering, for example, counterfeit goods, pirate content, or CSAM material as  ``content-related'' abuse, and therefore, fall outside the recently proposed definitions of DNS abuse. However, similarly to phishing or malware, the abusers may use the DNS infrastructure and in particular, maliciously registered domain names to distribute such content. In such cases, the registries and registrars  as well as content providers must react to these types of abuse.

\section{Role of Intermediaries in Abuse Handling}

This complex ecosystem requires a \textbf{bottom-up}  approach in handling DNS abuse. In respect to DNS abuse involving domain names, such as phishing, malware, IPR infringement, CSAM, etc., the DNS intermediary that detects or is notified about abuse must first assess if a given incident is related to DNS infrastructure and/or content abuse, identify and inform an appropriate party that might be in a better position to make such assessment, and address abuse.
 
Let us assume that a DNS operator (e.g., registry or registrar) receives an abuse notification and concludes that the domain name is registered for malicious purposes based on the collected evidence or the evidence provided by the (trusted) notifier. In this case, the domain name must be blocked by the DNS service operator, i.e., registry, registrar, domain reseller, or authoritative DNS service provider (if different from the registry or registrar) according to applicable policies. Assume that, in addition, the domain name reveals illegal/harmful content, i.e., is used to distribute malware, hosts CSAM material, or a phishing website. In this case, the DNS service operator should, in addition to the takedown at the DNS level, identify and contact a hosting provider using, for example, WHOIS information. The domain name must be blocked within the period specified by applicable laws, but also, the hosting is suspended. Otherwise, as mentioned earlier, after the domain name suspension, the attacker may register another domain name and map the newly registered domain name to the operational hosting service. The hosting operator must suspend the hosting or (if not possible) contact the responsible hosting reseller that must suspend the service. Note that if the domain name uses a reverse proxy service (e.g., Cloudflare), the proxy provider must suspend the service and contact the hosting provider of the back-end infrastructure, which must suspend the hosting service.
 
On the other hand, if a DNS operator receives an abuse notification and concludes, based on the collected evidence, or evidence provided by the notifier, that the domain name is legitimate but compromised (hacked), the DNS operator generally should not suspend the domain name.
The notified DNS service operator, i.e., registry, registrar, domain reseller, or authoritative DNS service provider (if different from the registry or registrar), should contact the hosting provider.  A provider should not suspend the hosting server (especially a shared hosting server) but it should block access to a malicious website used to host or facilitate the distribution of illegal content. The hosting provider must contact the webmaster (possibly the domain owner) and inform about the incident. For unmanaged services, the hosting provider must contact the webmaster directly to mitigate abuse. If the hosting provider is not in a direct relationship with the end user, it must identify the hosting reseller, who must take appropriate steps.

\newpage
\setstretch{1.0}
\part{Datasets}
\setstretch{1.5}
\rhead{Datasets}

In this part, we present the datasets as well as passive and 
active measurement methods used in the analysis of the scope and magnitude of DNS 
Abuse. 

We first leverage the list of abused domain names and URLs (Section~\ref{sec:blacklists}) and the complete list of all domain names for certain TLDs, or the large sample of domains for the TLDs that do not make their zone files available (Section \ref{sec:general_dataset}). We collect the registration information of the active domain names (Section \ref{sec:whois}), collect DNS records to evaluate the deployment of security technologies (Section \ref{sec:dnsscans}) and finally regularly collect Web content, DNS and WHOIS information for blacklisted domain names/URLs to assess the time required to mitigate abuse (Section \ref{sec:uptimemeasure}).

\section{Blacklists\label{sec:blacklists}}

To estimate the prevalence and persistence of DNS infrastructure and content abuse, we use \textbf{sixteen} distinct blacklists generously provided to us by six blacklist providers: Spamhaus~\cite{spamhaus},  SURBL~\cite{surbl}, ABUSE.ch \cite{abuse.ch}, the Anti-Phishing Working Group (APWG) \cite{apwg}, PhishTank \cite{phishtank}, and OpenPhish \cite{openphish}.
They represent URL, fully-qualified domain name (FQDN)/IP address or domain name blacklists of spam, malware distribution, command-and-control (C\&C), phishing, and IPR infringement. The blacklists are typically used by different intermediaries such as Internet Service Providers (ISPs) in operational environments to block malicious communications.

\begin{itemize}
  \item \textbf{Spamhaus} Domain Block List (DBL) provides malicious domains obtained from URLs enumerated in spam email payloads, senders of spam emails, known spammers, phishing, malware-related websites, or suspicious domain names sharing common patterns with domains involved in DNS infrastructure and content abuse \cite{dbl}. We use eight data feeds provided by Spamhaus grouped based on the type of malicious activity the domain is associated with\footnote{\url{https://www.spamhaus.org/faq/section/Spamhaus\%20DBL\#291}}:
  \begin{itemize}
    \item \textbf{Spam (SP)}: is a group of seemingly maliciously registered domains used for spam distribution (i.e., for sending unsolicited emails).
        
    \item \textbf{Phishing (PH)}: is a group of domains seemingly registered by malicious actors and used for phishing attacks.
    \item \textbf{Malware (MW)}: is a group of domains seemingly registered by malicious actors and used for distributing malicious software and infecting end users to gain access to private computer systems and sensitive information.
    \item \textbf{Botnet command-and-control (C\&C)}: is a group of seemingly maliciously registered domains used for  botnet C\&C communication between C\&C servers and compromised machines (bots).
    \vspace{+0.3cm}
    \item \textbf{Abused-legit spam (AL-SP)}: 
    is a group of domains that are mostly registered by benign users (businesses and individuals) but compromised and abused by malicious actors to distribute spam.
    \item \textbf{Abused-legit phish (AL-PH)}: is a group of domains that, similarly to abused-legit spam domains, are labeled as benign but compromised and abused by malicious actors in phishing attacks.
    \item \textbf{Abused-legit malware (AL-MW)}: is a group of domains registered by legitimate users but compromised and abused to distribute malware.
    \item \textbf{Abused-legit botnet C\&C (AL-C\&C)}: is a group of benign domains abused by malicious actors to perform botnet C\&C communication.

  \end{itemize}
  
  \item \textbf{SURBL} list is composed of domain names observed in unsolicited email messages and external blacklists~\cite{surbllists}. The combined list provides domain names categorized as phishing, malware, or spam. 
  
  \begin{itemize}
    \item \textbf{SURBL phishing (PH)} contains phishing domains supplied by several organizations including PhishTank, OITC phishing, PhishLabs, RiskAnalytics, and internal feeds maintained by SURBL. These blacklists collide maliciously registered domain names with domains of compromised websites abused to launch phishing attacks.
    
    \item \textbf{SURBL malware (MW)} contains data from multiple sources that cover domains used to host malware websites, payloads, or associated redirectors. Note that compromised domains may also be included in the list since there is no distinction between compromised websites and maliciously registered domain names.
    
    \item \textbf{SURBL abuse (SP)} comprises domain names involved in spam activity. The list is mostly based on SURBL internal research, however, ISPs, Email Service Providers (ESPs), and other entities also contribute to this list. It may contain both maliciously registered domain names and domains of compromised websites.

  \end{itemize}
  
  \item \textbf{APGW} contains phishing URLs submitted by accredited users through the eCrime Exchange (eCX) platform.\footnote{\url{https://apwg.org/ecx/}} Blacklisted URLs are accompanied by metadata, including the confidence level and the \textbf{target brand name}. There is no distinction between maliciously registered domains and compromised websites.
  
  \item \textbf{PhishTank} provides a community-based phishing verification system, which contains phishing URLs submitted and verified manually by its contributors as malicious. Blacklisted ULRs contain additional metadata such as the \textbf{target brand name} but no indication if the registered domain name is malicious or the website is compromised.
  
  \item \textbf{OpenPhish} feed contains phishing URLs (and \textbf{targeted brands}) identified by OpenPhish or reported to OpenPhish and then verified as phishing. The feed collides malicious registrations and compromised websites.
  
  \item \textbf{Abuse.ch} is a non-profit organization combating malware and helping ISPs and network operators protecting their infrastructure from malware activities. We use two following sources:

  \begin{itemize}
    \item \textbf{URLHaus}\footnote{\url{https://urlhaus.abuse.ch}} is a service provided by \texttt{abuse.ch} that publishes URLs (containing either domains or IP addresses) used for malware distribution.
    
    \item \textbf{ThreatFox}\footnote{\url{https://threatfox.abuse.ch}} is a platform from \texttt{abuse.ch} with the goal of sharing indicators of compromise (IOCs) associated with malware. The database contains URLs and domains used to communicate with command-and-control servers, the malware family and confidence level.
    
  \end{itemize}
  \end{itemize}

  \begin{table}[ht]
    \begin{center}
      \centering
      \resizebox{\textwidth}{!}{\begin{tabular}{ l c c c c c c}
        \toprule
        \thead{\textbf{Dataset}} & \thead{\textbf{Type}} & \thead{\textbf{\# Entries}} & \thead{\textbf{\# Entries}\\ \textbf{(with domains)}} & \thead{\textbf{\% Entries}\\ \textbf{(with domains)}} & \thead{\textbf{Unique}\\ \textbf{domains}} & \thead{\textbf{Period}} \\
        \midrule
        \textbf{SURBL}    & FQDN/IP & 797,162 & 595,383 & 74.69\% & 475,925 & Mar-June 2021 \\
        \multicolumn{2}{l}{\hspace{1 mm} SP} & 340,611 & 340,513 & 99.97\% & 338,363\\
        \multicolumn{2}{l}{\hspace{1 mm} MW}  & 205,839 & 8,933 & 4.34\% & 7,276\\
        \multicolumn{2}{l}{\hspace{1 mm} PH}  & 253,169 & 248,384 & 98.11\% & 132,838\\
        \midrule
        \textbf{Spamhaus} & Domain & 1,088,863 & 1,088,863 & 100.0\% & 1,087,770 & Apr-June 2021 \\
        \multicolumn{2}{l}{\hspace{1 mm} SP} & 946,068 & 946,067 & 100.0\% & 945,397\\
        \multicolumn{2}{l}{\hspace{1 mm} PH}  & 105,254 & 105,254 & 100.0\% & 105,025 \\
        \multicolumn{2}{l}{\hspace{1 mm} MW}  & 4,143 & 4,143 & 100.0\% & 4,143\\
        \multicolumn{2}{l}{\hspace{1 mm} C\&C}  & 6,798 & 6,798 & 100.0\% & 6,798\\
        \multicolumn{2}{l}{\hspace{1 mm} AL-SP}  & 25,166 & 25,165 & 100.0\% & 25,065\\
        \multicolumn{2}{l}{\hspace{1 mm} AL-PH}  & 6,147 & 6,147 & 100.0\% & 6,070\\
        \multicolumn{2}{l}{\hspace{1 mm} AL-MW}  & 4,258 & 4,258 & 100.0\% & 4,257\\
        \multicolumn{2}{l}{\hspace{1 mm} AL-C\&C}  & 139 & 139 & 100.0\% & 139\\
        \midrule
        \textbf{APGW}  & URL   & 248,457 & 248,432 & 99.99\% & 42,748 & Mar-June 2021 \\
        \midrule
        \textbf{OpenPhish} & URL & 177,330 & 170,055 & 95.9\% & 49,418 & Mar-June 2021 \\
        \midrule
        \textbf{PhishTank} & URL & 54,975 & 54,156 & 98.51\% & 22,717 & Mar-June 2021  \\
        \midrule
        \textbf{URLHaus} & URL & 349,834 & 14,085 & 4.03\% & 6,401 & Mar-June 2021  \\
        \midrule
        \textbf{ThreatFox} & URL & 3,724 & 2,930 & 78.68\% & 1,290 & Apr-June 2021  \\
        \bottomrule
      \end{tabular}}
    \end{center}
    \caption{Summary of blacklists including the total number of entries (FQDNs/IP, Domains, URLs), the number of entries including a domain name and its ratio in regard to the total number of entries, the number of unique domain names, 
    and the collection period per feed.}
    \label{feeds}
  \end{table}

  For this study, we collect newly blacklisted resources from March/April to June 2021 (second quarter of 2021) and gather relevant metadata related to registration or hosting infrastructure using active and passive methods. 
  
  Table \ref{feeds} shows the summary of sixteen datasets. The \textbf{Type} refers to one of three types of entries that the blacklists provide, i.e. FQDN/IP, Domain, or URL. The FQDN/IP type refers to fully-qualified domain names (e.g., \texttt{malicious.example.com}) or IP addresses used in attacks, whereas the Domain type refers to 2\textsuperscript{nd}–level domain names (e.g., \texttt{malicious.pl}) or sometimes 3\textsuperscript{rd}–level domains if a given TLD registry provides such registrations (e.g., \texttt{malicious.com.pl}). Finally, the URL type provides URLs that serves malicious content (e.g.,~\texttt{http://example.com.pl/paypal-account.php}). 
  The number of \textbf{Entries} corresponds to the identified abuse incidents (i.e., the number of unique blacklisted FQDNs/IPs, domains, or URLs). Note that the total number of entries for a given abuse feed (e.g., SURBL) might be different from the total number of entries from individual datasets combined (e.g., SURBL SP, SURBL MW, and SURBL PH) because some FQDNs 
  may appear in two or more blacklists (e.g., SURBL~PH and SURBL MW) at the same time.
      
  In this study, we analyzed over \textbf{2.7 M} incidents (almost \textbf{2.17 M involving domain names}) and \textbf{1.68 M abused domain names}.
  While for some types of attacks, such as phishing, malicious actors typically use domain names to induce the victim to reveal sensitive personal or financial information, other types of attacks, such as malware distribution, generally do not need to involve domain names to be successful. Therefore, Table \ref{feeds} presents the number and percentage of entries that involve (abuse) domain names (\textbf{Entries with domains}).  Note that SURBL~MW and URLHaus datasets, representing malware distribution incidents, contain \textbf{only 5\% and 7\% of entries with domain names}.
  Moreover, not all of those attacks  (involving domain names) abuse DNS infrastructure as both feeds collide maliciously registered and compromised domains (i.e., benign domain names whose websites have been compromised) or, in the case of URLHaus, free services such as online file storage services like Google drive.

 \section{Domain Names} \label{sec:general_dataset}
 To calculate the abuse rates per different actors involved in the domain name registration and hosting, and the deployment of DNS security technologies, we first need the list of domain names for each TLD. We use two sources of data: i) zone files whenever available, and ii) active web content crawling. 
 
 \begin{itemize}
     \item \textbf{DNS Zone Files}. For the purposes of the study, we collect .com, .net, .org, .biz, .tel, .info legacy gTLDs and new gTLDs made available to us by the ICANN Centralized Zone Data Service (CZDS),\footnote{\url{https://czds.icann.org/home}} as well as .se and .nu ccTLD zone  files.\footnote{\url{https://internetstiftelsen.se/en/domains/tech-tools/access-to-zonefiles-for-se-and-nu/}}
  Zone files provide the most complete list of registered domain names in each TLD and are therefore, the most accurate source of TLD sizes needed to  calculate abuse rates for individual TLDs reliably. 
  
  \item \textbf{Active Web Scans}. Apart from TLDs, for which we have access to their zone files through CZDS, to have a more exhaustive list of domains, especially for ccTLDs without having access to zone files, we develop the scanning platform to  crawl all the websites of the domains in our database actively (only scanning homepages looking for hyperlinks) to retrieve newly observed domain names.  
  
 \end{itemize}

Using the approaches, we can collect a total of \textbf{251,585,395} active domain names related to 1,376 distinct TLDs including new gTLDs (16,081,977 domains related to 1,057 distinct TLDs), legacy gTLDs (172,796,509 domains related to 22 distinct TLDs), European Union ccTLDs (26,280,324 domains related to 32 distinct TLDs), other European ccTLDs (10,516,931 domains related to 28 distinct TLDs), and non-European ccTLDs (25,909,654 domains related to 237 distinct TLDs).

Although our collection of domain names is not complete, it provides a representative sample of registered domains per TLD registry to evaluate the deployment of security technologies (DNSSEC, SPF, DMARC) that prevent DNS infrastructure abuse such as DNS cache poisoning attacks (DNSSEC) or domain/email spoofing (SPF, DMARC).

We also use the enumerated domain names to calculate sizes of domain registrars and hosting providers.
We further collect additional information about hosting infrastructure (by resolving the registered domains to their IP addresses and respective autonomous systems) and domain registrars (by collecting the registrar information using the WHOIS or RDAP protocols).
However, not having access to the full list of domain names is a limitation of this work: we do not have a complete picture of the market share of registrars and hosting providers, so we cannot precisely calculate their abuse rates. 
More generally, such data allows interested parties to conduct research and develop new insights into the security practices of hosting providers or domain registrars, verify their policies or create reliable reputation metrics.

On the other hand, the TLD size information, relevant to the calculation of abuse rates, does not necessarily involve the requirement of having access to the full list of domain names.
We use available zone files as the most accurate source of sizes.
If these are not available, we use the sizes of ccTLDs affiliated with the Council of European National Top-Level Domain Registries (CENTR),\footnote{\url{https://www.centr.org}} whose members  have explicitly agreed to make this information available for the purpose of this study.
For all other TLDs, we used the approximate sizes provided by DomainTools.\footnote{\url{https://research.domaintools.com/statistics/tld-counts}}

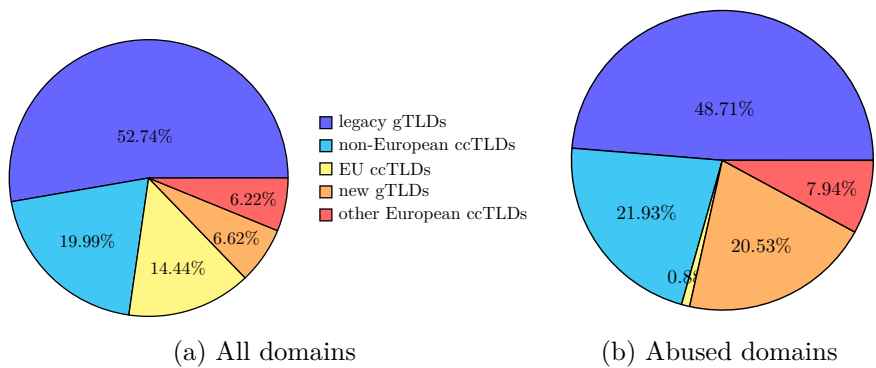
\begin{figure}[ht]
  \centering
  
  \subfloat[All domains]{
    \resizebox{0.3\textheight}{!}{  
      \begin{tikzpicture}
            \pie[
              text=legend
            ]{52.74/legacy gTLDs, 19.99/non-European ccTLDs, 14.44/EU ccTLDs, 6.62/new gTLDs, 6.22/other European ccTLDs}
        \end{tikzpicture}
    }
  }
  \subfloat[Abused domains]{
    \resizebox{0.3\textwidth}{!}{  
      \begin{tikzpicture}
            \pie[
            ]{48.71/, 21.93/, 0.88/, 20.53/, 7.94/}
        \end{tikzpicture}
    }
  }
  
  \caption{Divison of the domain namespace per TLD type}
  \label{fig:tld_distribution}
\end{figure}

Figure~\ref{fig:tld_distribution}a, shows the market share of five groups of TLDs: legacy gTLDs (53\%), new gTLDs (6\%), European Union ccTLDs (14\%), other European ccTLDs (7\%), and other non-European ccTLDs (20\%).
By comparing the market share with the distribution of blacklisted domains (Figure~\ref{fig:tld_distribution}b), we conclude that EU ccTLDs are the least abused in both absolute and relative terms to market share.
In relative terms, new gTLDs (with the estimated market share of 6.6\%) are the most abused group of TLDs (20.5\% of abused domain names).

\pagelabel{text:czds}As one of the safeguards introduced by ICANN intended to mitigate the abusive and criminal activity, all new gTLDs and some legacy gTLDs make their zone files available to third parties using the Centralized Zone Data Service (CZDS).
Some ccTLD registry operators make their zone files available and easily accessible to the public either voluntarily (e.g., .se),\footnote{\url{https://internetstiftelsen.se/en/domains/tech-tools/access-to-zonefiles-for-se-and-nu/}} or as a result of local regulations (e.g., .ch).\footnote{\url{https://www.switch.ch/de/open-data/\#tab-c5442a19-67cf-11e8-9cf6-5254009dc73c-3}} 
Both zones are available through DNS zone transfer (AXFR).\footnote{\url{https://zonedata.iis.se/}}$^,$\footnote{\url{https://www.switch.ch/de/open-data/}}
Some others make zone files available to the vetted parties (e.g., .nl), for example, for the research purposes\footnote{\url{https://openintel.nl/coverage}} and after signing a data sharing agreement.
However, the majority of ccTLD registry operators are against making their zone files available to third parties.
One argument is that open access to a list of domain names may have unforeseen negative consequences for their security and stability.
\newline

 \begin{tcolorbox}[enhanced,colback=blue!5!white,colframe=blue!75!black,colbacktitle=red!80!black]
 \textbf{Recommendation}: The registry operators of ccTLDs should consider publishing DNS zone file data through DNS zone transfer or a system similar to the Centralized Zone Data Service (CZDS) maintained by ICANN.
  \end{tcolorbox}

  \section{Registration  Information}\label{sec:whois}

  To calculate the reputation of each registrar, first, we have to calculate their sizes, i.e., the number of domain names registered with each registrar. By the term `registrar', we refer to the ICANN-accredited list of registrars.\footnote{\url{https://www.icann.org/en/accredited-registrars}} It is a challenging task since the only publicly available data source that can be queried and obtained at scale   is the WHOIS information of the registered domain names. Using our data collection method described in Section \ref{sec:general_dataset}, we enumerated 251,585,395 domain names   for different TLDs.
  
  Two general approaches are available for collecting WHOIS information at scale: i) the Registration Data Access Protocol (RDAP),\footnote{\url{https://datatracker.ietf.org/doc/html/rfc3912}} and ii) the WHOIS protocol.   Whenever the registry operator supports the RDAP protocol, we use it to collect registration information since the result of RDAP is in the JSON format (easy for machines to parse) and generally does not need an extra parsing step to extract the information.
  Occasionally, we observe cases in which RDAP implementations are not fully compliant with the RFC specification and recommended format.
  We informed one of the European ccTLD registries (also managing a new gTLD) about such a problem. We received a reply confirming the existence of the issue and assuring that the operator will fix it.
  At the time of writing, only 17 ccTLDs (such as .vg, .no, .cz) have support for RDAP.\footnote{\url{https://data.iana.org/rdap/dns.json}} Therefore, for most of the ccTLDs, we had to retrieve the data from a TLD WHOIS server where there is no specific and consistent format for serving data. 
Therefore, we parsed the collected data and used regular expressions to extract relevant fields, including the registrar name, IANA ID (for ICANN-accredited registrars only), and the registration date. The approach required extensive data processing as the formats differ among different entities.

In total, we collected WHOIS data for 241,618,279 (approximately 96\%) domain names in our list.
Note that for some TLDs, it is not feasible to collect WHOIS information since either there is no TLD WHOIS server (e.g., in case of .gr) or the access is restricted to authorized IP addresses (e.g., in case of .es TLD). 
Moreover, for some TLDs (e.g., .de), the information displayed in the public WHOIS is very limited. It does not contain the registrar name, IANA ID, administrative or abuse email address or the domain name creation date. Instead, the registry operator provides the web-based WHOIS service that cannot be queried at scale.
Access to the registration data \textit{at scale} via WHOIS or (preferably via) RDAP is essential for different security tasks. For example, large-scale vulnerability or domain name abuse notifications require mapping abused or vulnerable resources to their respective intermediaries, such as registrars, that are in the best position to mitigate specific vulnerabilities or can takedown maliciously registered domain names. 
It also requires collecting (at scale) e-mail information to contact registrars directly.
One limitation of this work is that we have no information about specific TLD ecosystems, such as for .es, .gr, or .de domain namespaces, for which the WHOIS information is not available, restricted, or provided only via the web-based service.
In these cases, we cannot map domain names (particularly maliciously registered) to relevant registrars in specific TLD ecosystems and calculate abuse rates.
\pagelabel{text:whois_rec}
\newline

\begin{tcolorbox}[enhanced,colback=blue!5!white,colframe=blue!75!black,colbacktitle=red!80!black]
 \textbf{Recommendation}:  Likewise gTLDs, ccTLDs registries should provide a scalable and unified way of accessing complete registration (WHOIS) information (in compliance with data protection laws), using the RDAP protocol, necessary to attribute abused and vulnerable domain names to their respective registrars and obtain their contact information.

 \end{tcolorbox}

After collecting and parsing WHOIS information, for each record, we use the following algorithm to identify the registrar name:

\begin{enumerate}
 \item if the IANA ID field exists and is valid, automatically extract the corresponding registrar name from the ICANN-accredited registrars list.
    \item in the absence of the IANA ID or the presence of an invalid value (not present in the ICANN list), extract the raw registrar name from the WHOIS record and match with the registrar names in the ICANN list to get the correct IANA ID. The second step required manual labeling.
\end{enumerate}
  
While the approach mentioned above provides us with the common registrar names, some cases still exist that cannot be handled in this way. For example, looking at Table~\ref{tbl_invalid_reg_names} for the Godaddy\footnote{https://godaddy.com/} registrar, there are three ICANN-accredited registrars with IANA ID 146, 3786, and 1659\footnote{Note that we do not merge entities whenever the IANA ID is different, even though they may belong to one parent company.} all with different registrar names in WHOIS data (the first column of the table). This case is the result of selling domain names by re-sellers or merging registrar companies to expand their businesses \cite{godaddy_merge}. For such cases, we manually labeled registrar names based on the ICANN list and checked the registrar websites. Our final list consists of 3,733 registrar names, which have 1,222 variants of registrar names for those in the ICANN-accredited list. 
\vspace{+0.4cm} 
\begin{table}[h]
     \centering
     \scalebox{0.7}{
         \begin{tabular}{|c|c|c|c|}
         \hline
         \#&Registrar name & ICANN Registrar name & IANA ID  \\ [0.5ex] 
         \hline\hline
         1 &Go Daddy, LLC & GoDaddy.com, LLC & 146  \\ 
         \hline
         2&GoDaddy.com, LLC & GoDaddy.com, LLC & 146  \\
         \hline
         3&GoDaddy & GoDaddy.com, LLC & 146  \\
         \hline
         4&GoDaddy.com & GoDaddy.com, LLC & 146  \\
         \hline
         5&Godaddy.com, LLC & GoDaddy.com, LLC & 146  \\
         \hline
         6&GoDaddy.com, LLC. & GoDaddy.com, LLC & 146  \\
         \hline
         7&GoDaddy.com LLC & GoDaddy.com, LLC & 146  \\
         \hline
         8&GoDaddy.com, Inc. & GoDaddy.com, LLC & 146  \\
         \hline
         9&GODADDY & GoDaddy.com, LLC & 146  \\
         \hline
         10&GoDaddy.com,LLC & GoDaddy.com, LLC & 146  \\
         \hline
         11&GoDaddy.com LLC trading as GoDaddy.com & GoDaddy.com, LLC & 146  \\
         \hline
         12&GoDaddy Corporate Domains LLC & GoDaddy Corporate Domains, LLC & 3786  \\
         \hline
         13&GoDaddy Corporate Domains, LLC & GoDaddy Corporate Domains, LLC & 3786 \\
         \hline
         14&GoDaddy Online Services Cayman Islands & GoDaddy Online Services Cayman Islands Ltd. & 1659 \\
         \hline
         \end{tabular}
     }
     \caption{Fourteen different registrar names in the WHOIS data for three ICANN-accredited registrars all belong to godaddy.com.}    
     \label{tbl_invalid_reg_names}
 \end{table}

While there are still many non-standard registrar names that we cannot reliably label (such as personal names or companies accredited by ccTLD registries locally), we can identify registrars for about 85\% of WHOIS records.

We also map each domain name found in an abuse feed to its corresponding registrar name using WHOIS records collected as soon as the domain name gets blacklisted. Having the size of each registrar and the number of the maliciously registered domain for that registrar, we can calculate the abuse rate for each of them.

\section{Active DNS Scans\label{sec:dnsscans}}

This section explains the method for collecting information regarding different types of DNS abuse, vulnerabilities, and security technologies.
We actively collect the following data: i) `A' resource records to calculate the reputation metrics for hosting providers, ii) information about the deployment of DNSSEC (`DS', `DNSKEY', and `RRSIG' resource records), iii) open DNS resolvers, and, iv) SPF and DMARC entires in `TXT' records to measure the deployment of email security extensions. 
  \begin{itemize}
    \item \textbf{DNS Resource Records: `A' Records and Geolocation}
    
To calculate the reputation of hosting providers, we need to collect all the `A' records of all the enumerated benign and malicious domain names using the open-source ZDNS\footnote{\url{https://github.com/zmap/zdns}} tool---a fast DNS resource record scanner. 
Using the MaxMind\footnote{\url{https://www.maxmind.com/}} database, we convert each `A' record to its corresponding AS number and a country code.

    \item \textbf{DNSSEC-related Records}
    
To measure the deployment of DNSSEC, we actively collect three DNS resource records for each domain in our database: `DNSKEY', `DS', and `RRSIG'. The presence of such records does not necessarily mean that domains are correctly signed but rather signifies that domain owners attempted to do so. If the `DNSKEY' and `DS' are present, we make an attempt to validate DNSSEC chain using our recursive (validating) resolver. If the validation succeeds, the domain name is correctly signed.  
    
    \item \textbf{Email Security Extensions: SPF and DMARC}
    
To measure the deployment of the Sender Policy Framework (SPF) and Domain-based Message Authentication, Reporting and Conformance (DMARC), we collect the `TXT' record of each registered domain (e.g., example.com) for SPF and the `TXT' record of the `\_dmarc' subdomain (e.g., \_dmarc.example.com). Note that the presence of the SPF and DMARC does not necessarily mean that the domain name is safe against email spoofing since these records should be configured correctly to protect the domain name. However, the absence of even one of these records makes the domain name vulnerable to email spoofing attacks~\cite{maroofi2020defensive}.

    \item \textbf{Identifying Open Resolvers} 
    
Various open services (such as DNS, NTP, Memcached, etc.) have long been known as efficient DDoS reflectors and powerful amplificators~\cite{amplificationHell,exitFromHell}. Open DNS resolvers accept DNS requests from any end host (instead of the predefined ranges of IP addresses). They can be used to either target authoritative nameservers by sending the excessive number of incoming requests or, if combined with IP address spoofing, used to redirect responses to victim end hosts (e.g., high-profile websites). 
Reducing the number of open resolvers will increase the barriers for cybercriminals to launch reflective DDoS attacks (DRDoS).

We actively scan for open DNS resolvers in IPv4 and IPv6 address spaces and analyze their distribution across organizations and countries.
      
Scanning for open resolvers requires sending DNS requests to end hosts and inspecting the received responses. The response codes (\texttt{RCODE}), defined in RFC~1035 \cite{rfc1035}, signal whether the DNS server is configured to process the incoming request. If the query resolution is successful, open resolvers send back the responses to end clients along with \texttt{NOERROR} status code. 

We use the following three datasets to scan for open resolvers: IPv4 BGP prefixes~\cite{routeviews}, IPv6 Hitlist Service~\cite{hitlist}, and IPv6 addresses enumerated in our previous study~\cite{korczyski2020closed}. All three datasets contain globally reachable IP addresses that may be operational recursive resolvers. Each end host from the list receives an `A' request for the unique domain name maintained by us. 
We developed software that allows us efficiently sending DNS packets on a large scale \cite{AXFR}. 

  \end{itemize}

  \section{Uptime Measurements\label{sec:uptimemeasure}}
  
  The goal of the uptime measurement is to find the amount of time a malicious URL is accessible from the time it appeared in one of the blacklists.
  The uptimes (or persistence of abuse) will be calculated for different intermediaries and abuse types (phishing, malware and botnet C\&C).

First, we collect a variety of information related to each URL when it gets blacklisted. We download the content of the malicious URL, the content of the homepage of the registered domain name, and the WHOIS information of the domain name for each malicious URL we find in the blacklist feeds. 
The difficulty of the uptime measurements compared with a single-time measurement is that the platform repeats the single-time measurement twelve times for each blacklisted URL.

  \begin{figure}[h]
    \centering
      \includegraphics[width=1\textwidth]{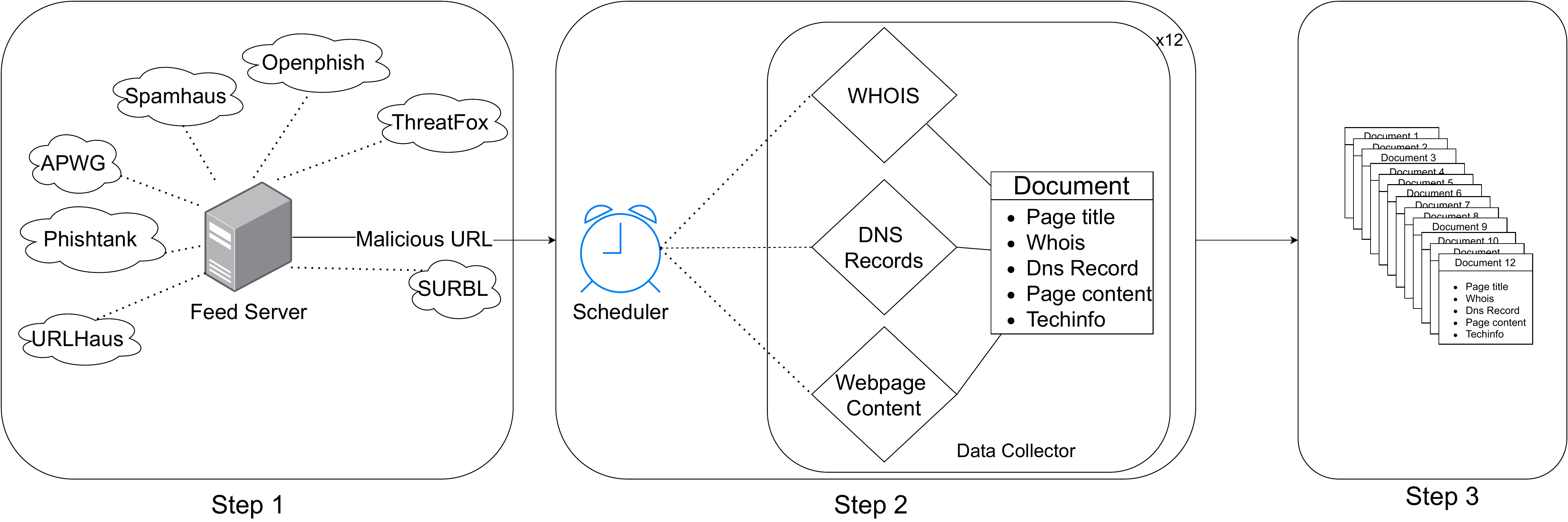}
      \caption{Uptime measurement platform}
      \label{fig:uptime_platform}
  \end{figure}

Figure \ref{fig:uptime_platform} shows the diagram of the uptime measurement platform designed for the purpose of this study. The ``feed server'' (step \textcircled{\small{1}}) is synchronized with all the blacklist feeds introduced in Section \ref{sec:blacklists}. Whenever a newly blacklisted URL appears in one of the feeds, we receive the update on our feed server. 
We automatically collect all data for each URL (step \textcircled{\small{2}}) once it gets blacklisted, 5 minutes after blacklisting, 30 minutes, 1 hour, 6 hours, 12 hours, 24 hours, 48 hours after 
blacklisting, and then once per week (12 measurements in total).

  After having all the 12 measurements for each URL (step \textcircled{\small{3}}), we can compare the 12 data collections. 
  The difference in some key features of each data collection (IP address of the registered domain, WHOIS, page content, etc.) can answer when the URL/domain name has been suspended.
  We show the result of the uptime measurements in Part~\ref{chapter:experiments}.

\newpage
\setstretch{1.0}
\newpage

\part{Domain Name Abuse Landscape}
\rhead{Domain Name Abuse Landscape}

\setstretch{1.5}
\label{chapter:experiments}

In this part, we show the results of the domain name abuse measurements and analysis performed in this study. 
We start with statistics describing the distributions of the malicious resources and abuse rates per TLDs (Section \ref{sec:TLDmet}).
We distinguish between compromised and maliciously registered domain names (Section \ref{sec:malcom}).
Since registrars cannot prevent abuse of vulnerable hosting software (unless they provide registration and hosting services), we calculate registrar reputation metrics based on domain names categorized as maliciously registered (Section \ref{sec:regrep}).
Then, we present reputation metrics for hosting providers, and countries for different abuse types (Section \ref{sec:rephos}). 
Finally, we discuss abuse of special domains (Section \ref{sec:special}) and targeted brands in phishing attacks (Section \ref{sec:targeted}).

\section{TLD Reputation Metrics \label{sec:TLDmet}}

\subsection{Methodology}
To measure the reputation of each TLD \cite{reputation-encyclopedia}, we use two security metrics: i) \textit{occurrence} and ii) \textit{ratio}. 
The \textit{occurrence} metric refers to the number of unique domains extracted from blacklisted URLs.
Note that we distinguish the reputation metric from measuring the ``security performance'' of TLD registries. While some domain names may be maliciously registered (and therefore they abuse DNS and hosting infrastructure), some websites are hacked and used to distribute harmful or illegal content (they abuse hosting but not necessarily DNS infrastructure).
In these latter cases, TLD registries generally should not intervene at the DNS level to avoid collateral damage to legitimate owners and visitors of benign but compromised domain names (unless attackers have exploited DNS vulnerabilities).
However, note that even hacked websites, distributing malicious content, impair the reputation of legitimate domain names and all intermediaries involved in domain name registration and hosting.  
Therefore, TLD reputation metrics express the ``overall health'' of TLD ecosystems consisting of many types of intermediaries such as domain registrars, re-sellers, hosting, content providers, etc.

Previous work \cite{korczynski2018cybercrime,tld_maciej} proposed the use of two additional complimentary security metrics to measure TLD health, i.e., the number of blacklisted fully qualified domain names (FQDNs) and unique blacklisted URLs. The reason is that one registered domain name can be used in a single attack while another can be used to distribute different malicious content using subdomains or multiple URLs.
Therefore, two complementary metrics provide a more complete picture of the DNS abuse extent.
However, they must be interpreted with caution because a single phishing campaign that uses hundreds or even thousands of subdomains can affect the reputation of an entire TLD.
More importantly, this research revealed an emerging trend among attackers: one maliciously registered domain name serves the same content but dynamically generates a unique URL each time a potential victim visits it. 
As such, the complementary reputation metrics may be biased and overestimate the number of security incidents.
For example, we observed one malicious domain redirecting users to the same page impersonating the Chase banking website and 5,192 unique URLs blacklisted by APWG, OpenPhish, and Phishtank.
Therefore, in this study, we only use the \textit{occurrence} of unique domain names as a TLD reputation metric.

Note that, reliable reputation metrics need to account for sizes, as bigger providers tend to suffer from more security incidents \cite{tld_maciej,noroozian2015}.
To calculate the abuse rates of each TLDs, we use the following formula:
\begin{equation}
\label{eq:ratio}
    ratio = \frac{occurrence}{size\ of\ a\ TLD} \times 10,000 
\end{equation}
This rate expresses the number of abused domains per 10,000 registrations.

We extract registered domain names from blacklisted URLs using the modified version of the public suffix list maintained by Mozilla.\footnote{https://publicsuffix.org} For the purpose of this study, we only use the public part of the suffix list.

\subsection{Results}\label{sec:tld_rep_results}

\begin{table}[hbt!]
    \centering
    \resizebox{\textwidth}{!}{\begin{tabular}{ l r  || l r  || l r  || l r || l r }
      \toprule
      \multicolumn{2}{c}{APWG} & \multicolumn{2}{c}{OpenPhish} & \multicolumn{2}{c}{PhishTank} & \multicolumn{2}{c}{URLHaus} & \multicolumn{2}{c}{ThreatFox} \\
      \midrule
      TLD & \# Domains & TLD & \# Domains  & TLD & \# Domains  & TLD & \# Domains & TLD & \# Domains \\
      \midrule
         com & 18,662 &   com & 23,253 &   com & 9,161 &com & 3,265 &   com &  424 \\
          ir &  3,509 &   xyz &  2,418 &   xyz & 1,324 & in &  249 &   top &  213 \\
         xyz &  2,375 &    tk &  2,125 &    tk & 1,139 & br &  244 &   xyz &   61 \\
         top &  1,456 &    ml &  1,473 &   top &  793  & net &  219 &    ru &   56 \\
         net &  1,123 &    ga &  1,318 &    cn &  662  & org &  207 &    in &   40 \\
          ru &    946 &   net &  1,178 &   net &  525  &  top &  113 &   net &   38 \\
         org &    835 &    cf &  1,093 &  shop &  492  &  uk &  112 &    br &   35 \\
        shop &    806 &   org &  1,055 &   org &  420  &  xyz &  111 &  club &   29 \\
        site &    753 &    gq &  1,016 &    ml &  332  &   za &  111 &   org &   26 \\
          ml &    561 &   top &    897 &    ga &  319  &   ru &   90 & space &   22 \\
      \toprule
      \multicolumn{2}{c}{SURBL combined} & \multicolumn{2}{c}{SURBL mw} & \multicolumn{2}{c}{SURBL ph} & \multicolumn{2}{c}{SURBL abuse} \\
      \midrule
        com & 146,955 &    com & 3,551 &  com & 52,134 &   com & 91,818 \\
         ru &  48,202 &    xyz &  729 &   ml &  9,142 &    ru & 46,086 \\
         cn &  42,738 &    top &  668 &   tk &  7,455 &    cn & 41,105 \\
         su &  30,713 &     za &  210 &  xyz &  7,425 &    su & 30,691 \\
        xyz &  25,065 &    net &  200 &   ga &  5,223 &   xyz & 17,007 \\
       work &  16,736 &   shop &  180 &   cf &  5,102 &  work & 16,130 \\
        top &  10,807 &    org &  128 &   gq &  4,368 &    us &  9,598 \\
         us &  10,193 &     in &  111 &  net &  2,784 &   top &  8,258 \\
         ml &   9,541 &     ru &  106 &   cn &  2,443 &    me &  6,888 \\
         tk &   8,520 & online &   94 &   ru &  2,184 &  info &  6,269 \\
      \toprule
      \multicolumn{2}{c}{Spamhaus combined} & \multicolumn{2}{c}{Spamhaus SP} & \multicolumn{2}{c}{Spamhaus PH} & \multicolumn{2}{c}{Spamhaus MW} &  \multicolumn{2}{c}{Spamhaus C\&C} \\
      \midrule
         com & 470,877 &    com & 425,632 &   com & 27,547 &     com &  853 &   com & 2,594 \\
          cn & 144,785 &     cn & 136,284 &    cn &  8,015 &    buzz &  664 &  info &  673 \\
         net &  66,077 &    net &  63,123 &   xyz &  7,047 &     xyz &  634 &   net &  497 \\
        work &  58,094 &   work &  57,673 &    ga &  5,511 &     top &  235 &   top &  442 \\
         xyz &  43,353 &    xyz &  35,390 &    tk &  5,505 &     vip &  172 &   org &  225 \\
          ru &  23,025 &     ru &  20112 &  shop &  4,959 &   cloud &  136 &    ru &  198 \\
        info &  16,625 &   info &  13,944 &    ml &  4,029 &      br &   98 &    me &  189 \\
          tk &  15,658 & online &  12,280 & press &  3,971 &      ru &   91 &    su &  142 \\
      online &  13,666 &    biz &  12,239 &    cf &  3,434 & monster &   83 &    eu &  142 \\
         biz &  12,709 &     tk &  10,139 &   bar &  3,008 &    live &   82 &    tk &  137 \\
      \toprule
      \multicolumn{2}{c}{Spamhaus AL-SP} & \multicolumn{2}{c}{Spamhaus AL-PH} & \multicolumn{2}{c}{Spamhaus AL-MW} & \multicolumn{2}{c}{Spamhaus AL-C\&C} \\
      \midrule
        com & 11,927 &    com & 3,399 &   com & 2,253 &   com &   54  & & \\
         ru &  1,468 &    org &  196 &    in &  207 &    ru &   32  & & \\
        net &   884 &     br &  195 &    br &  190 &   org &    7  & & \\
        org &   748 &    net &  167 &   org &  156 &    kr &    5  & & \\
         tk &   546 &     in &  146 &   net &  146 &    cz &    3  & & \\
         br &   452 &     uk &  136 &    za &   72 &   net &    3  & & \\
         cn &   433 &     cn &  129 &    uk &   71 &    pl &    3  & & \\
         in &   403 &     au &  126 &    au &   54 &    it &    3  & & \\
         de &   392 &     za &   76 &    co &   46 &    ua &    3  & & \\
         uk &   256 &     co &   69 &    cn &   41 &    ir &    2  & & \\
      \bottomrule
    \end{tabular}}
    \caption{Top 10 TLDs with the highest occurrence of blacklisted domains for APWG, OpenPhish, PhishTank, URLHaus, SURBL, and Spamhaus datasets.}
    \label{top10tlds-counts}
      \end{table}

Table \ref{top10tlds-counts} shows the occurrence of the top 10 most abused TLDs in seven blacklist feeds (and the corresponding sixteen blacklists), namely: APWG, OpenPhish, PhishTank, Abuse.ch (URLHaus, ThreatFox), SURBL, and Spamhaus. The table shows all TLDs, including new gTLDs, legacy gTLDs, and ccTLDs. 
In Table \ref{top10tlds-counts-aggregated}, we aggregate different blacklists per abuse type (botnet C\&C, malware, phishing, and spam). 

In different types of abuse, .com and (generally) .net legacy gTLDs are among the most abused TLDs. The distribution of abused legacy gTLDs domains is mainly driven by .com domains, which is not surprising given the TLD market share.
For phishing abuse, five out of ten most abused TLDs (.ml, .tk, .ga, .cf, and .gq) are operated by Freenom.\footnote{https://www.freenom.com}
Freenom provides domain registrations for free for .ml, .tk, .ga, .cf, and .gq ccTLDs.
Freenom is also an ICANN-accredited registrar and provides registrations for most of the other top-level domains.
Registrants of free domain names have no ``licensee'' status, but they are considered as domain name ``users''.
Therefore, Freenom can take down any domain name without further notice.
The lack of registration fees is the most likely reason Freenom's TLDs are widely abused by malicious actors.

\begin{table}[hbt!]
    \centering
    \resizebox{\textwidth}{!}{\begin{tabular}{ l r  || l r  || l r  || l r  }
      \toprule
      \multicolumn{2}{c}{Botnet C\&C} & \multicolumn{2}{c}{Malware} & \multicolumn{2}{c}{Phishing} & \multicolumn{2}{c}{Spam} \\
      \midrule
      TLD & \# Domains & TLD & \# Domains  & TLD & \# Domains  & TLD & \# Domains  \\
      \midrule
     com &  2,763 &  com &  6,788 &  com &  91,587 &  com & 519,965 \\
    info &    676 &  xyz &  1,373 &  xyz &  15,658 &   cn & 177,374 \\
     top &    567 &  top &    894 &   ml &  12,644 &  net &  66,565 \\
     net &    523 & buzz &    668 &   tk &  11,912 &   ru &  64,462 \\
      ru &    261 &  net &    405 &   ga &  10,459 & work &  64,235 \\
     org &    238 &   br &    371 &   cn &  10,259 &  xyz &  48,790 \\
      me &    189 &  org &    357 &   cf &   8,241 &   su &  32,974 \\
     xyz &    174 &   in &    298 & shop &   6,511 & info &  19,460 \\
      su &    145 &   za &    270 &   gq &   6,235 &  top &  15,724 \\
      eu &    142 & shop &    191 &  top &   5,652 &   us &  15,499 \\
     
    \end{tabular}}
    \caption{Top 10 TLDs with the highest occurrence of blacklisted domains per abuse type.   
    \label{top10tlds-counts-aggregated}}
      \end{table}

\begin{table}[hbt!]
    \centering
    \resizebox{0.9\textwidth}{!}{\begin{tabular}{ l r  || l r  || l r  || l r  }
      \toprule
      \multicolumn{2}{c}{Botnet C\&C} & \multicolumn{2}{c}{Malware} & \multicolumn{2}{c}{Phishing} & \multicolumn{2}{c}{Spam} \\
      \midrule
      TLD & \# Domains & TLD & \# Domains  & TLD & \# Domains  & TLD & \# Domains \\
      \midrule
              ru &    261  &          ru &    190  &          ru &  3,787 &          ru &  64,462 \\
              me &    189  &          uk &    143  &          me &  1,918 &          su &  32,974 \\
              su &    145  & \textbf{de} &     57  &          uk &  1,826 &          me &   8,900 \\
     \textbf{eu} &    142  & \textbf{ro} &     48  & \textbf{de} &    857 &          \textcyrillic{рф} &   4,129 \\
     \textbf{at} &     91  &          me &     39  & \textbf{fr} &    600 & \textbf{eu} &   1,948 \\
              ua &     18  &          su &     38  & \textbf{eu} &    513 &          uk &   1,518 \\
     \textbf{pl} &     15  & \textbf{it} &     35  & \textbf{nl} &    444 & \textbf{nl} &   1,067 \\
     \textbf{nl} &     15  & \textbf{fr} &     34  & \textbf{it} &    413 & \textbf{de} &   1,026 \\
     \textbf{de} &     15  & \textbf{es} &     32  & \textbf{pl} &    386 & \textbf{pl} &    853 \\
     \textbf{it} &     14  & \textbf{nl} &     32  & \textbf{es} &    271 & \textbf{fr} &    778 \\
     \textbf{fr} &     12  & \textbf{eu} &     29  & \textbf{ro} &    204 & \textbf{it} &    488 \\
     \textbf{hu} &     10  & \textbf{pl} &     27  & \textbf{be} &    171 &          ua &    380 \\
     \textbf{be} &     10  & \textbf{gr} &     21  &          ch &    157 & \textbf{be} &    329 \\
              uk &      4  &          rs &     20  &          ua &    135 & \textbf{se} &    236 \\
     \textbf{cz} &      3  & \textbf{hu} &     20  & \textbf{hu} &    135 & \textbf{es} &    218 \\
     \textbf{pt} &      3  & \textbf{pt} &     20  & \textbf{gr} &    121 &          ch &    193 \\
              im &      1  &          ch &     18  & \textbf{pt} &     98 & \textbf{at} &    191 \\
     \textbf{bg} &      1  & \textbf{hr} &     12  & \textbf{cz} &     94 & \textbf{ro} &    147 \\
              by &      1  & \textbf{be} &     12  & \textbf{dk} &     88 & \textbf{cz} &    119 \\
     \textbf{lt} &      1  & \textbf{at} &      7  &          su &     85 & \textbf{dk} &    110 \\
      
    \end{tabular}}
    \caption{Top 20 European ccTLDs with the highest occurrence of blacklisted domains per abuse type.   
    \label{top20-eucctlds-counts}}
  \end{table}

\begin{table}[th]
  \centering
  \resizebox{\textwidth}{!}{\begin{tabular}{ l r r || l r r || l r r  || l r r }
    \toprule
    \multicolumn{3}{c}{Botnet C\&C} & \multicolumn{3}{c}{Malware} & \multicolumn{3}{c}{Phishing} & \multicolumn{3}{c}{Spam} \\
    \midrule
    TLD & \# Domains & Rate & TLD & \# Domains & Rate & TLD & \# Domains & Rate & TLD & \# Domains & Rate \\
    \midrule
         su       &    145 &  8.08  &         \textbf{ax} &      1 &  2.78   &                 me &   1,918 & 17.50   &                 su &  32,974 & 1,837.10 \\
         me       &    189 &  1.72  &                  su &     38 &  2.12   &                 al &     26 & 11.63   &                 ru &  64,462 &  131.48 \\
\textbf{at}       &     91 &  0.66  &                  rs &     20 &  1.73   &                 im &     23 &  8.06   &                 me &   8,900 &   81.21 \\
         ru       &    261 &  0.53  &         \textbf{hr} &     12 &  1.14   &                 ru &   3,787 &  7.72   &  \textcyrillic{рф} &   4,129 &   63.11 \\
\textbf{eu}       &    142 &  0.39  &  \textcyrillic{бел} &      1 &  0.98   &                 md &     19 &  7.33   &                 va &      1 &   24.39 \\
         im       &      1 &  0.35  &         \textbf{ro} &     48 &  0.90   &                 rs &     77 &  6.67   &                 al &     25 &   11.18 \\
         ua       &     18 &  0.31  &         \textbf{gr} &     21 &  0.49   &                 mk &     16 &  5.29   &                 im &     31 &   10.87 \\
\textbf{bg}       &      1 &  0.15  &         \textbf{ee} &      6 &  0.45   &        \textbf{hr} &     50 &  4.74   &                 by &     87 &    6.73 \\
\textbf{hu}       &     10 &  0.12  &                  al &      1 &  0.45   &                 su &     85 &  4.74   &                 ua &    380 &    6.65 \\
         by       &      1 &  0.08  &                  ru &    190 &  0.39   &                 gg &     11 &  3.84   &                 md &     16 &    6.17 \\
\textbf{pl}       &     15 &  0.06  &                  me &     39 &  0.36   &        \textbf{ro} &    204 &  3.84   &        \textbf{eu} &   1,948 &    5.33 \\
\textbf{be}       &     10 &  0.06  &                  gg &      1 &  0.35   &                 li &     21 &  3.72   &                 ad &      1 &    4.78 \\
\textbf{lt}       &      1 &  0.05  &                  mk &      1 &  0.33   &        \textbf{bg} &     22 &  3.39   &                 sm &      1 &    4.49 \\
\textbf{it}       &     14 &  0.05  &                  by &      4 &  0.31   &        \textbf{si} &     36 &  2.84   &                 li &     24 &    4.25 \\
\textbf{fr}       &     12 &  0.03  &         \textbf{hu} &     20 &  0.25   &        \textbf{gr} &    121 &  2.82   &        \textbf{fo} &      2 &    3.84 \\
\textbf{nl}       &     15 &  0.02  &         \textbf{ie} &      6 &  0.19   &                 by &     35 &  2.71   &        \textbf{pl} &    853 &    3.44 \\
\textbf{cz}       &      3 &  0.02  &         \textbf{lv} &      2 &  0.17   &        \textbf{mt} &      4 &  2.46   &        \textbf{lv} &     36 &    3.11 \\
\textbf{pt}       &      3 &  0.02  &         \textbf{es} &     32 &  0.16   &                 ua &    135 &  2.36   &                 mk &      9 &    2.97 \\
\textcyrillic{рф} &      1 &  0.02  &         \textbf{bg} &      1 &  0.15   &                 is &     18 &  2.36   &                 rs &     34 &    2.95 \\
\textbf{de}       &     15 &  0.01  &         \textbf{pt} &     20 &  0.15   & \textcyrillic{бел} &      2 &  1.96   &        \textbf{bg} &     19 &    2.93 \\

  \end{tabular}}
  \caption{Top 20 European ccTLDs with the highest abuse rate per abuse type.   
  \label{top20-eucctlds-rates}}
  \end{table}

We also observe that malicious actors extensively misuse .xyz,  .work (for spam) and .top new gTLDs.
One of the plausible reasons might be their pricing, as suggested already in the previous studies (e.g., \cite{korczynski2018cybercrime}).
The retail pricing of .xyz and .top domains oscillates around US \$1.
In 2020, we observed that the price of .xyz domains proposed by Alibaba Cloud was as low as US \$0.18.
Figure \ref{fig:tld_distribution} in Section \ref{sec:general_dataset} shows that new gTLDs suffer from the highest concentration of abused domains relative to their market share.
It does not mean that the entire market of new gTLDs is extensively abused.
\pagelabel{text:abused_new_gtlds}The two most abused new gTLDs (.work and .xyz) together account for 41\% of all abused new gTLD names in the second quarter of 2021, while abused domain names from the top five most abused new gTLDs (.work, .xyz, .top, .online, .site, .icu) account for about 60\% of all abused new gTLD domains.

Note that European Union ccTLDs are not among the most abused TLDs (with the exception of .eu domains for the botnet C\&C feed). 
In general, EU ccTLDs are proactive in deploying (voluntarily) preventive and reactive security measures to combat DNS abuse more effectively (e.g., .eu, .nl, .dk, .se), and do not need to aggressively compete in pricing (as opposed to some new gTLDs such as .top, or .xyz).

Table \ref{top20-eucctlds-counts} shows the top 20 European ccTLDs (European Union ccTLDs are marked in bold) with the highest occurrence of abused domains for four different abuse types (botnet C\&C, malware, phishing, and spam). The Russian Federation top-level domain (.ru) is the most abused European ccTLD of all different abuse types (in terms of occurrence).
As many as 64 K .ru and 33 K .su (Soviet Union) ccTLD domain names were labeled as spam.
In comparison, for the fifth most abused .eu ccTLD, significantly less (2 K) domains were blacklisted in the second quarter of 2021.

If we consider only non-European Union ccTLDs and combined abuse counts (C\&C, malware, phishing and spam), abused .ru and .su second-level domain names account for 75\% (50.4\% and 24.8\%, respectively) of all abused domains among non-EU ccTLDs.
In comparison, abused domain names in EU ccTLDs, apart from significantly lower absolute abuse counts, are more distributed among different TLDs. Seven most abused EU ccTLDs (.eu, .de, .nl, .fr., .pl, .it, .es, .be) account for about 76\% of all abused domains among EU ccTLDs.

The .dk TLD is at the other extreme: only 110 .dk domain names were blacklisted and marked as spam, 88 as phishing. In total, our dataset contains 168 unique, abused .dk domain names over a three-month period.
Some domains were reused in both phishing and spam attacks.
Therefore, the sum of spam and phishing domains is higher than the number of unique domains.
Using the method described in Section \ref{sec:malcom}, we automatically labeled .dk domain names as compromised or maliciously registered.
One limitation of the classification is that we could not automatically identify patterns of mass registrations indicating malicious registrations because WHOIS of .dk domains does not provide the registrar name or IANA ID.
Only 15 domains were found to be registered with malicious intent.
The cheapest registration price for .dk domains during the study oscillated between US \$9 and US \$10.
The price is just one of the possible factors that make attackers not to choose to maliciously register domain names with .dk TLD.
There may be other reasons as well, such as verifying the accuracy of the registrant information.
If the registry operator verifies the registrant identity (as is the case with .dk domain names\footnote{\url{https://www.dk-hostmaster.dk/en/id-check}}), a cybercriminal may choose to register a malicious domain name with a different TLD.

\begin{tcolorbox}[enhanced,colback=blue!5!white,colframe=blue!75!black,colbacktitle=red!80!black]
  \pagelabel{text:kybc_rec}
  \textbf{Recommendation}: TLD registries, registrars, and resellers should verify the accuracy of the domain registration (WHOIS) data. The identification of the registrants could be implemented through possibly harmonised Know Your Business Customer (KYBC) procedures. In case of registrants from the EU, KYBC could be carried out through eID authentication in accordance with the eIDAS Regulation, as amended by the forthcoming Regulation on the European Digital Identity. KYBC procedure shall use cross-checks in other publicly available and reputed databases.
 \end{tcolorbox}

This study provides anecdotal evidence indicating that cybercriminals choose to abuse TLDs that offer low domain name registration prices and avoid TLDs that strictly verify the registrant identity.
However, there is a need for a very comprehensive statistical analysis of factors driving DNS abuse.
One malicious actor may prefer lower registration prices. The other may choose to abuse a registrar that offers specific payment methods (e.g, cryptocurrencies) or a free API allowing for domain registration~in~bulk.

\begin{table}[t!]
  \centering
  \resizebox{\textwidth}{!}{
  
  \begin{tabular}{ c  l c r || l c r || l c r || l c r }
    
    \cmidrule[0.25ex]{1-13}
    \multirow{12}{*}{\rotatebox[origin=c]{90}{$(0,10k]$}} 
    & \multicolumn{3}{c}{Botnet C\&C} & \multicolumn{3}{c}{Malware} & \multicolumn{3}{c}{Phishing} & \multicolumn{3}{c}{Spam} \\
    \cmidrule{2-13}
    & TLD & \# Domains & Rate & TLD & \# Domains & Rate & TLD & \# Domains & Rate & TLD & \# Domains & Rate \\
    \cmidrule{2-13}
    &      ug &     29 &    37  &  date &     33 &    33  &     date &    591 &   596 &    ryukyu &    151 &  2,654 \\
    &    bond &      2 &     7  &    ug &     20 &    25  &      gle &      1 &   556 &   okinawa &    295 &   701 \\
    &download &      5 &     6  &    lr &      1 &    24  &       ci &    303 &   305 &      date &    525 &   529 \\
    &      sb &      1 &     6  &   krd &      1 &    16  &       ss &      1 &   286 & xn--nqv7f &     10 &   316 \\
    &      sx &      1 &     4  &    sd &      3 &     8  &     saxo &      1 &   118 &        ss &      1 &   286 \\
    &    army &      1 &     3  & bingo &      1 &     7  &       aq &      1 &   104 &       sbs &     73 &   253 \\
    &          &        &       &    tg &      1 &     5  &      int &      2 &    97 &   recipes &    106 &   209 \\
    &          &        &       &    cd &      3 &     5  &   makeup &      6 &    70 &    webcam &     44 &   125 \\
    &          &        &       &    mz &      3 &     4  &       ki &      3 &    54 &  cleaning &     32 &   103 \\
    &          &        &       &    cv &      1 &     4  & delivery &     52 &    54 &   exposed &     32 &    99 \\
    \cmidrule[0.25ex]{1-13}
    \multirow{12}{*}{\rotatebox[origin=c]{90}{$(10k,100k]$}} 
    & \multicolumn{3}{c}{Botnet C\&C} & \multicolumn{3}{c}{Malware} & \multicolumn{3}{c}{Phishing} & \multicolumn{3}{c}{Spam} \\
    \cmidrule{2-13}
    & TLD & \# Domains & Rate & TLD & \# Domains & Rate & TLD & \# Domains & Rate & TLD & \# Domains & Rate \\
    \cmidrule{2-13}
    &support &     28 &    10  &     trade &     54 &    28    &  press &   3,984 &  1,420  &    surf &   1,979 &  1,529 \\
    &    uno &     22 &     9  &       bid &     49 &    27    &support &    514 &   180  &     cam &   5,143 &   989 \\
    &    bid &     12 &     7  & institute &     10 &     7    &   casa &    349 &    71  &    casa &   3,675 &   751 \\
    &   casa &     30 &     6  &        pk &     54 &     6    &   help &    113 &    67  &    rest &   2,459 &   701 \\
    &    gdn &      3 &     3  &      casa &     26 &     5    &   cyou &    608 &    55  &     fit &   4,563 &   549 \\
    &  rocks &     19 &     2  &      best &     34 &     4    & review &     61 &    53  &    cyou &   5,195 &   467 \\
    &   golf &      3 &     2  &        py &     10 &     4    &   surf &     60 &    46  &     uno &    864 &   335 \\
    &   city &      5 &     2  &       win &     24 &     4    &finance &    118 &    33  &    host &   1,450 &   325 \\
    &   band &      2 &     1  &    review &      4 &     3    &digital &    301 &    33  &   trade &    428 &   221 \\
    &digital &     13 &     1  &        bo &      4 &     3    &    cam &    168 &    32  & support &    526 &   184 \\
    \cmidrule[0.25ex]{1-13} 
    \multirow{12}{*}{\rotatebox[origin=c]{90}{$(100k,1M]$}} 
    & \multicolumn{3}{c}{Botnet C\&C} & \multicolumn{3}{c}{Malware} & \multicolumn{3}{c}{Phishing} & \multicolumn{3}{c}{Spam} \\
    \cmidrule{2-13}
    & TLD & \# Domains & Rate & TLD & \# Domains & Rate & TLD & \# Domains & Rate & TLD & \# Domains & Rate \\
    \cmidrule{2-13}
    &      su &    145 &  8.00 &    buzz &    668 &    27  &    buzz &    668 &    27   &     su &  32,974 &  1,837 \\
    &      pw &     75 &  2.00 & monster &     85 &     7  & monster &     85 &     7   &   work &  64,235 &  1,123 \\
    &   space &     39 &  1.00 &   cloud &    137 &     7  &   cloud &    137 &     7   &  email &   2,878 &   261 \\
    & website &     33 &  1.00 &   email &     43 &     4  &   email &     43 &     4   &   buzz &   5,523 &   219 \\
    &     fun &     21 &  1.00 &      ke &     31 &     3  &      ke &     31 &     3   &   life &   3,387 &   154 \\
    &      cc &     69 &  0.91 &    link &     32 &     3  &    link &     32 &     3   &    fun &   3,094 &   154 \\
    &    link &      9 &  0.77 &      ng &     30 &     2  &      ng &     30 &     2   &  world &   1,906 &   148 \\
    &      kz &     11 &  0.70 &     vip &    178 &     2  &     vip &    178 &     2   &   asia &   2,587 &   131 \\
    &    mobi &     21 &  0.65 &    live &    104 &     2  &    live &    104 &     2   &website &   3,791 &   129 \\
    &    life &     11 &  0.50 &      su &     38 &     2  &      su &     38 &     2   &   tech &   4,182 &   128 \\
    \cmidrule[0.25ex]{1-13}
    \multirow{12}{*}{\rotatebox[origin=c]{90}{$(1M,\infty)$}} 
    & \multicolumn{3}{c}{Botnet C\&C} & \multicolumn{3}{c}{Malware} & \multicolumn{3}{c}{Phishing} & \multicolumn{3}{c}{Spam} \\
    \cmidrule{2-13}
    & TLD & \# Domains & Rate & TLD & \# Domains & Rate & TLD & \# Domains & Rate & TLD & \# Domains & Rate \\
    \cmidrule{2-13}
    &    top &    567 &  5.00  &    top &    567 &  5.00 & xyz &  15,658 &    52   &    cn & 177,374 &   170 \\
    &     me &    189 &  2.00  &     me &    189 &  2.00 & top &   5,652 &    48   &    xyz &  48,790 &   161 \\
    &   info &    676 &  2.00  &   info &    676 &  2.00 &  ir &   3,865 &    34   &    top &  15,724 &   132 \\
    &     at &     91 &  0.66  &     at &     91 &  0.66 &  ml &  12,644 &    32   &     ru &  64,462 &   131 \\
    &   club &     69 &  0.66  &   club &     69 &  0.66 &  tk &  11,912 &    22   &    biz &  14,456 &   105 \\
    &    xyz &    174 &  0.57  &    xyz &    174 &  0.57 &site &   2226 &    20   &     us &  15,499 &    92 \\
    &     ru &    261 &  0.53  &     ru &    261 &  0.53 &  ga &  10,459 &    19   &     me &   8,900 &    81 \\
    & online &     94 &  0.53  & online &     94 &  0.53 &  cf &   8,241 &    19   & online &  13,561 &    76 \\
    &    biz &     62 &  0.45  &    biz &     62 &  0.45 &  gq &   6,235 &    18   &   site &   7,313 &    66 \\
    &    net &    523 &  0.40  &    net &    523 &  0.40 &club &   1,898 &    18   &   club &   5,706 &    55 \\
    \cmidrule[0.25ex]{1-13}
  \end{tabular}}
  \caption{Top 10 TLDs with the highest relative concentrations of blacklisted domains classified by their corresponding TLD size and abuse type.}
  \label{top10tlds-rates}
  \end{table}

While the absolute number of abused domains is essential, yet it does not show the (relative to the size) \textit{overall health} of each TLD. Table \ref{top20-eucctlds-rates} shows the abuse rate for the top 20 European ccTLDs per abuse type (European Union ccTLDs are highlighted in bold). While the number of malicious domains with .ru TLD is the highest (see Table~\ref{top20-eucctlds-counts}), .su suffers from the highest abuse rate in terms of spam and botnet C\&C because the size of the .ru zone file is approximately 45 times bigger than .su zone file.
Security researchers observe that for years, .su domains are abused to control botnets, spread malware, send spam, or infect end-users to steal banking data.
While hosting companies and registrars usually suspend such domains instantly, malicious .su domains run for months.\footnote{\url{https://www.theguardian.com/technology/2013/may/31/ussr-cybercriminals-su-domain-space}}
The uptime (the time between the blacklisting/notification and the takedown) is another plausible reason why certain TLDs are abused more than others.

Table \ref{top10tlds-rates} shows the top 10 TLDs with the highest abuse rates in four abuse categories (botnet C\&C, malware, phishing, and spam). We divide all TLDs in five different groups of TLDs, corresponding to 5 different size ranges. The reason for grouping TLDs is to make a more appropriate comparison by considering the estimated size of each TLD. For example, in the spam category, the abuse rate of the .cleaning TLD is 103 while the number of abused domains for this TLD is 32, which is not comparable to the .xyz TLD with 48,790 abused domains for spam with abuse rate of 161. 
Abuse rates, especially for the smallest TLDs, should be considered with caution. Even a small number of abused domains (or sometimes false positives, i.e., domain names blacklisted by security companies by mistake) can strongly affect a TLD reputation.

Abuse rates should be monitored on an ongoing basis by independent researchers on behalf of regulatory bodies and should not exceed predetermined thresholds (e.g., 30\%, e.g., 3 K abused per 10 K registered domain names). Such recommendations were already proposed in recent years \cite{sunrise,sadag-cct,sadag-ssr2}.
If a contracted party (TLD registry or registrar) exceeds the acceptable abuse rates and does not improve within a given time period, accreditation may be revoked~\cite{sunrise,sadag-ssr2}.
Note that abuse thresholds should be carefully designed, as even a low absolute number of incidents for small TLD registries may result in thresholds being exceeded. On the other hand, large registries (in terms of registered domain names) may not face elevated abuse rates even if the number of abused domains is high (e.g., .com).

\pagelabel{text:reg_monitor_rec}
 \vspace{+0.5cm}
\begin{tcolorbox}[enhanced,colback=blue!5!white,colframe=blue!75!black,colbacktitle=red!80!black]
 \textbf{Recommendation}: The study recommends that the abuse rates of TLD registries or registrars be monitored on an ongoing basis by independent researchers in cooperation with institutions and regulatory bodies (e.g., ICANN, European Commission, European Union Agency for Cybersecurity – ENISA or national authorities). Abuse rates should not exceed predetermined thresholds. If thresholds are exceeded and the abuse rates do not improve within a given time period, accreditation may be revoked.
 \end{tcolorbox}

Previous reports \cite{sunrise,sadag-cct,sadag-ssr2} indicate that TLD registries (and registrars) facing disproportionately high rates of abused domain names could be charged higher fees for domain registrations. In contrast, intermediaries with lower abuse rates could be financially rewarded, e.g., by reducing domain registration fees to align incentives with raising barriers to abuse. 

Such incentive structures must be very carefully designed to encourage the development of security practices by TLD registries with high abuse rates and those that affect the reputation of the overall market. For example, .xyz does not suffer from high abuse rates (spam rate in Q2 2021: 161 per 10,000 registrations; spam domains: 48,790, Table~\ref{top10tlds-rates}). However, new gTLDs as a class of registry operators suffer from disproportionately high abuse rates (Figure~\ref{fig:tld_distribution}), with .xyz accounting for about 20\% of all abused domains among new gTLDs.

\pagelabel{text:reg_reward_rec}
\vspace{+0.5cm}
\begin{tcolorbox}[enhanced,colback=blue!5!white,colframe=blue!75!black,colbacktitle=red!80!black]
 \textbf{Recommendation}:  TLD registries and registrars with lower abuse rates may be financially rewarded, e.g., through a reduction in domain registration fees, to align economic incentives and raise barriers to abuse. Price incentives could be implemented by ICANN for TLD registries, or by TLD registries for locally accredited registrars.
\end{tcolorbox}

\begin{figure}
    \centering
    \includegraphics[width=0.9\textwidth]{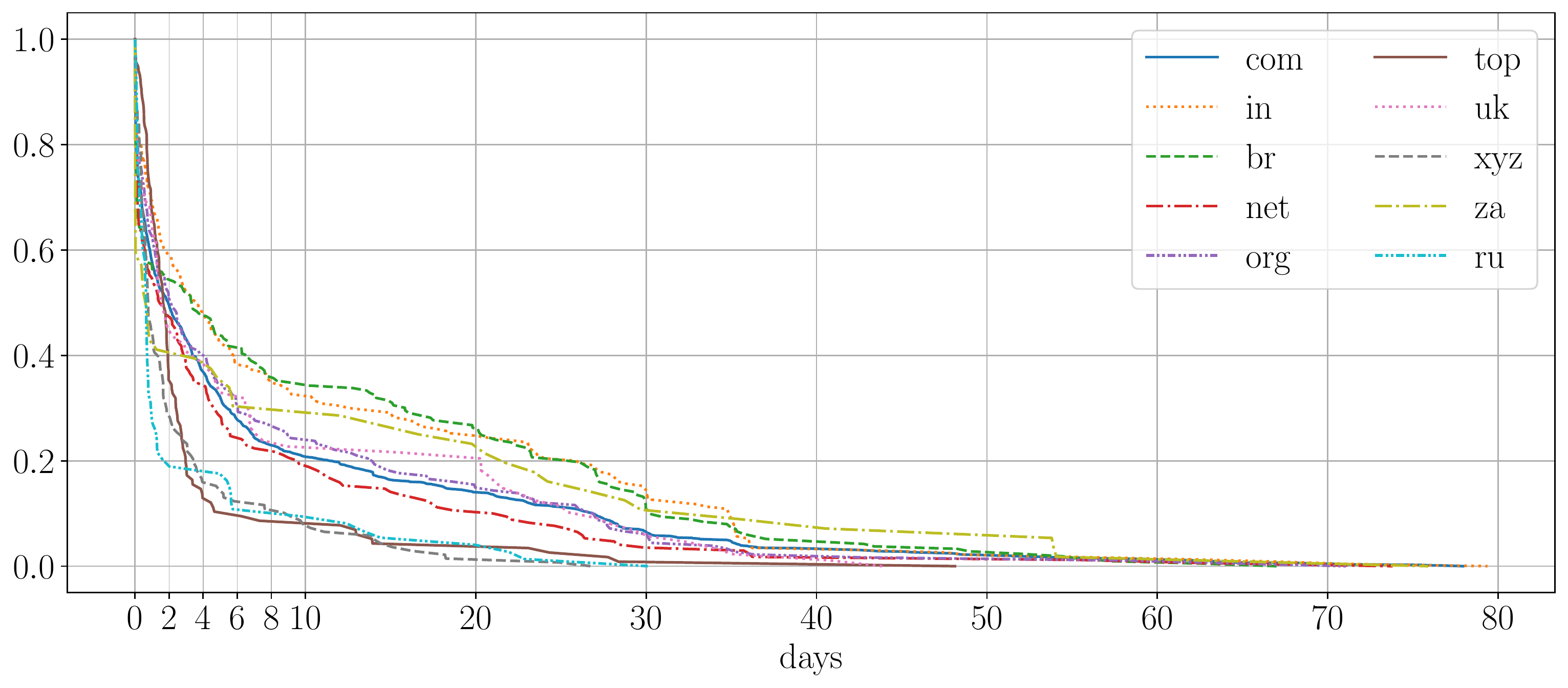}
    \caption{Survival analysis of the top 10 most abused TLDs for malware delivery.}
    \label{fig:survive_top10}
\end{figure}

We now analyze uptimes (persistence of abuse), i.e., the time between blacklisting a URL and taking mitigation action. 
Figure \ref{fig:survive_top10} shows the so-called survival analysis of the ten most abused TLDs (in terms of occurrence) associated with malware delivery URLs. 
The data results from an uptime analysis for 13,249 URLs (5,547 unique domains) identified by URLhaus. The survival analysis and the curves shown illustrate the chance of survival (not being taken down by the intermediaries involved) over a certain period of time. For example, in Figure \ref{fig:survive_top10}, the chance for an abused domain name with the \texttt{.net} TLD to stay alive after four days is about 32\%, while the probability for a domain with the \texttt{.br} TLD to survive after four days is about 47\%. Therefore, the intermediaries involved in the \texttt{.net} TLD ecosystem mitigate faster abuse compared to the \texttt{.br} TLD.

\begin{figure}
    \centering
    \includegraphics[width=0.9\textwidth]{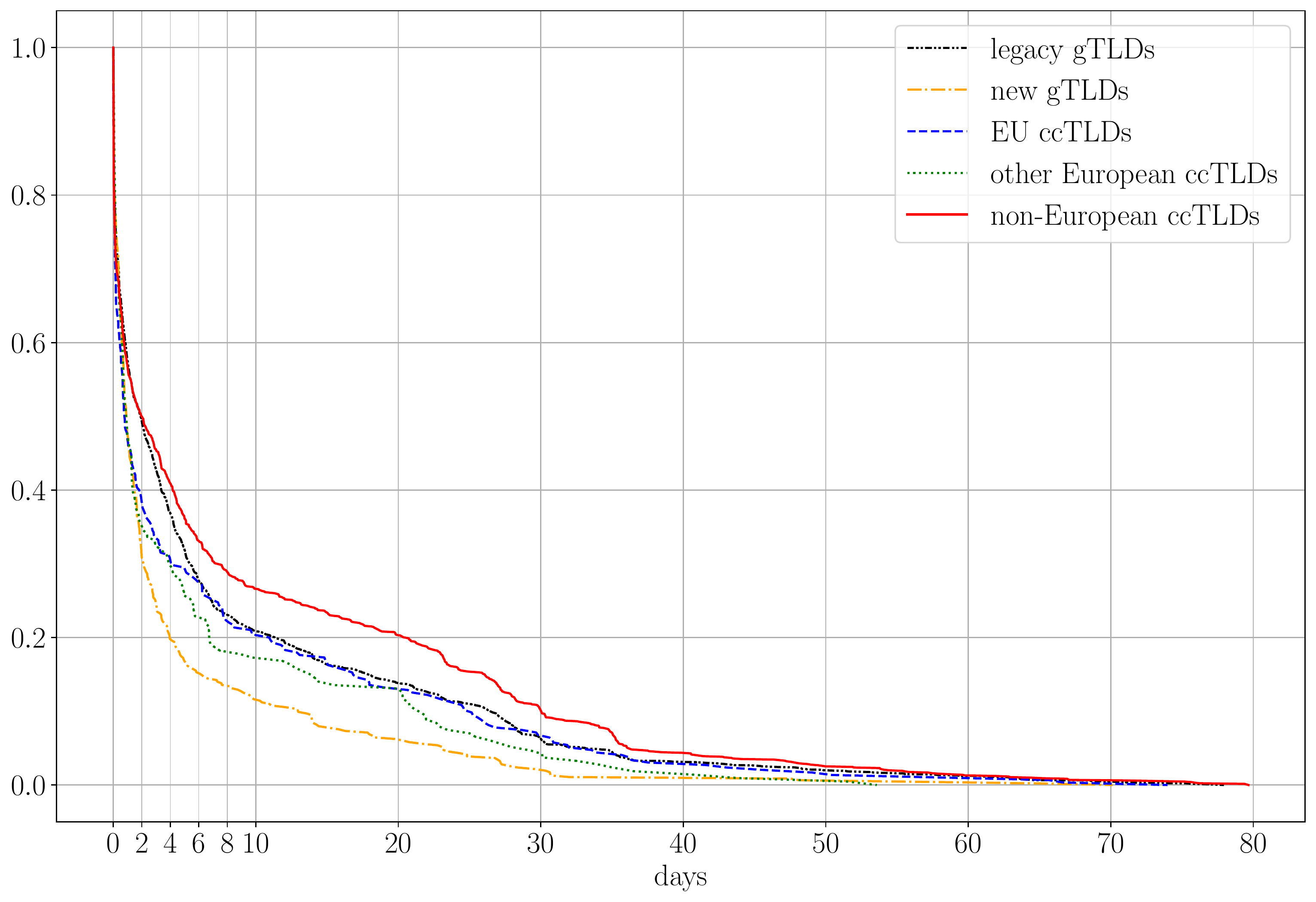}
    \caption{Survival analysis for different types of TLDs for malware delivery.}
    \label{fig:survive_tldtype}
\end{figure}

Figure \ref{fig:survive_tldtype} shows the survival analysis for different TLD types. In general, new gTLDs are the most efficient at cleaning up malicious content compared to legacy gTLDs and (European/non-European) ccTLDs.  For example, the chance of accessing a malicious payload on a non-European ccTLD ten days after it has been blacklisted is about 15\% higher compared to registered domains of new gTLDs.

\begin{figure}
    \centering
    \includegraphics[width=0.9\textwidth]{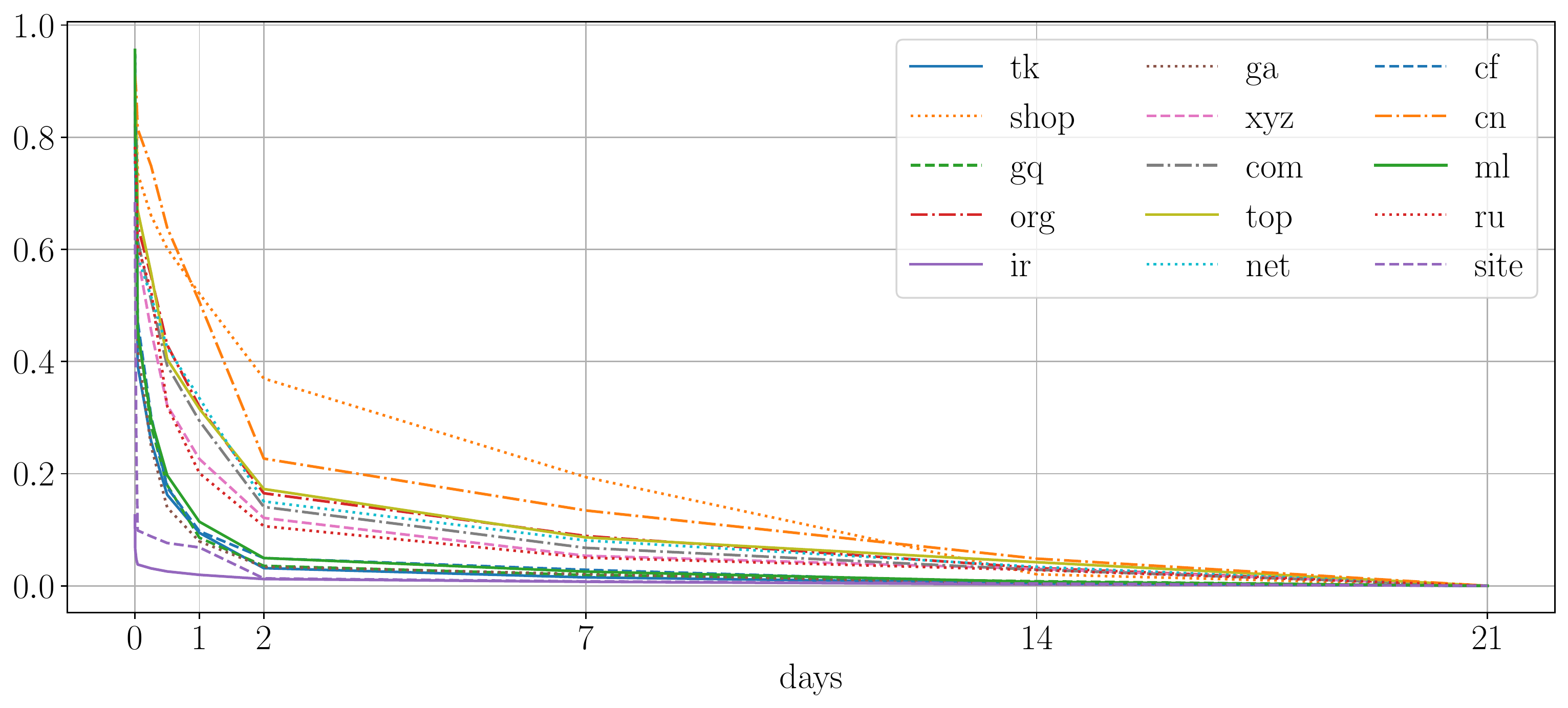}
    \caption{Survival analysis of the 15 most abused TLDs for phishing.}
    \label{fig:survive_phish_tld_1}
\end{figure}

Figure \ref{fig:survive_phish_tld_1} shows the survival analysis of the 15 most used TLDs in phishing (APWG, OpenPhish, and PhishTank data), while Table \ref{tab:top_15_phish_survival} shows the mean and median uptimes. In Table \ref{tab:top_15_phish_survival}, all Freenom TLDs (.tk, .ml, .cf, .ga, and .gq) are in the top 15 TLDs most commonly used for phishing. However, the median for cleaning actions is 1 hour for all of them, which is 12 times faster than for the \texttt{.net} and \texttt{.com} TLDs. It is probably since, as mentioned earlier, registrants of free domain names do not have the status of ``licensee'' but are treated as ``users'' of the domain name.  Therefore, Freenom can take down any domain name without notice (see Section \ref{sec:TLDmet} for more details).

Interestingly, in the case of the \texttt{.ir} ccTLD (the second most abused TLD for phishing) the median is 0, while the average uptime is about 4 hours. It is because for at least 50 percent of the phishing URLs in the \texttt{.ir} TLD, the takedown action was carried out within the first 5 minutes after the URL was blacklisted, while the median value is skewed by a few samples that took up to 21 days to be removed. Overall, the \texttt{.ir} TLD demonstrated the highest efficiency (Figure \ref{fig:survive_phish_tld_1}) in cleaning up phishing content compared to the other abused TLDs in the top 15 list (Table \ref{tab:top_15_phish_survival}).

\begin{table}[hbt!]
    \centering
    \begin{tabular}{lrrr}
        \toprule
        TLD &  \# of domains &                mean &       median \\
        \midrule
        com  &  20,577 & 2 days 03:44:20 & 0 days 12:00:00 \\
        ir   &   3,476 & 0 days 04:18:27 & 0 days 00:00:00 \\
        xyz  &   2,578 & 1 days 20:22:06 & 0 days 06:00:00 \\
        tk   &   1,460 & 0 days 15:34:26 & 0 days 01:00:00 \\
        net  &   1,150 & 2 days 09:14:22 & 0 days 12:00:00 \\
        ml   &    989 & 0 days 21:19:05 & 0 days 01:00:00 \\
        org  &    945 & 2 days 12:11:45 & 0 days 12:00:00 \\
        ga   &    859 & 0 days 16:06:09 & 0 days 01:00:00 \\
        site &    760 & 0 days 07:33:18 & 0 days 00:30:00 \\
        shop &    733 & 4 days 12:34:25 & 2 days 00:00:00 \\
        ru   &    677 & 1 days 17:30:50 & 0 days 12:00:00 \\
        cf   &    667 & 0 days 20:55:12 & 0 days 01:00:00 \\
        gq   &    618 & 0 days 17:56:30 & 0 days 01:00:00 \\
        top  &    475 & 2 days 14:16:02 & 0 days 12:00:00 \\
        cn   &    454 & 3 days 15:13:22 & 2 days 00:00:00 \\
        \bottomrule
    \end{tabular}
    \caption{Uptimes of the phishing URLs for top 15 most abused TLDs.}
    \label{tab:top_15_phish_survival}
\end{table}

 \section {Malicious versus Compromised Domains\label{sec:malcom}}
 
 \subsection{Motivation}
The distinction between compromised and malicious domains is essential for intermediaries involved in the registration and hosting.
When a malicious URL/domain name is encountered (i.e., a URL/domain serving harmful/illegal content), it is crucial to assess the intentions of the domain registrant, as mitigation actions may vary if the registration is done for malicious purposes or not.
While some domains are registered exclusively for malicious purposes, others are benign but get compromised (usually exploiting web vulnerabilities) and misused to serve malicious content.
In many cases, there is no clear-cut distinction between DNS technical versus content-related abuse.

In the remainder of this section, we describe the method used to distinguish between compromised and maliciously registered domain names and present key statistics for TLDs and types of abuse.
Since registrars cannot prevent the abuse of vulnerable hosting software (unless they provide both services), in the following section, we present registrar reputation metrics based on domain names categorized as maliciously registered.

\subsection{Methodology}

There exist two main approaches used in distinguishing compromised from maliciously registered domain names. 
COMAR (Classification of
COmpromised versus MAliciously Registered Domains) \cite{comar} is a tool based on a machine-learning approach developed at Grenoble INP in close collaboration with and funded by two registry operators SIDN (registry of .nl domains) and AFNIC (registry of .fr domains).
COMAR uses a set of 38 features and can achieve high accuracy without using any non-publicly available data, which makes it suitable for use by any organization. 

Another alternative approach \cite{APWG2016} proposed by Greg Aaron and Rod Rasmussen, used a set of heuristics to distinguish maliciously registered domains from compromised websites. A domain name is considered malicious if it was reported shortly after registration or contained a brand name or was registered in a batch, or there was a pattern indicating common owner or intent.
Similar approaches have been used by researchers in various academic papers and industry reports \cite{korczynski2018cybercrime,sadag,landscape}.

In this study, we use a method similar to the second approach. We automatically flag a domain as maliciously registered if it was registered in a batch (i.e., among the blacklisted URLs, there are at least two domain names registered with the same registrar and at precisely the same time). We also automatically flag a domain name as maliciously registered if the time between registration and blacklisting does not exceed three months.
The period was determined based on a sample of manually labeled spam, phishing, malware, and C\&C domains and by tuning the parameter. The parameter giving the highest accuracy is equal to 98 days. We set the threshold to three months because it does not degrade the accuracy on the sample of manually labeled data and is consistent with previous studies \cite{sadag,korczynski2018cybercrime}. 
For phishing attacks, we build a list of 230 brand names. 
From the list of enumerated brand names, 
we generate a list of misspelled versions of brand names using standard methods such as omission, insertion, character substitution, and homographs. If a given FQDN contains a brand name or its misspelled version, the registered domain is also considered to be maliciously registered.
We exclude all free service provider domains from the classification because they are neither compromised nor maliciously registered, and the domains for which we were unable to collect registration information.

\begin{figure}[ht]
  \centering
  
  \subfloat[Phishing]{
    \resizebox{0.2\textwidth}{!}{  
      \begin{tikzpicture}
            \pie[
            /tikz/every pin/.style={align=center},
            color={red!70, cyan!70}
            ]{75.01/, 24.99/}
        \end{tikzpicture}
    }
  }
  \subfloat[Spam]{
    \resizebox{0.2\textwidth}{!}{  
      \begin{tikzpicture}
            \pie[
            color={red!70, cyan!70}
            ]{93.80/, 6.20/}
        \end{tikzpicture}
    }
  }
   \subfloat[Botnet C\&C]{
    \resizebox{0.2\textwidth}{!}{  
      \begin{tikzpicture}
            \pie[
            color={red!70, cyan!70}
            ]{86.67/, 13.33/}
        \end{tikzpicture}
    }
  }
  \subfloat[Malware]{
    \resizebox{0.2\textwidth}{!}{  
      \begin{tikzpicture}
            \pie[
            color={red!70, cyan!70}
            ]{59.46/, 40.54/}
        \end{tikzpicture}
    }
  }

  \caption{Distribution of compromised (blue) and maliciously registered (red) domain names per abuse type. \label{fig:malcompabuse}}
\end{figure}
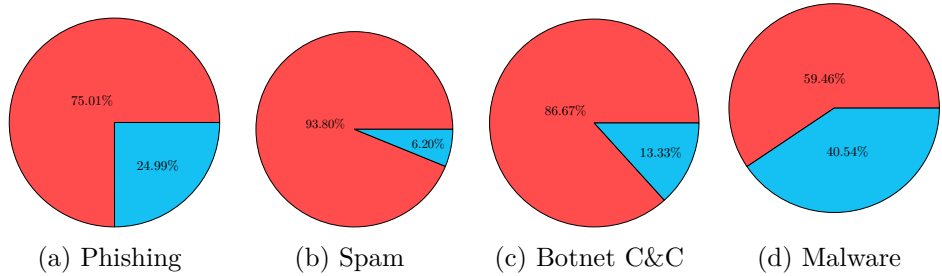
 
\subsection{TLD Statistics: Malicious vs. Compromised}\label{sec:tld_statistics_mal_vs_com}

Figure \ref{fig:malcompabuse} shows the distribution of compromised (blue) and maliciously registered (red) domain names per abuse type.
The vast majority of spam domains (94\%) are presumably maliciously registered, which is expected since they are typically used only to send an email, and attackers need to have complete control over them (maliciously registered domains ensure this).
Alternatively, attackers can gain access to a legitimate email server and use it to send emails on behalf of benign (reputed) domain names.
Domain name (email) spoofing is another attack vector described in Section \ref{sec:spfdmarc}.

We observe similar results for botnet C\&C domain names. Most of them (87\%) are maliciously registered.
The remaining 13\% of alleged compromised domains is more than we expected. That said, there are cases where compromised machines (domain names) are used in C\&C communications as intermediate proxy components that act as an interface between the main C\&C panel and bots (e.g. LokiBot\footnote{\url{https://www.virusbulletin.com/virusbulletin/2020/02/lokibot-dissecting-cc-panel-deployments/}}).

About 25\% and 41\% of phishing and malware domains are presumably benign but compromised at the hosting level.
These proportions may not fully reflect the attacker preferences, as some blacklist providers may develop methods to detect maliciously registered domains rather than compromised websites.
Therefore, the second group may be underrepresented.

In the cases of compromised domains, TLD operators generally should not mitigate abuse at the DNS level because it can cause collateral damage to their benign owners and visitors.
Cybercriminals, when looking for the suitable means to, for example, distribute malicious content such as phishing or malware, must choose the most convenient strategy depending on their skills and available financial resources.
Malicious (and legitimate) users have a vast choice of TLDs whose registrars compete on price, available options such as payment methods using (stolen) PayPal accounts or cryptocurrencies, etc. Therefore, it is not surprising that malicious registrations remain a more attractive option than exploiting website vulnerabilities.

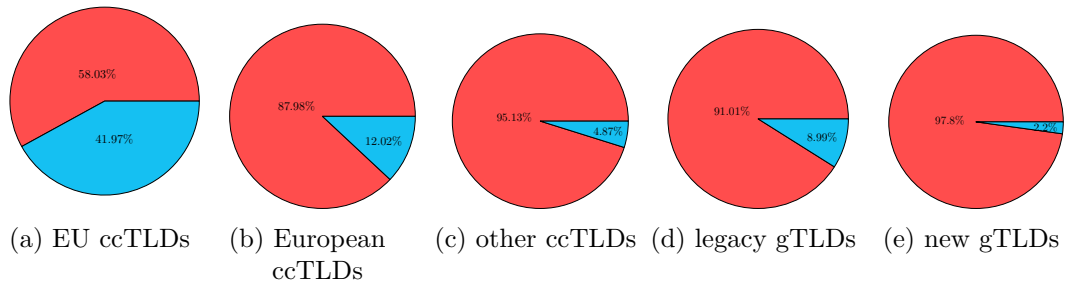
\begin{figure}[ht]
  \centering
  \captionsetup[subfigure]{justification=centering}
  
  \subfloat[EU ccTLDs]{
    \resizebox{0.18\textwidth}{!}{  
      \begin{tikzpicture}
            \pie[
            /tikz/every pin/.style={align=center},
            color={red!70, cyan!70}
            ]{58.03/, 41.97/}
        \end{tikzpicture}
    }
  }
  \subfloat[European ccTLDs][European \newline ccTLDs]{
    \resizebox{0.18\textwidth}{!}{  
      \begin{tikzpicture}
            \pie[
            color={red!70, cyan!70}
            ]{87.98/, 12.02/}
        \end{tikzpicture}
    }
  }
   \subfloat[other ccTLDs]{
    \resizebox{0.18\textwidth}{!}{  
      \begin{tikzpicture}
            \pie[
            color={red!70, cyan!70}
            ]{95.13/, 4.87/}
        \end{tikzpicture}
    }
  }
  \subfloat[legacy gTLDs]{
    \resizebox{0.18\textwidth}{!}{  
      \begin{tikzpicture}
            \pie[
            color={red!70, cyan!70}
            ]{91.01/, 8.99/}
        \end{tikzpicture}
    }
  }
  \subfloat[new gTLDs]{
    \resizebox{0.18\textwidth}{!}{  
      \begin{tikzpicture}
            \pie[
            color={red!70, cyan!70}
            ]{97.8/, 2.2/}
        \end{tikzpicture}
    }
  }
 
  \caption{Distribution of compromised (blue) and maliciously registered (red) domain names per TLD type. \label{fig:malcomptld}}
\end{figure}

Figure \ref{fig:malcomptld} shows the distribution of compromised (blue) and maliciously registered (red) domains in different TLD markets.
The vast majority of new gTLD domain names (97.8\%) were flagged as maliciously registered.
One likely reason is the low price of some of the new gTLDs. 
However, previous research \cite{korczynski2018cybercrime, Halvorson} also shows that in general, for new gTLDs, a relatively large proportion of domain names are either parked or have no content (DNS or HTTP errors), compared to legacy gTLDs.
Intuitively, only domain names
serving content are vulnerable to certain types of exploits and
can be compromised.

For ccTLDs in the EU, compromised websites account for 42\% of all abused domain names.
ccTLDs do not generally aggressively compete on price.
Moreover, ccTLDs are proactive in introducing preventive and reactive security measures to combat DNS abuse more effectively (e.g. .eu, .nl, .dk, .se, .fr, .cz, .sk). Consequently, they generally suffer from much less abuse (see Figure \ref{fig:tld_distribution}), in particular malicious registrations.
To the best of our knowledge, no research has systematically classified the type of content served by ccTLDs (parked domains, HTTP/DNS errors, etc.).
We hypothesize that speculative registrations in EU ccTLDs are less prevalent (fewer parked domains, fewer domains that fail to resolve) and thus more sites serving content and potentially vulnerable to exploits.

\section{Registrar Reputation Metrics \label{sec:regrep}}
\subsection{Methodology}

To measure the reputation of each registrar, similarly to TLDs, we use the \textit{occurrence} of registered domain names and \textit{rate} as security metrics.
We collected and parsed the registration data of domain names of blacklisted URLs (or domains if URLs were not present) as soon as they were blacklisted. As mentioned in the previous section, registrars can only prevent malicious registrations (proactive measures) and remove maliciously registered domains (reactive measures) at the DNS level. They have no control over the hosting infrastructure (unless they also provide a hosting service). Therefore, we compute reputation metrics for domain names that we have determined to be malicious.

To compute abuse rates for each registrar, we had to estimate their sizes.
As described in detail in Section \ref{sec:whois}, using the RDAP and WHOIS protocols, we collected registration information for approximately 241 million domain names (96\% of all active domains we enumerated). 
We were able to parse the registration information and match the domain names of about 85\% of the RDAP/WHOIS records to their respective registrars.
We calculate rates as the number of maliciously registered domains per 10,000 registrations.

\begin{table}[ht!]
\centering
\begin{tabular}{lcc}
\toprule
Name &     Size &  Market share (\%) \\
\midrule
GoDaddy.com, LLC & 63,522,904 &        30.84 \\
NameCheap, Inc. & 10,901,924 &         5.29 \\
Tucows Domains Inc. &  9,492,927 &         4.61 \\
Network Solutions, LLC &  6,393,947 &         3.10 \\
Alibaba Cloud Computing (Beijing) Co., Ltd. &  5,668,641 &         2.75 \\
Google LLC &  5,342,956 &         2.59 \\
1\&1 IONOS SE &  4,861,279 &         2.36 \\
eNom, LLC &  4,650,888 &         2.26 \\
PDR Ltd. d/b/a PublicDomainRegistry.com &  4,564,240 &         2.22 \\
TurnCommerce, Inc. DBA NameBright.com &  3,583,210 &         1.74 \\
GMO Internet, Inc. d/b/a Onamae.com &  3,403,676 &         1.65 \\
OVH sas &  3,208,371 &         1.56 \\
NameSilo, LLC &  3,166,460 &         1.54 \\
Wild West Domains, LLC &  2,842,400 &         1.38 \\
FastDomain Inc. &  2,272,984 &         1.10 \\
\bottomrule
\end{tabular}
\caption{Top 15 registrars based on the estimated overall domain market share.}
\label{tbl_registrar_size}
\end{table}

\subsection{Results}\label{sec:regrep_results}

Table \ref{tbl_registrar_size} shows the 
estimated market share and the top 15 registrars based on the collected and parsed registration data.
Therefore, the reported sizes of each registrar constitute the lower bound of the actual number of registered domain names mainly because of no access to all the domain names globally and not available registration information (WHOIS/RDAP).
Godaddy\footnote{\url{https://godaddy.com}} owns 30.84\% of the market by having more than 63 M domains registered, followed by NameCheap\footnote{\url{https://www.namecheap.com}}, Tucows\footnote{\url{https://tucows.com}} and Network Solutions.\footnote{\url{https://www.networksolutions.com}} 
The presented size estimations are in agreement with the independent market share analysis preformed by \texttt{domainstate.com}.\footnote{\url{https://www.domainstate.com/registrar-stats.html}}

Table \ref{tab:top-10-registrars} shows the top 30 registrars with the highest number of registered (abused) domains and the corresponding abuse rates. 
NameCheap suffers from the highest amount of abused domains followed by GMO Internet, Inc. d/b/a Onamae.com (with IANA ID 49), and GoDaddy.com, LLC (IANA ID 146). The top five most abused registrars account for \pagelabel{text:registrar_mal_reg} 48\% of all maliciously registered domain names. However, in terms of abuse rates, Xi’an Qianxi Network Technology Co. Ltd and EIMS (Shenzhen) Culture \& Technology Co., Ltd have the highest abuse rates of 6,921 and 2,366 maliciously registered domains per 10,000 registrations, respectively (see Table \ref{tab:top-10-registrars-rate}).

The metrics presented here cannot be directly compared to those described in the SADAG study \cite{sadag} due to differences in data collection and aggregation, and the much shorter time frame of the data analyzed.
The SADAG report revealed registrars with disproportionately high concentrations (absolute and relative) of abused domains, such as Nanjing Imperiosus Technology or Alpnames Limited,\footnote{\url{http://domainincite.com/22659-tech-giants-gunning-for-alpnames-over-new-gtld-abuse}} whose accreditation was revoked by ICANN.
Both registrars willingly or unwillingly facilitated cybercrimes.
Note that we focused on ICANN-accredited registrars in this study  because mapping of domain names to the appropriate registrars is complex and error-prone if they are not ICANN-accredited (there is no IANA ID) or WHOIS data is missing.
Consequently, it is much more challenging to identify domain resellers and non-ICANN accredited registrars that potentially facilitate domain name abuse.

\begin{table}[hbt!]
\centering
\resizebox{0.9\textwidth}{!}{
    \begin{tabular}{l c c c}
    \toprule
    Name & IANA ID &  \# of domains &  Rate      \\
    \midrule
                                   NameCheap, Inc. &    1068 & 131,925 &         121 \\
               GMO Internet, Inc. d/b/a Onamae.com &      49 &  93,905 &         276 \\
                                  GoDaddy.com, LLC &     146 &  53,185 &           8 \\
                                     NameSilo, LLC &    1479 &  52,188 &         165 \\
           PDR Ltd. d/b/a PublicDomainRegistry.com &     303 &  38,804 &          85 \\
       Alibaba Cloud Computing (Beijing) Co., Ltd. &     420 &  35,242 &          62 \\
                    PSI-USA, Inc. dba Domain Robot &     151 &  23,485 &         181 \\
  ALIBABA.COM SINGAPORE E-COMMERCE PRIVATE LIMITED &    3775 &  22,139 &         321 \\
                    Xin Net Technology Corporation &     120 &  18,497 &         110 \\
Hongkong Domain Name Information Management Co.... &    2251 &  16,000 &         800 \\
                                  Key-Systems GmbH &     269 &  15,056 &          87 \\
                                      Dynadot, LLC &     472 &  14,835 &          69 \\
 Web Commerce Communications Limited dba WebNic.cc &     460 &  11,700 &         324 \\
                                Launchpad.com Inc. &     955 &  11,251 &         154 \\
                      Eranet International Limited &    1868 &  10,097 &         623 \\
                                          REGRU-RU &     --- &   9,598 &          71 \\
                            Wild West Domains, LLC &     440 &   9,389 &          33 \\
      Hong Kong Juming Network Technology Co., Ltd &    3855 &   8,478 &         721 \\
              Registrar of Domain Names REG.RU LLC &    1606 &   7,396 &         151 \\
                                    Cloud Yuqu LLC &    3824 &   7,025 &         298 \\
                                    Hostinger, UAB &    1636 &   7,012 &         173 \\
                                Register.com, Inc. &       9 &   6,926 &          37 \\
                                         eNom, LLC &      48 &   6,534 &          14 \\
                                      DNSPod, Inc. &    1697 &   6,389 &          84 \\
Chengdu West Dimension Digital Technology Co., ... &    1556 &   6,385 &          80 \\
                                           OVH sas &     433 &   6,318 &          20 \\
                                      Sav.com, LLC &     609 &   6,162 &         118 \\
                                         Beget LLC &    3806 &   5,283 &         465 \\
                                    Name.com, Inc. &     625 &   5,040 &          27 \\
                Xiamen 35.Com Technology Co., Ltd. &    1316 &   4,531 &         247 \\
    \bottomrule
    \end{tabular}
}
\caption{Top 30 registrars with the highest occurrence and rates of abuse.}
\label{tab:top-10-registrars}
\end{table}

\begin{table}[hbt!]
\centering
\resizebox{0.9\textwidth}{!}{
    \begin{tabular}{l c c c}
    \toprule
    Name & IANA ID &  \# of domains &  Rate      \\
    \midrule
          Xi'an Qianxi Network Technology Co. Ltd. &    3825 &    454 &            6,921 \\
     EIMS (Shenzhen) Culture \& Technology Co., Ltd &    2485 &   2,337 &            2,366 \\
Tencent Cloud Computing (Beijing) Limited Liabi... &    3755 &   2,315 &            2,351 \\
             Global Domain Name Trading Center Ltd &    3792 &    892 &            1,231 \\
                               FLAPPY DOMAIN, INC. &    1872 &   1,538 &            1,097 \\
                                  DotMedia Limited &    1863 &    925 &            1,037 \\
                             DOMAINNAME BLVD, INC. &    1870 &    903 &            1,001 \\
                           DOMAIN ORIENTAL LIMITED &    3252 &    428 &             972 \\
                              DOMAINNAME FWY, INC. &    1871 &    715 &             907 \\
                                      MainReg Inc. &    1917 &    182 &             836 \\
          Hefei Juming Network Technology Co., Ltd &    3758 &   3,180 &             798 \\
Hongkong Domain Name Information Management Co.... &    2251 &  16,000 &             800 \\
          NICENIC INTERNATIONAL GROUP CO., LIMITED &    3765 &    987 &             726 \\
      Hong Kong Juming Network Technology Co., Ltd &    3855 &   8,478 &             721 \\
                       Shinjiru Technology Sdn Bhd &    1741 &    908 &             601 \\
                      Eranet International Limited &    1868 &  10,097 &             623 \\
                    BR domain Inc. dba namegear.co &    1898 &    143 &             586 \\
AppCroNix Infotech Private Limited, d/b/a VEBON... &    3844 &     35 &             585 \\
                          Intracom Middle East FZE &    1875 &      1 &             500 \\
                               Taka Enterprise Ltd &    1726 &      2 &             455 \\
                                         Beget LLC &    3806 &   5,283 &             465 \\
                             Topnets Group Limited &    3805 &      2 &             426 \\
                             Vertex Names.com, LLC &    1665 &      2 &             408 \\
                             DomainName Path, Inc. &    1907 &   1,351 &             401 \\
                       DOMAIN NAME NETWORK PTY LTD &    1527 &    795 &             380 \\
                                      Dynadot4 LLC &    1652 &    523 &             377 \\
Atak Domain Hosting Internet ve Bilgi Teknoloji... &    1601 &   3,710 &             348 \\
                             Net-Chinese Co., Ltd. &    1336 &   2,265 &             345 \\
                                     Dynadot10 LLC &    1865 &    480 &             333 \\
                                 CyanDomains, Inc. &    1899 &    900 &             323 \\
    \bottomrule
    \end{tabular}
}
\caption{Top 30 registrars with the highest rate of abuse.}
\label{tab:top-10-registrars-rate}
\end{table}

We manually analyzed the registration patterns of maliciously registered domain names for a few the most abused registrars.
For NameCheap, Inc., for example, we found no registered domain with the keyword ``facebook'', which may be a consequence of the disagreement between Facebook and NameCheap last year regarding the sale of suspicious domains similar to that of Facebook.\footnote{\url{https://www.zdnet.com/article/facebook-sues-namecheap-to-unmask-hackers-who-registered-malicious-domains}}
Such proactive anti-cybersquatting measures significantly increase the barriers to abuse, especially for less-skilled cybercriminals.
Nevertheless, such measures do not eliminate the problem of deceptive subdomains.
An attacker can register a random second-level domain name and then create a subdomain containing deceptive keywords.
However, assuming that the registrar controls the zone file, such a subdomain name can be easily detected when added to the zone file.
Therefore, to avoid detection, the malicious user must delegate the registered domain name to its authoritative name server and add the deceptive subdomain.
In this way, the registrar is unable to detect the malicious subdomain at the DNS level.

Interestingly, for NameCheap, Inc., we identified only three likely maliciously registered domains with subdomains containing the deceptive keyword ``facebook'' (e.g.,  \texttt{facebook.com.marketplace-item-id-135921470.com}).
A manual analysis of phishing data indicates that NameCheap prevents malicious registration of selected brand names.
We found 113 domain names containing the keywords ``wells'' and ``fargo'' (e.g., \texttt{wells-fargo-page.mobi}, or \texttt{wells-fargo-login-security.work}). 
In comparison, for GoDaddy.com, LLC (IANA ID 146), we found maliciously registered domain names containing ``paypal'', ``facebook'' and ``wellsfargo'' in our dataset.
\vspace{+0.5cm}
\pagelabel{text:tld_ipr_rec}
\begin{tcolorbox}[enhanced,colback=blue!5!white,colframe=blue!75!black,colbacktitle=red!80!black]
 \textbf{Recommendation}: TLD registries are encouraged to offer, directly or through the registrars or resellers, services allowing intellectual property rights (IPR) holders to preventively block infringing domain name registrations (similar to services already existing on the gTLD market). 
 \end{tcolorbox}

To effectively reduce domain name abuse, TLD registry operators should maintain access to existing domain/URL blacklists and partner with trusted notifiers.
TLD registries should identify the registrars with the highest concentration and rates of DNS abuse in their ecosystems.

One limitation of this study is that we focus on ICANN-accredited registrars and have limited access to domain names, especially ccTLDs, and limited access to WHOIS.
Operators of ccTLD registries that work with ICANN-accredited registrars, but also accredit registrars locally, have a complete picture of their local market. They should consider developing incentive structures to encourage their registrars to develop methods to  prevent malicious registrations effectively, such as reduced domain registration fees for registrars with the least amount of abuse.

\pagelabel{text:tld_blacklist_rec}
\vspace{+0.5cm}
\begin{tcolorbox}[enhanced,colback=blue!5!white,colframe=blue!75!black,colbacktitle=red!80!black]
 \textbf{Recommendation}: TLD registry operators are encouraged to:
 \vspace{-0.3cm}
 \begin{itemize}
     \item maintain access to existing domain/URL blacklists,
    \vspace{-0.3cm}
     \item identify the registrars with the highest and lowest concentrations and rates of DNS abuse in their ecosystems,
      \vspace{-0.3cm}
     \item develop incentive structures to encourage their registrars to develop methods to  prevent and mitigate malicious registrations effectively.
 \end{itemize}

 \end{tcolorbox}

\begin{figure}
    \centering
    \includegraphics[width=0.9\textwidth]{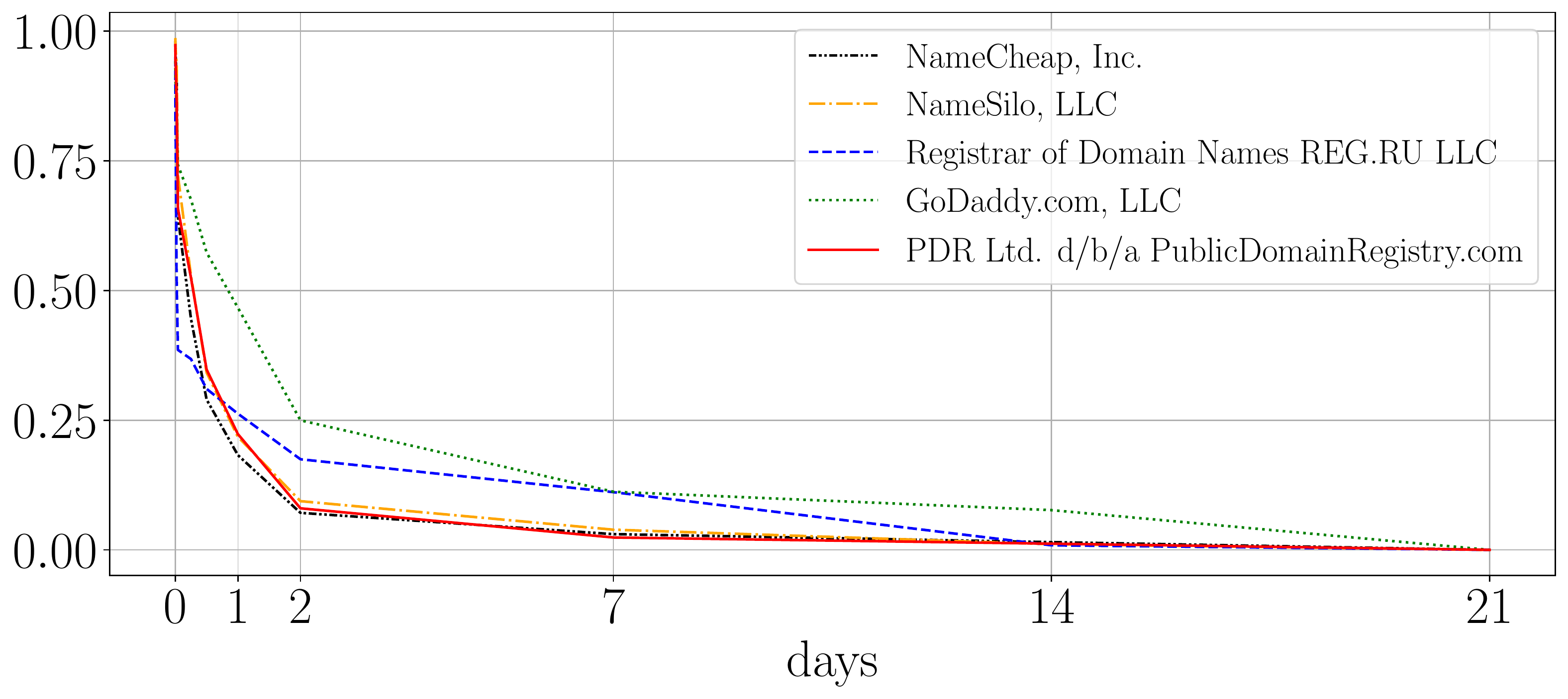}
    \caption{Survival analysis of the top five most abused registrars for phishing.}
    \label{fig:survive_phish_registrars}
\end{figure}

Table \ref{tab:survival-top20-registrars}, shows uptimes (medians and means) of \textit{maliciously registered} phishing domains for the top 30 most abused registrars (for APWG, OpenPhish, and PhishTank data sources). Figure \ref{fig:survive_phish_registrars} shows the survival analysis of the top 5 most abused registrars, corresponding to Table \ref{tab:survival-top20-registrars}. NameCheap, Inc. suffers from the highest number of maliciously registered domain names (5,774). However, Figure \ref{fig:survive_phish_registrars}, shows that in the second quarter of 2021, it mitigated abuse faster compared to other registrars. The median uptime for NameCheap is 6 hours, and the average is one day and 6 hours.  Although it accounts for the highest number of maliciously registered domain names, the survival analysis shows that it is one of the fastest registrars in terms of remediation.

\begin{table}[]
    \centering
    \resizebox{\textwidth}{!}{\begin{tabular}{lrrr}
        \toprule
                                            Registrar &  count &                 mean &        median \\
        \midrule
                                           NameCheap, Inc. &  5,774 &  1 days 06:50:06 &  0 days 06:00:00 \\
                                             NameSilo, LLC &  1,928 &  1 days 12:41:29 &  0 days 12:00:00 \\
                      Registrar of Domain Names REG.RU LLC &  1,025 &  2 days 07:57:14 &  0 days 01:00:00 \\
                                          GoDaddy.com, LLC &    705 &  3 days 16:22:11 &  1 days 00:00:00 \\
                   PDR Ltd. d/b/a PublicDomainRegistry.com &    587 &  1 days 08:29:26 &  0 days 12:00:00 \\
                       GMO Internet, Inc. d/b/a Onamae.com &    475 &  2 days 00:55:39 &  1 days 00:00:00 \\
                                       Tucows Domains Inc. &    409 &  1 days 07:43:38 &  0 days 12:00:00 \\
                                    Wild West Domains, LLC &    392 &  1 days 22:08:03 &  1 days 00:00:00 \\
                                                  REGRU-RU &    186 &  1 days 07:44:13 &  0 days 12:00:00 \\
               Alibaba Cloud Computing (Beijing) Co., Ltd. &    169 &  4 days 16:44:01 &  2 days 00:00:00 \\
                                            Hostinger, UAB &    162 &  0 days 06:43:49 &  0 days 01:00:00 \\
                                   Squarespace Domains LLC &    151 &  0 days 15:58:04 &  0 days 12:00:00 \\
                                            Name.com, Inc. &    146 &  1 days 05:15:45 &  1 days 00:00:00 \\
                                                Google LLC &    129 &  2 days 14:35:48 &  2 days 00:00:00 \\
         Web Commerce Communications Limited dba WebNic.cc &    122 &  1 days 00:03:31 &  0 days 06:00:00 \\
        Alibaba Cloud Computing Ltd. d/b/a HiChina (www... &    110 &  7 days 03:42:40 &  2 days 00:00:00 \\
                                          Key-Systems, LLC &    109 &  6 days 20:44:35 &  2 days 00:00:00 \\
                  Hosting Concepts B.V. d/b/a Registrar.eu &    101 &  0 days 21:17:55 &  0 days 06:00:00 \\
                             West263 International Limited &     95 & 10 days 19:38:31 & 14 days 00:00:00 \\
                                               Porkbun LLC &     93 &  2 days 13:49:01 &  0 days 12:00:00 \\
        \bottomrule
        \end{tabular}}
    \caption{Uptimes of maliciously registered domain names used in phishing for the top 20 most abused registrars (in terms of abuse counts).}
    \label{tab:survival-top20-registrars}
\end{table}

\section{Hosting Providers Reputation \label{sec:rephos}}

\subsection{Motivation}

Malware distribution domains, CSAM, hate speech, intellectual property infringement, or phishing attacks spread harmful and illegal content. For example, as mentioned earlier, phishing attacks contain deceptive content asking for sensitive information, usually infringing on the trademarks of companies such as banks, payment services, and social networking sites. Malware attacks are used to distribute and infect end users with malicious software. CSAM attacks facilitate access to files that harm minors. Therefore, in addition to potential DNS infrastructure abuse, they often represent content abuse.

Some other domains, such as spam (used to send emails containing, for example, phishing content), botnet C\&C, IP fast-flux infrastructures, or domain names redirecting users to other malicious websites (for example, phishing or malware distribution), may not host any malicious or illegal content. Nevertheless, they are involved in attacks and generally resolve to IP addresses and thus, abuse the hosting infrastructure.

Therefore, we resolve each domain name to its IP address and build reputation metrics for hosting providers, more specifically, information society service providers (ISSPs), including access, hosting, and online platform providers.\footnote{\url{https://eur-lex.europa.eu/legal-content/EN/TXT/?uri=CELEX\%3A52016DC0872}} We consider both maliciously registered and compromised domains and exclude legitimate services such as free subdomain providers, which we discuss later.

\subsection{Methodology}
Reliable reputation metrics must account for the commonly observed trend that larger hosting providers also experience more abuse \cite{arman}. A common way to measure the ``size'' of hosting providers is the number of IP addresses routed through the corresponding AS (Autonomous System). However, not all IP addresses routed through an AS are used to host content. IP addresses may be leased and used for other purposes. Inaccuracies in the size estimation can adversely affect the validity of the metric and lead to erroneous results \cite{arman}. Due to the simplicity of computation, the advertised IP space remains a popular choice for size estimation. Another approach is to calculate the portion of the routed IP address space used for hosting as the size estimate. However, this approach is also not free from bias because it favors hosting providers with a disproportionately large volume of shared hosting (e.g., 10,000 domains hosted at a single IP address). Therefore, we use the number of hosted second-level domains as an estimator, which treats shared hosting fairly.

Our approach is also not without bias. For example, infrastructure providers may lease their IP space to other, smaller providers such as hosting services. There may be a chain of resellers difficult to identify even for AS operators. 
For example, Leaseweb provides services to businesses and generally cannot directly control the end users who may host malicious content. However, victims typically contact them in case of abuse, and AS operators should directly contact the reseller, which should mitigate hosting abuse.

To estimate the reputation of hosting providers, we first collect the `A' records of each domain. To get the location information of providers we use the MaxMind dataset.\footnote{\url{https://www.maxmind.com}} It provides us with both the country code and the AS number associated with the `A' record of a domain name. If the domain name has more than one `A' record then we collect the set of ASes from which the domain name is hosted.

We also explore the difference between the set of ASes associated with the FQDN and the registered domains. We found that 97\% of abused domains have the same set of ASes for FQDN and registered domain names, which means that if an attacker registers a domain name maliciously, the chance of managing different subdomains in different ASes is as low as 3\%, as this may increase the cost of, for example, renting servers in different locations.

\begin{table}
  \centering
  \begin{adjustbox}{totalheight=\textheight-2\baselineskip}
    \begin{tabular}{ c  l r r || l r r }
      \cmidrule[0.25ex]{1-7}
      \multirow{15}{*}{\rotatebox[origin=c]{90}{$(0,10k]$}} 
      & \multicolumn{3}{c}{Botnet C\&C} & \multicolumn{3}{c}{Malware} \\
      \cmidrule{2-7}
      & AS & \# Domains & Rate & AS & \# Domains & Rate \\
      \cmidrule{2-7}
      &                                    ACCESS INTERNET &      1 & 10,000  &   Dinas Komunikasi dan Informatika Kota Medan &      1 & 10,000 \\
      &                       LEGON TELECOMUNICACIONES SAS &      1 & 10,000  &VietServer Services technology company limited &      1 & 10,000 \\
      &                                   GLOBALWEB S.R.L. &      1 & 10,000  &              Rajabhat Mahasarakham University &      1 &  5,000 \\
      &                                          SURF B.V. &    352 &  6,132  &     Dm Lot Infotech Solutions Private Limited &      1 &  5,000 \\
      &                    UNICOM JiangSu WuXi IDC network &      3 &  4,286  &                     PT Widara Media Informasi &      1 &  3333 \\
      &                             Trans Tel Services SRL &      1 &  3,333  &               WideBand Communications(Pvt)Ltd &      1 &  2,500 \\
      & TAQUARANET SERVICOS DE PROVEDOR DE INTERNET LTDA M &      1 &  3,333  &                           university of dhaka &      1 &  2,500 \\
      &                                        NETCOM Ltd. &      3 &  3,000  &                         Universitas Brawijaya &      1 &  2,500 \\
      &                                          ERITEL-AS &      5 &  2,500  &                                     SAYDA LLC &     23 &  1,667 \\
      & Bangladesh University of Engineering and Techno... &      1 &  2,000  &                                Electrosim Srl &      3 &  1,667 \\

      \cmidrule[0.25ex]{1-7}
      \multirow{24}{*}{\rotatebox[origin=c]{90}{$(10k,100k]$}} 
      & \multicolumn{3}{c}{Botnet C\&C} & \multicolumn{3}{c}{Malware} \\
      \cmidrule{2-7}
      & AS & \# Domains & Rate & AS & \# Domains & Rate \\
      \cmidrule{2-7}
      &                                 MICROSOFT-CORP-AS &     90 &    69   &   DataWeb Global Group B.V. &    213 &   116 \\
      &CHINATELECOM JiangSu YangZhou IDC networkdescr:... &     11 &     9   &                        CDSI &     58 &    54 \\
      &                                         HURRICANE &     82 &     9   &                         WII &    119 &    25 \\
      &                         Leaseweb Deutschland GmbH &     27 &     7   &                  HOSTROCKET &     30 &    18 \\
      &                                            Fuzhou &      7 &     6   &          AZDIGI Corporation &     27 &    15 \\
      &                                     FRONTIER-FRTR &      7 &     5   &              Host-Africa-AS &     18 &    14 \\
      &                                  Claranet Limited &     17 &     4   &                   ZA-1-Grid &     39 &    13 \\
      &                          FEDERAL-ONLINE-GROUP-LLC &      9 &     4   &                     IOFLOOD &     82 &    12 \\
      &  AS Number for CHINANET jiangsu province backbone &      7 &     4   &               Telepoint Ltd &     40 &    12 \\
      &                                  G-Core Labs S.A. &      4 &     4   & Shinjiru Technology Sdn Bhd &     13 &     9 \\
      \cmidrule[0.25ex]{2-7}
      & \multicolumn{3}{c}{Phishing} & \multicolumn{3}{c}{Spam} \\
      \cmidrule{2-7}
      & AS & \# Domains & Rate & AS & \# Domains & Rate \\
      \cmidrule[0.25ex]{2-7}
      &                 PT. Jupiter Jala Arta &   1,573 &  1,115 &                        GROUP-IID-01 &  12,282 &  3,430 \\
      &           Shinjiru Technology Sdn Bhd &    971 &   703 &       Equinix Japan Enterprise K.K. &   8,205 &  3,305 \\
      &                      G-Core Labs S.A. &    658 &   603 &            FEDERAL-ONLINE-GROUP-LLC &   7,139 &  3,292 \\
      &    Asiatech Data Transmission company &   1,973 &   571 &  EONIX-COMMUNICATIONS-ASBLOCK-62904 &   9,165 &  3,009 \\
      &                         LLC Smart Ape &    605 &   518 &                     Network-Transit &   5,592 &  1979 \\
      &                               CONTABO &    772 &   267 &                 SANREN DATA LIMITED &   8,065 &  1,605 \\
      &Guangdong Mobile Communication Co.Ltd. &    608 &   267 &           DataWeb Global Group B.V. &   2,740 &  1,488 \\
      &    Netmihan Communication Company Ltd &    592 &   203 &                            TIER-NET &   2,577 &  1,331 \\
      &                             HOSTWINDS &    979 &   191 &                        SERVER-MANIA &   2,133 &  1,312 \\
      &                        Host-Africa-AS &    219 &   169 &                    H4Y-TECHNOLOGIES &   1,332 &  1,275 \\
      \cmidrule[0.25ex]{1-7}
      \multirow{24}{*}{\rotatebox[origin=c]{90}{$(100k,1M]$}} 
      & \multicolumn{3}{c}{Botnet C\&C} & \multicolumn{3}{c}{Malware} \\
      \cmidrule{2-7}
      & AS & \# Domains & Rate & AS & \# Domains & Rate \\
      \cmidrule{2-7}
      &                                        STEADFAST &   1,490 &   117   &        AS-COLOCROSSING &    112 &    10 \\
      &                                Private Layer INC &     57 &     4   & PUBLIC-DOMAIN-REGISTRY &    424 &     8 \\
      &                        LeaseWeb Netherlands B.V. &     50 &     2   &                  NOCIX &    207 &     7 \\
      &                   CHINA UNICOM China169 Backbone &     23 &     2   &                DIMENOC &    124 &     6 \\
      &                          Gigabit Hosting Sdn Bhd &     16 &     2   &           Contabo GmbH &    146 &     5 \\
      &             Tencent Building, Kejizhongyi Avenue &     47 &     1   &                 xneelo &     74 &     4 \\
      &                                        CNSERVERS &     42 &     1   &       GO-DADDY-COM-LLC &     94 &     3 \\
      &                      MICROSOFT-CORP-MSN-AS-BLOCK &     73 &     1   &          SERVERCENTRAL &     54 &     3 \\
      &                                         Chinanet &     16 &     1   &          SINGLEHOP-LLC &     64 &     2 \\
      &Shenzhen Tencent Computer Systems Company Limited &     21 &     1   &              STEADFAST &     28 &     2 \\
      \cmidrule[0.25ex]{2-7}
      & \multicolumn{3}{c}{Phishing} & \multicolumn{3}{c}{Spam} \\
      \cmidrule{2-7}
      & AS & \# Domains & Rate & AS & \# Domains & Rate \\
      \cmidrule[0.25ex]{2-7}
      &            AS-COLOCROSSING &   2,319 &   198 &              Clayer Limited &  75,874 &  1,992 \\
      &       ASN-QUADRANET-GLOBAL &   4,399 &    65 &      PUBLIC-DOMAIN-REGISTRY &  49,259 &   919 \\
      &               Contabo GmbH &   1,670 &    53 &            HENGTONG-IDC-LLC &  10,029 &   694 \\
      &               DEDIPATH-LLC &    618 &    53 &    acens Technologies, S.L. &   8,784 &   684 \\
      &              JSC The First &    716 &    46 &                  PEGTECHINC &  28,749 &   561 \\
      &          BGPNET Global ASN &    644 &    41 &         LEASEWEB-USA-LAX-11 &  14,805 &   427 \\
      &MICROSOFT-CORP-MSN-AS-BLOCK &   2,530 &    40 &         LEASEWEB-USA-SFO-12 &   5,819 &   347 \\
      &          Private Layer INC &    597 &    38 &       POWER LINE DATACENTER &  22,405 &   330 \\
      &                  AS-CHOOPA &   1,575 &    32 & ABCDE GROUP COMPANY LIMITED &   6,615 &   310 \\
      &              Online S.a.s. &    629 &    27 &             AS-COLOCROSSING &   3,051 &   261 \\
      \cmidrule[0.25ex]{1-7}
      \multirow{24}{*}{\rotatebox[origin=c]{90}{$(1M,\infty)$}} 
      & \multicolumn{3}{c}{Botnet C\&C} & \multicolumn{3}{c}{Malware} \\
      \cmidrule{2-7}
      & AS & \# Domains & Rate & AS & \# Domains & Rate \\
      \cmidrule{2-7}
      &     Alibaba (US) Technology Co., Ltd. &    119 &     1  &                 UNIFIEDLAYER-AS-1 &   1,369 &     2 \\
      &                            AMAZON-AES &    149 &     1  &                     CLOUDFLARENET &   3,131 &     2 \\
      & Hangzhou Alibaba Advertising Co.,Ltd. &     58 &     1  &                     NAMECHEAP-NET &    644 &     1 \\
      &                           Linode, LLC &     39 &     0  & Alibaba (US) Technology Co., Ltd. &     92 &     1 \\
      &                      DIGITALOCEAN-ASN &     51 &     0  &                        AMAZON-AES &    201 &     1 \\
      &                             Strato AG &     62 &     0  &               Hetzner Online GmbH &    192 &     1 \\
      &                         1\&1 Ionos Se &    147 &     0  &                  DIGITALOCEAN-ASN &    102 &     1 \\
      &                            EGIHOSTING &     21 &     0  &                           OVH SAS &    231 &     1 \\
      &                         CLOUDFLARENET &    241 &     0  &                  Host Europe GmbH &     66 &     0 \\
      &                                GOOGLE &    390 &     0  &         AS-26496-GO-DADDY-COM-LLC &    314 &     0 \\
      \cmidrule[0.25ex]{2-7}
      & \multicolumn{3}{c}{Phishing} & \multicolumn{3}{c}{Spam} \\
      \cmidrule{2-7}
      & AS & \# Domains & Rate & AS & \# Domains & Rate \\
      \cmidrule[0.25ex]{2-7}
      &Alibaba (US) Technology Co., Ltd. &   4,895 &    47   &       DXTL Tseung Kwan O Service &  37,214 &   329 \\
      &                    NAMECHEAP-NET &  19,649 &    40   &                       EGIHOSTING &  20,592 &   186 \\
      &                UNIFIEDLAYER-AS-1 &  15,947 &    29   &Alibaba (US) Technology Co., Ltd. &   6,584 &    63 \\
      &                 DIGITALOCEAN-ASN &   4,564 &    23   &                       AMAZON-AES &  15,180 &    56 \\
      &                    CLOUDFLARENET &  15,217 &    12   &                        AMAZON-02 &  35,236 &    49 \\
      &              Hetzner Online GmbH &   2,894 &    11   &                          OVH SAS &  21,839 &    49 \\
      &                          OVH SAS &   2,755 &     6   &                 DIGITALOCEAN-ASN &   8,275 &    42 \\
      &       Verotel International B.V. &   1,602 &     6   &                 GMO Internet,Inc &   3,822 &    34 \\
      &                       Linode, LLC &    603 &     5   &                      Linode, LLC &   3,784 &    32 \\
      &                        AMAZON-02 &   3,381 &     5   &                    CLOUDFLARENET &  40,367 &    31 \\
      \cmidrule[0.25ex]{1-7}
    \end{tabular}
  \end{adjustbox}
  \caption{Top 10 AS with the highest absolute (\# Domains) relative concentrations (Rate) of blacklisted domains grouped by their corresponding AS size and abuse type.}
  \label{top10as-rates}
  \end{table}

\subsection{Results}\label{sec:reghos_results}

Table \ref{top10as-rates} shows the top 10 ASes grouped into 5 different sizes i.e., ASes with less than 10 K domains, between 10 K and 100 K domains, between 100 K and 1 M domains, and finally greater than 1 M domains. The intuition behind grouping ASes is to compare relatively equal size ASes (in terms of hosted domains) with each other.

\begin{figure}[hbt!]
  \centering
  \subfloat[Occurrence of abuse]{
    \includegraphics[width=0.85\textwidth]{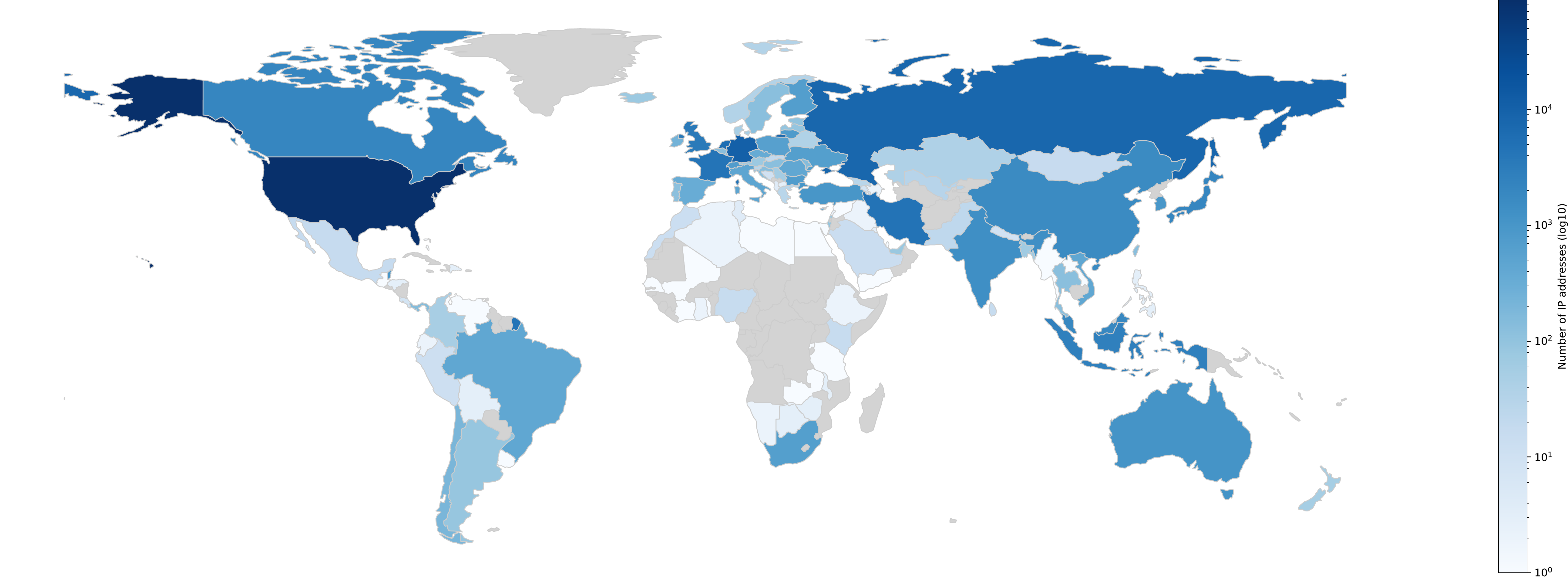}
  }
  \\
  \subfloat[Rate of abuse]{
    \includegraphics[width=0.85\textwidth]{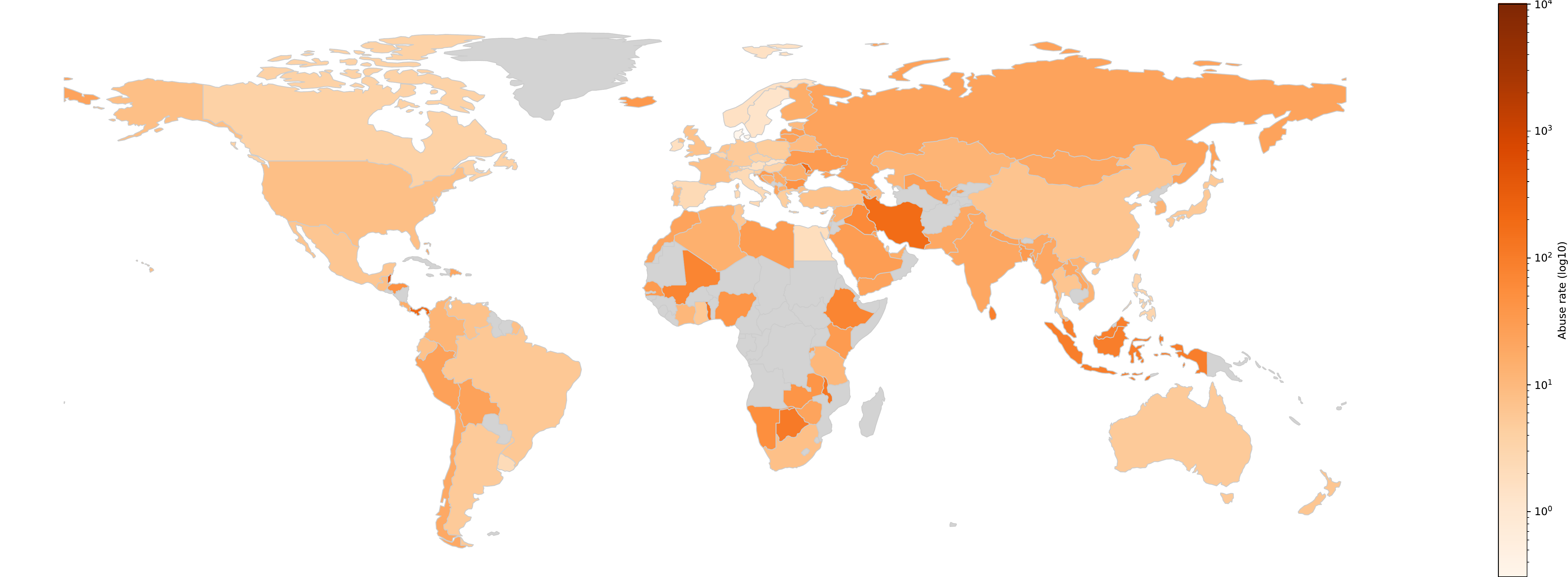}
  }
  \caption{Distribution and rates of IP addresses related to phishing.}
    \label{phishing}
    \end{figure}

\begin{figure}[hbt!]
  \centering
  \subfloat[Occurrence of abuse]{
    \includegraphics[width=0.85\textwidth]{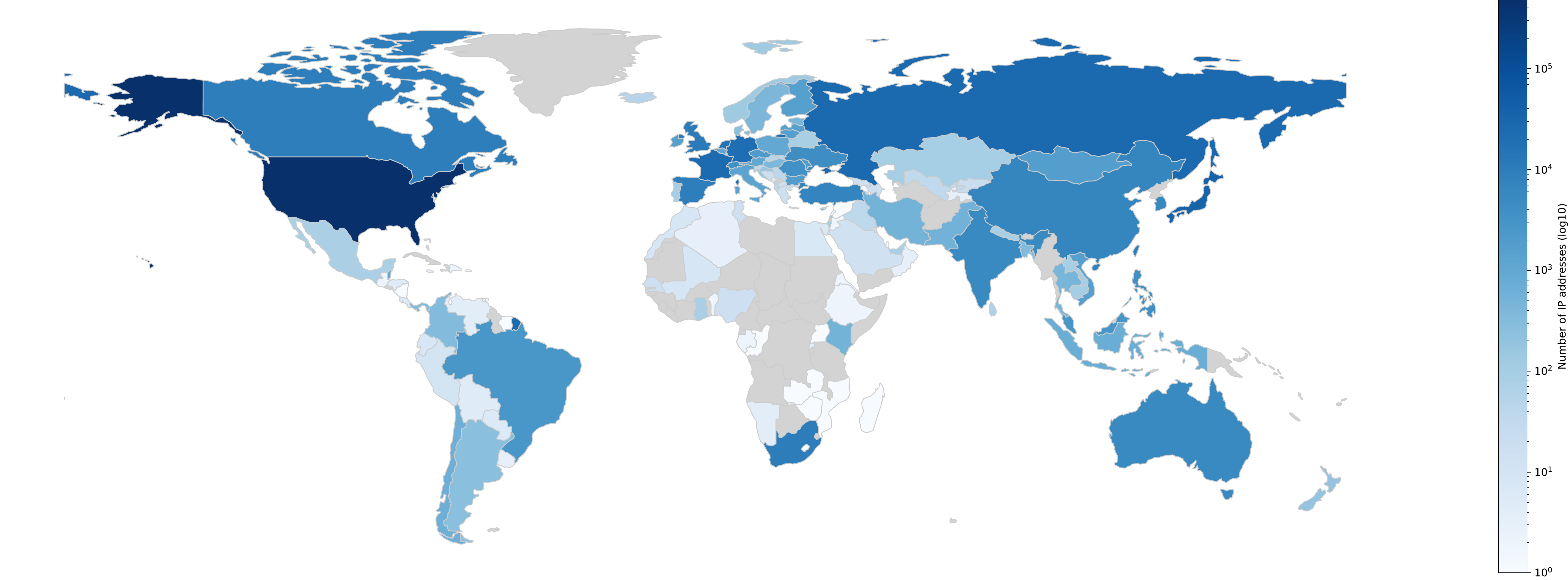}
  }
  \\
  \subfloat[Rate of abuse]{
    \includegraphics[width=0.85\textwidth]{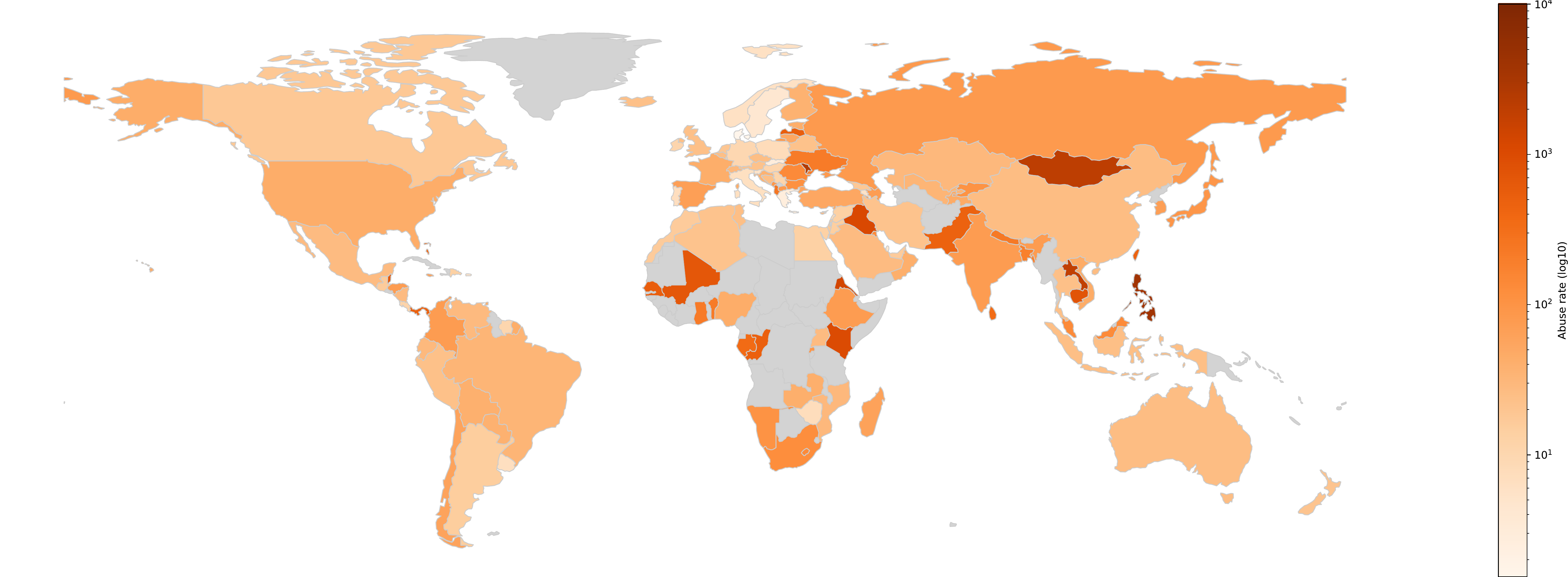}
  }
  \caption{Distribution and rates of IP addresses related to spam.}
  \label{spam}
  \end{figure}

\begin{figure}[hbt!]
  \centering
  \subfloat[Occurrence of abuse]{
    \includegraphics[width=0.85\textwidth]{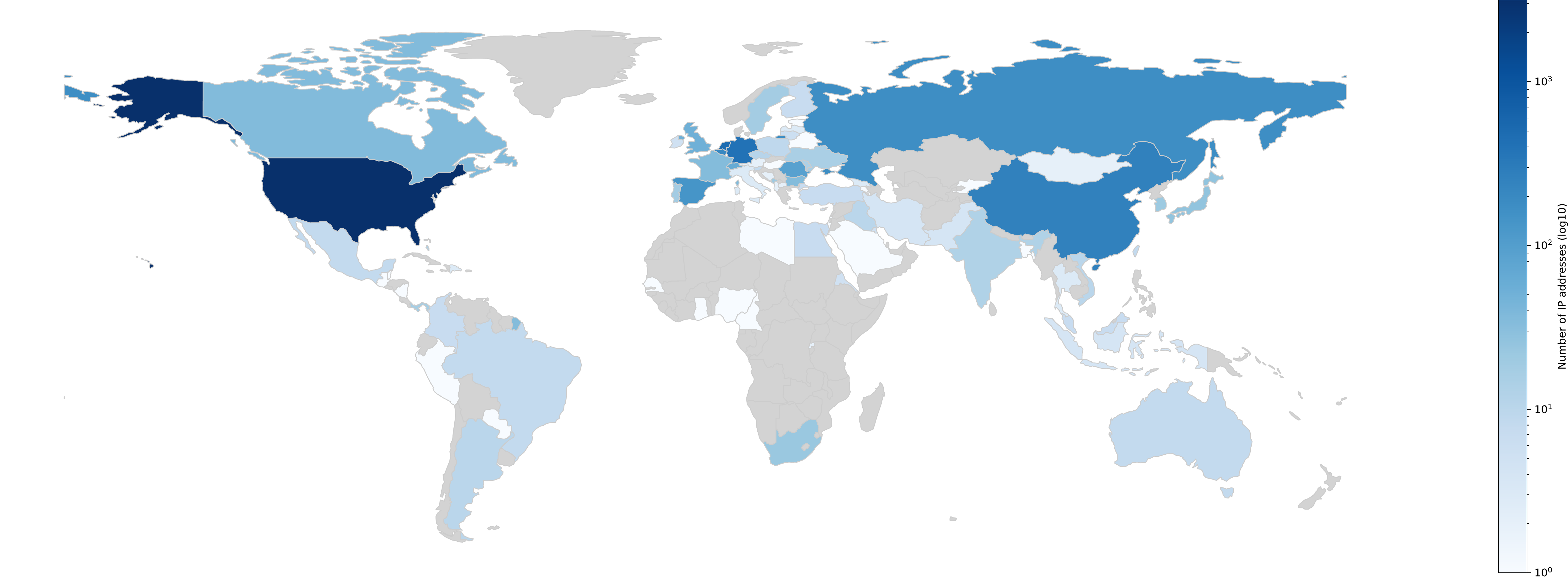}
  }
  \\
  \subfloat[Rate of abuse]{
    \includegraphics[width=0.85\textwidth]{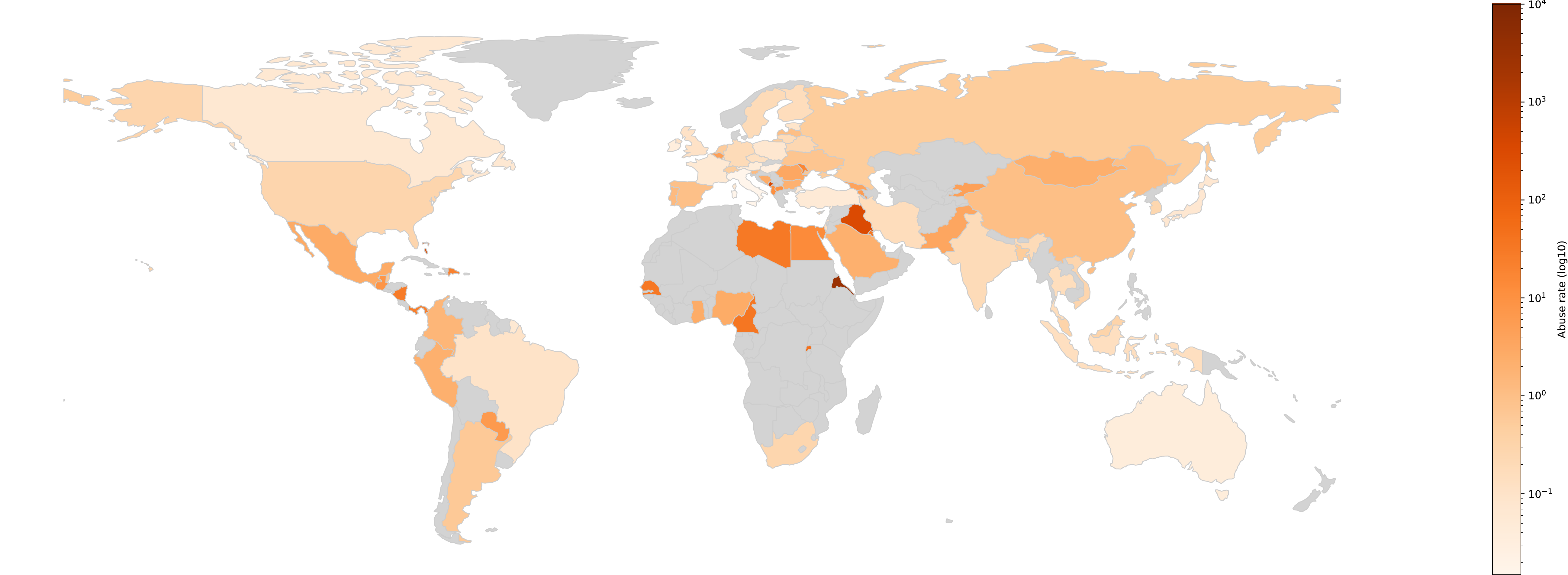}
  }
  \caption{Distribution and rates of IP addresses related to botnet C\&C.}
    \label{cc}
  \end{figure}

\begin{figure}[hbt!]
  \centering
  \subfloat[Occurrence of abuse]{
    \includegraphics[width=0.85\textwidth]{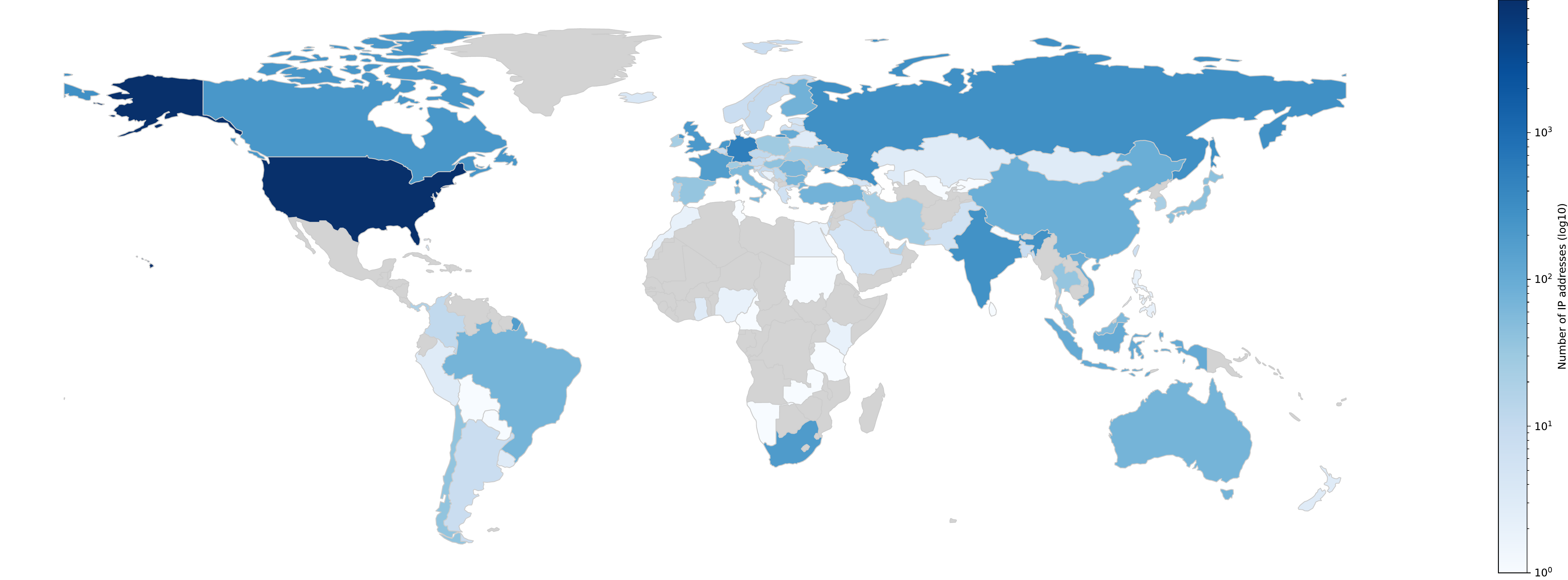}
  }
  \\
  \subfloat[Rate of abuse]{
    \includegraphics[width=0.85\textwidth]{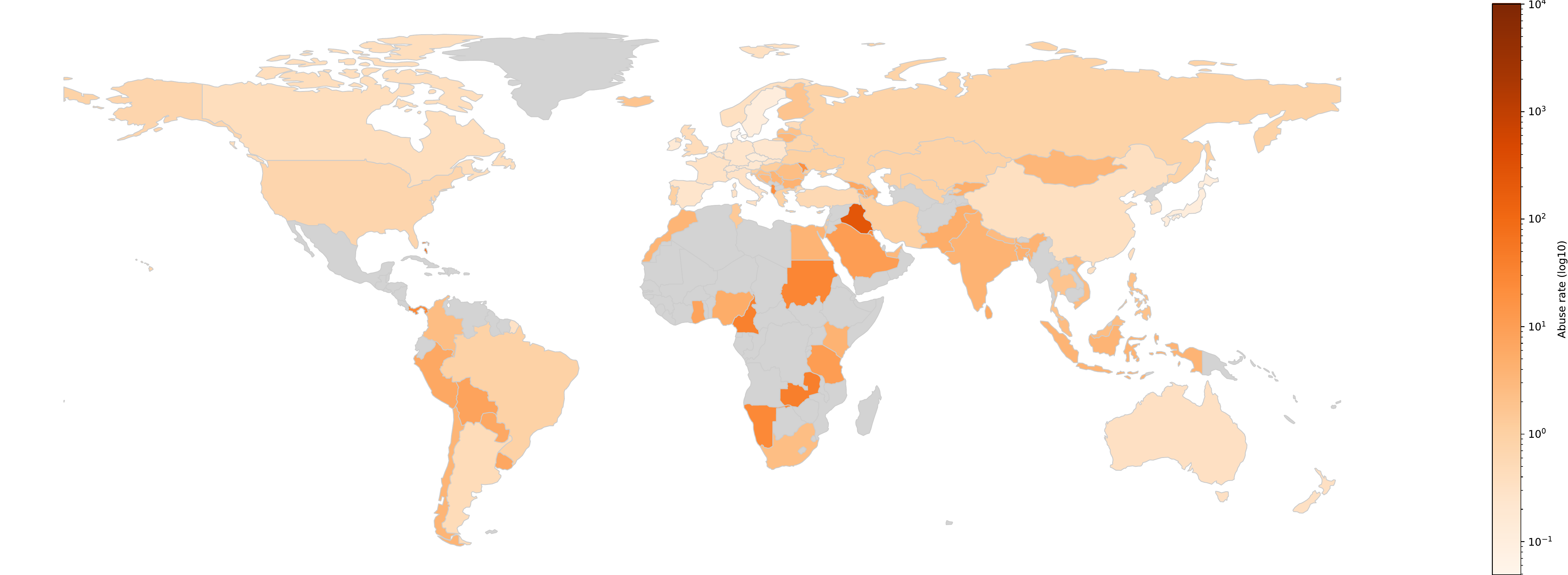}
  }
  \caption{Distribution and rates of IP addresses related to malware distribution.}
\label{mal}
  \end{figure}

Abuse rates for the smallest providers should be considered with caution. Similarly to TLDs or registrars, even a small number of abused domains can significantly affect their reputation.
Among the smallest providers, we may notice universities and educational networks.
Security researchers may deploy honeypots or, for example, harmless sites simulating phishing that can get blacklisted.
For example, SURF B.V.\footnote{\url{https://www.surf.nl}} is an ICT collaborative organization
in Dutch education and research.
Researchers often use their infrastructure in cybersecurity research (e.g., to study phishing \cite{human}).

The second group of providers, in terms of the size, is widely used by spammers. In the top ten most abused providers, \label{text:tld_rep_max} the percentage of spam domains ranges from 1,275 to 3,430 per 10,000 registrations (12,75\% to 34,3\% of all domains). Interestingly, some offshore providers (hosting resellers) claim to use the hosting infrastructure of some of these providers.\footnote{e.g., \url{http://bbar.sottoglialberi.it/free-vps-for-ddosing.html}}
Note that abused hosting providers often offer very cheap servers, such as Contabo, and therefore attract legitimate and malicious users.

Among the largest and most abused hosting providers, we observe companies that also offer cheap domain registration, such as NAMECHEAP-NET, Alibaba (US) Technology Co., Ltd, or OVH SAS. Similarly, competitive pricing may be one of the main reasons why these companies suffer from abuse.
CLOUDFLARENET is among the top ten most abused providers in all four abuse categories (malware, spam, phishing, and botnet C\&C). Note that Cloudflare provides a reverse proxy service that is widely used by cybercriminals. 
When a criminal (or legitimate) site is behind a reverse proxy, the IP address of the backend server hosting, for example, a phishing site, is not visible. Instead, the domain resolves to the Cloudflare IP address, which affects its reputation.

The IP address distributions for each type of abuse and each country (phishing, spam, botnet C\&C, and malware) are shown in Figures \ref{phishing}, \ref{spam}, \ref{cc}, \ref{mal}, respectively.
While in absolute terms, the hosting infrastructure in the US, the Russian Federation, and Germany are generally among the countries with the highest number of abuses, in terms of rates, these countries are relatively less affected.

\pagelabel{text:monitor_abuse_rate_rec}
\vspace{+0.5cm}
\begin{tcolorbox}[enhanced,colback=blue!5!white,colframe=blue!75!black,colbacktitle=red!80!black]
 \textbf{Recommendation}: In a similar manner with respect to the TLD registries and the registrars, the abuse rates of hosting providers should be monitored on an ongoing basis by independent researchers in cooperation with institutions and regulatory bodies (e.g., European Commission, European Union Agency for Cybersecurity – ENISA or national authorities). Abuse rates should not exceed predetermined thresholds. Incentive structures should be studied to induce hosting providers to develop technical solutions that effectively curb hosting and content abuse.
\end{tcolorbox}
\pagelabel{text:tld_rep_max_end}

\pagebreak

\begin{table}[ht!]
    \centering
   \begin{tabular}{lrrr}
\toprule
                            AS &  count &                mean &       median \\
\midrule
                 UNIFIEDLAYER-AS-1 &  3,853 & 3 days 09:19:31 & 1 days 00:00:00 \\
                     CLOUDFLARENET &  3,291 & 2 days 04:16:19 & 0 days 01:00:00 \\
Asiatech Data Transmission company &  1,933 & 0 days 03:14:37 & 0 days 00:00:00 \\
                     NAMECHEAP-NET &  1,671 & 2 days 02:01:42 & 0 days 06:00:00 \\
               Hetzner Online GmbH &  1,633 & 0 days 11:49:26 & 0 days 00:00:00 \\
                           OVH SAS &    802 & 2 days 01:17:45 & 0 days 00:30:00 \\
Netmihan Communication Company Ltd &    577 & 0 days 00:26:55 & 0 days 00:00:00 \\
         AS-26496-GO-DADDY-COM-LLC &    426 & 3 days 09:42:22 & 1 days 00:00:00 \\
                         AMAZON-02 &    392 & 2 days 12:01:41 & 0 days 12:00:00 \\
                  DIGITALOCEAN-ASN &    387 & 3 days 05:09:37 & 1 days 00:00:00 \\
                      Contabo GmbH &    369 & 3 days 04:47:39 & 1 days 00:00:00 \\
            PUBLIC-DOMAIN-REGISTRY &    365 & 3 days 21:38:50 & 1 days 00:00:00 \\
              ASN-QUADRANET-GLOBAL &    257 & 3 days 18:07:48 & 1 days 00:00:00 \\
                          INMOTION &    251 & 3 days 07:41:06 & 1 days 00:00:00 \\
                  DDOS-GUARD CORP. &    226 & 0 days 11:33:01 & 0 days 00:00:00 \\
      Faraso Samaneh Pasargad Ltd. &    214 & 0 days 00:22:27 & 0 days 00:00:00 \\
                            GOOGLE &    211 & 3 days 15:05:42 & 0 days 06:00:00 \\
                   AS-COLOCROSSING &    207 & 2 days 06:37:49 & 0 days 12:00:00 \\
                         LIQUIDWEB &    206 & 2 days 04:41:44 & 0 days 09:00:00 \\
                           IMH-IAD &    204 & 2 days 04:36:19 & 0 days 06:00:00 \\
\bottomrule
\end{tabular}
    \caption{Uptimes of the phishing URLs for top 20 most abused ASes.}
    \label{tab:survival-phish-top20}
\end{table}

\begin{figure}[]
    \centering
    \includegraphics[width=0.9\textwidth]{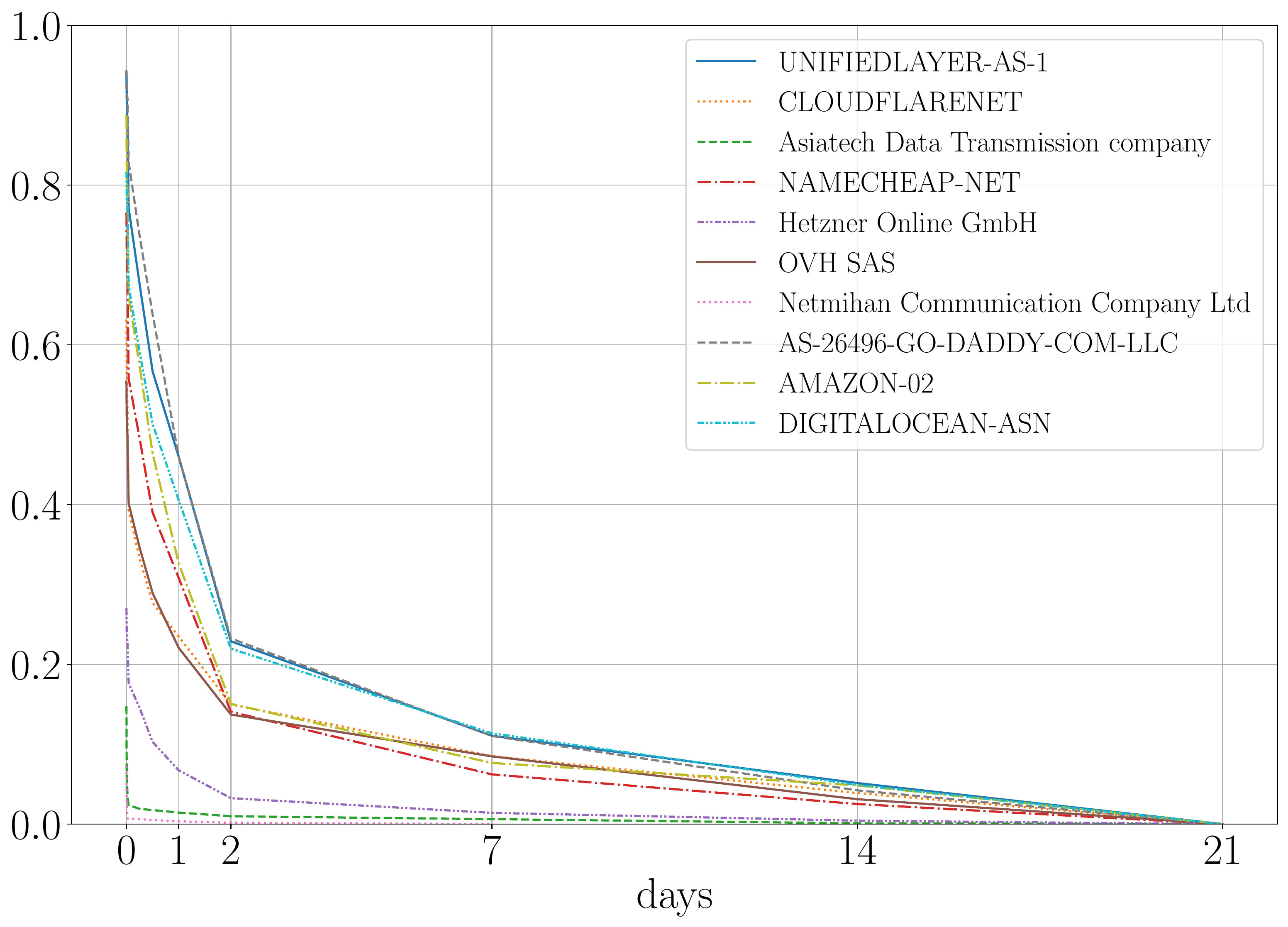}
    \caption{Survival analysis of the 10 most commonly abused ASes for phishing.}
    \label{fig:survive_phish_hp}
\end{figure}

Table \ref{tab:survival-phish-top20}, shows the uptime analysis (mean and median) of phishing URLs for the 20 most abused ASes. `UNIFIEDLAYER-AS-1' suffers from the highest number of phishing domains (3,853 domains), followed by `CLOUDFLARENET' (3,291 domains; as mentioned earlier, Cloudflare provides a proxy service, not hosting) and `Asiatech Data Transmission company' (1,933 domains). Figure \ref{fig:survive_phish_hp} shows the survival analysis of the ten most abused hosting providers shown in Table \ref{tab:survival-phish-top20}. Among these ten, `Netmihan communication Company Ltd' demonstrated the fastest remediation in Q2 2021 with a median of zero and an average of 26 minutes, followed closely by `Asiatech Data Transmission company. Results indicate that `UNIFIEDLAYER-AS-1', which suffered from the highest number of phishing attacks, seems to be among the slowest in terms of remediation efforts.

  \section{Special Domains \label{sec:special}}

Special domains are those whose original intent of registration is not malicious. They are registered by companies to provide legitimate services. Examples of such services include dynamic DNS services, free subdomains, and URL shorter systems. Due to the nature of these services, they are very enticing to cybercriminals. For example, free service operators, such as \texttt{000webhost}\footnote{\url{https://000webhost.com}}, provide free subdomains and free hosting, which is very conducive especially to phishers. 

Table \ref{top10-service-providers} shows the top 10 service providers with the most blacklisted fully qualified domain names (FQDNs) by abuse type. 
\texttt{ngrok}\footnote{\url{https://ngrok.com}} -- the most abused service in phishing -- is a tunneling reverse proxy that establishes secure tunnels from a public endpoint to a locally running web service. A malicious user can hide their backend server (e.g., a compromised end-host with a local IP address) and can use a subdomain under ngrok.io to launch a phishing campaign (e.g., user-account-verification.ngrok.io).
As we can see, phishers make heavy use of such services because, in most cases, they incur no (or meager) cost, which makes them practical for serving malicious content. However, these services are less suitable for spam attacks because criminals use domains to send fake/deceptive emails rather than directly deliver content. Similarly, for botnet C\&C communications, attackers generally need to control domains to ensure communication between compromised end-hosts and malicious servers.

  \begin{table}[h!]
    \centering
    \resizebox{\textwidth}{!}{\begin{tabular}{ l r  || l r  || l r  || l r }
      \toprule
      \multicolumn{2}{c}{Botnet C\&C} & \multicolumn{2}{c}{Malware} & \multicolumn{2}{c}{Phishing} & \multicolumn{2}{c}{Spam} \\
      \midrule
     Provider & \# Domains  & Provider & \# Domains & Provider & \# Domains & Provider & \# Domains \\
      \midrule
      Duck DNS          &  9 & dns.army          &  208 & ngrok             &  23,531 & Google Cloud    &  118 \\
      ChangeiP          &  3 & NoIP              &   92 & 000webhost        &  16,867 & Google Firebase &   30 \\
      000webhost        &  2 & 000webhost        &   41 & Google Firebase   &  13,371 & NoIP            &   14 \\
                        &    & Duck DNS          &   32 & Duck DNS          &   7,252 & amazonaws.com   &   12 \\
                        &    & amazonaws.com     &   23 & Google Cloud      &   5,440 & wixsite.com     &   11 \\
                        &    & soundcast.me      &   14 & NoIP              &   4,004 & blogspot.com    &    6 \\
                        &    & DynuDNS           &   10 & weebly.com        &   3,853 & IBM cloud       &    6 \\
                        &    & tmweb.ru          &    4 & ChangeiP          &   3,340 & glitch.me       &    5 \\
                        &    & weebly.com        &    3 & tmweb.ru          &   3,125 & Duck DNS        &    4 \\
                        &    & blogspot.com      &    2 & yolasite.com      &   1,952 & netlify.app     &    4 \\
      
    \end{tabular}}
    \caption{Top 10 special service providers with the highest occurrence of blacklisted FQDNs per abuse type.   
    \label{top10-service-providers}}
      \end{table}

Subdomains typically contain brand names to lure their victims into entering, for example, credentials (in the case of phishing attacks).
Since such services are commonly abused, providers should employ advanced preventive and reactive solutions to curb subdomain name abuse and hosting infrastructure. They should proactively detect suspicious domain names containing brand keywords most commonly targeted by attacks.
They should also work closely with the most attacked companies and develop trusted notifier programs to remove subdomain names with offensive and abusive content effectively and timely.

\begin{table}[hbt!]
    \centering
    \begin{tabular}{lcc}
    \toprule
                 name &                   mean &          median \\
    \midrule
          appspot.com & 5 days 07:46:21 & 2 days 00:00:00 \\
         blogspot.com & 3 days 02:19:00 & 0 days 12:00:00 \\
            IBM cloud & 1 days 19:38:31 & 0 days 12:00:00 \\
             tmweb.ru & 1 days 10:26:32 & 0 days 12:00:00 \\
            glitch.me & 1 days 02:13:37 & 1 days 00:00:00 \\
      Google Firebase & 1 days 01:31:41 & 0 days 12:00:00 \\
    000webhostapp.com & 0 days 21:20:56 & 0 days 06:00:00 \\
         yolasite.com & 0 days 20:56:32 & 0 days 12:00:00 \\
           weebly.com & 0 days 14:05:16 & 0 days 01:00:00 \\
             ngrok.io & 0 days 00:19:15 & 0 days 00:05:00 \\
    \bottomrule
    \end{tabular}
    \caption{Mean and median uptime of top 10 most abused subdomain service providers.}
    \label{tab:meanmedianspecial}
\end{table}

\begin{figure}[hbt!]
\subfloat[appspot.com]{
\begin{tikzpicture}
        \begin{axis}[
            minor y tick num = 3,
            ymin=0,
            ybar,
            area style,
            symbolic x coords={start, 5min, 30min, 1h, 6h, 12h, 1d, 2d, 1w, 2w, 3w, 4w},
            enlarge x limits={0.01},
            ymajorgrids = true,
            bar width=0.5em,
            x tick label style={rotate=45, anchor=north east},
            xticklabel style={font=\small},
            xtick=data,
            enlarge x limits=0.05
            ]
        \addplot plot coordinates { (start, 0) (5min, 2) (30min, 2) (1h, 4) (6h, 5) (12h, 7) (1d, 8) (2d, 26) (1w, 3) (2w, 7) (3w, 10) (4w, 0)  };
        \end{axis}
        \end{tikzpicture}
}
\subfloat[Firebase (web.app, firebaseapp.com)]{
\begin{tikzpicture}
        \begin{axis}[
            minor y tick num = 3,
            ymin=0,
            ybar,
            area style,
            symbolic x coords={start, 5min, 30min, 1h, 6h, 12h, 1d, 2d, 1w, 2w, 3w, 4w},
            enlarge x limits={0.01},
            ymajorgrids = true,
            bar width=0.5em,
            x tick label style={rotate=45},
            xticklabel style={font=\small},
            xtick=data,
            enlarge x limits=0.05
            ]
        \addplot plot coordinates { (start, 4) (5min, 9) (30min, 26) (1h, 60) (6h, 87) (12h, 394) (1d, 103) (2d, 34) (1w, 15) (2w, 14) (3w, 5) (4w, 0)   };
        \end{axis}
        \end{tikzpicture}
}
\caption{Uptimes subdomain providers (appspot.com, Firebase). \label{fig:uptimessubdomain3}}
\end{figure}
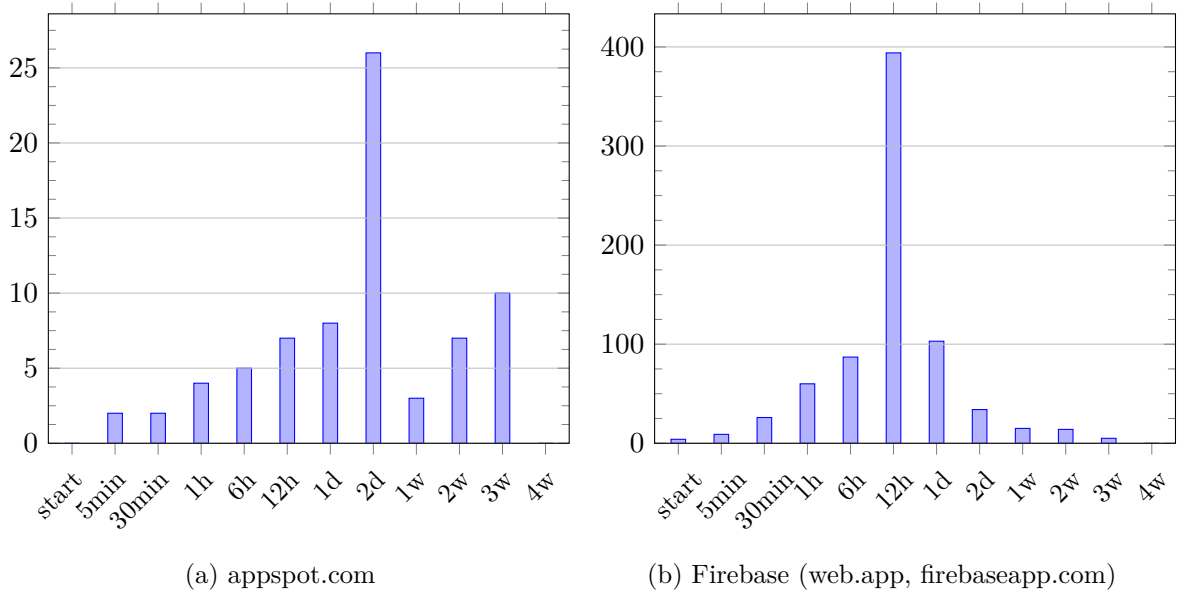

Table \ref{tab:meanmedianspecial} and Figures \ref{fig:uptimessubdomain1}, \ref{fig:uptimessubdomain2}, and \ref{fig:uptimessubdomain3}, show mean, median, and the distribution of uptimes of the 10 most abused subdomain providers.
While the mean uptime is the most intuitive metric, it comes with its limitation: it may be skewed by long-lived malicious domains. To overcome this limitation, we also consider the use of median uptime of abused subdomains. 
The median uptime (Table \ref{tab:meanmedianspecial}) may indicate the policies imposed by different providers.
ngrok.io suspends the majority of malicious subdomains within first 5 minutes after blacklisting, 000webhostapp.com within 6 hours, blogspot.com, IBM Cloud (appdomain.cloud), tmweb.ru, Google Firebase (web.app, firebaseapp.com), and yolasite.com within 12h, glitch.me within one day, whereas appspot.com within two days after blacklisting.
Figure \ref{fig:uptimessubdomain1} shows that ngrok.io (the most abused service according to the collected data) is also the fastest to remediate: 99\% of abused domains are suspended within an hour of being blacklisted.
It may indicate that they actively detect and suspend abusive domains and/or collaborate with trusted notifiers.

\begin{figure}[hbt!]

\subfloat[ngrok.io]{
\begin{tikzpicture}
        \begin{axis}[
            minor y tick num = 3,
            ymin=0,
            ybar,
            area style,
            symbolic x coords={start, 5min, 30min, 1h, 6h, 12h, 1d, 2d, 1w, 2w, 3w, 4w},
            enlarge x limits={0.01},
            ymajorgrids = true,
            bar width=0.5em,
            x tick label style={rotate=45, anchor=north east},
            xticklabel style={font=\small},
            xtick=data,
            enlarge x limits=0.05
            ]
        \addplot plot coordinates { (start, 90) (5min, 94) (30min, 31) (1h, 24) (6h, 1) (12h, 0) (1d, 1) (2d, 0) (1w, 0) (2w, 0) (3w, 0) (4w, 0)  };
        \end{axis}
        \end{tikzpicture}
}
\subfloat[IBM cloud (appdomain.cloud)]{
\begin{tikzpicture}
        \begin{axis}[
            minor y tick num = 3,
            ymin=0,
            ybar,
            area style,
            symbolic x coords={start, 5min, 30min, 1h, 6h, 12h, 1d, 2d, 1w, 2w, 3w, 4w},
            enlarge x limits={0.01},
            ymajorgrids = true,
            bar width=0.5em,
            x tick label style={rotate=45},
            xticklabel style={font=\small},
            xtick=data,
            enlarge x limits=0.05
            ]
        \addplot plot coordinates { (start, 14) (5min, 21) (30min, 34) (1h, 102) (6h, 56) (12h, 68) (1d, 72) (2d, 71) (1w, 39) (2w, 19) (3w, 5) (4w, 0) };
        \end{axis}
        \end{tikzpicture}
}\par
\subfloat[000webhostapp.com]{
\begin{tikzpicture}
        \begin{axis}[
            minor y tick num = 3,
            ymin=0,
            ybar,
            area style,
            symbolic x coords={start, 5min, 30min, 1h, 6h, 12h, 1d, 2d, 1w, 2w, 3w, 4w},
            enlarge x limits={0.01},
            ymajorgrids = true,
            bar width=0.5em,
            x tick label style={rotate=45},
            xticklabel style={font=\small},
            xtick=data,
            enlarge x limits=0.05
            ]
        \addplot plot coordinates { (start, 183) (5min, 250) (30min, 128) (1h, 491) (6h, 464) (12h, 296) (1d, 226) (2d, 199) (1w, 87) (2w, 34) (3w, 5) (4w, 0)  };
        \end{axis}
        \end{tikzpicture}
}
\subfloat[glitch.me]{
\begin{tikzpicture}
        \begin{axis}[
            minor y tick num = 3,
            ymin=0,
            ybar,
            area style,
            symbolic x coords={start, 5min, 30min, 1h, 6h, 12h, 1d, 2d, 1w, 2w, 3w, 4w},
            enlarge x limits={0.01},
            ymajorgrids = true,
            bar width=0.5em,
            x tick label style={rotate=45},
            xticklabel style={font=\small},
            xtick=data,
            enlarge x limits=0.05
            ]
        \addplot plot coordinates { (start, 7) (5min, 19) (30min, 36) (1h, 119) (6h, 69) (12h, 51) (1d, 130) (2d, 224) (1w, 3) (2w, 4) (3w, 1) (4w, 0)  };
        \end{axis}
        \end{tikzpicture}
}
    \caption{Uptimes subdomain providers (ngrok.io, IBM cloud, 000webhostapp.com, glitch.me). \label{fig:uptimessubdomain1}}
\end{figure}
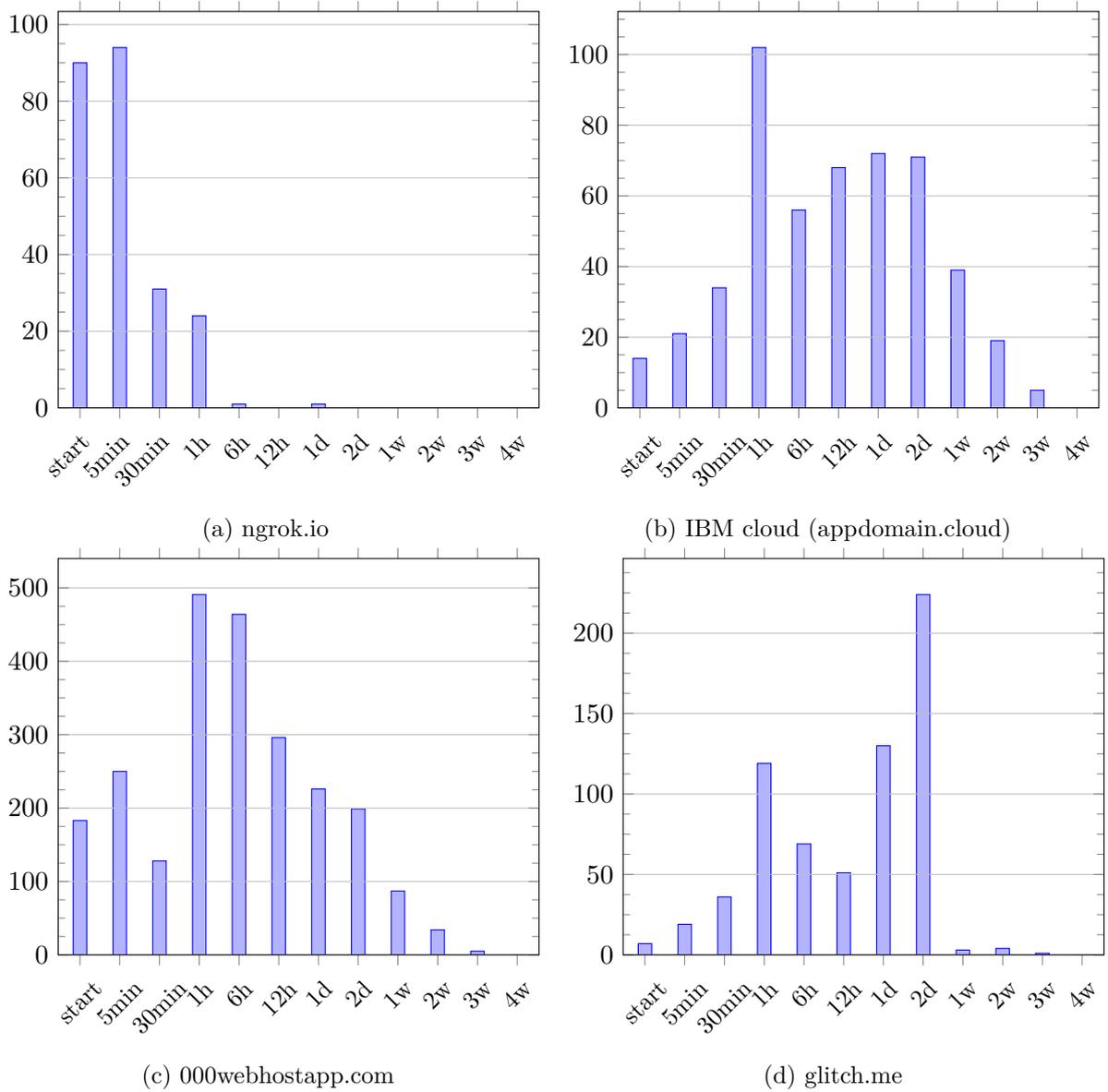

\begin{figure}[hbt!]
\subfloat[blogspot.com]{
\begin{tikzpicture}
        \begin{axis}[
            minor y tick num = 3,
            ymin=0,
            ybar,
            area style,
            symbolic x coords={start, 5min, 30min, 1h, 6h, 12h, 1d, 2d, 1w, 2w, 3w, 4w},
            enlarge x limits={0.01},
            ymajorgrids = true,
            bar width=0.5em,
            x tick label style={rotate=45, anchor=north east},
            xticklabel style={font=\small},
            xtick=data,
            enlarge x limits=0.05
            ]
        \addplot plot coordinates { (start, 4) (5min, 5) (30min, 4) (1h, 22) (6h, 22) (12h, 47) (1d, 39) (2d, 15) (1w, 11) (2w, 22) (3w, 6) (4w, 0)   };
        \end{axis}
        \end{tikzpicture}
}
\subfloat[yolasite.com]{
\begin{tikzpicture}
        \begin{axis}[
            minor y tick num = 3,
            ymin=0,
            ybar,
            area style,
            symbolic x coords={start, 5min, 30min, 1h, 6h, 12h, 1d, 2d, 1w, 2w, 3w, 4w},
            enlarge x limits={0.01},
            ymajorgrids = true,
            bar width=0.5em,
            x tick label style={rotate=45},
            xticklabel style={font=\small},
            xtick=data,
            enlarge x limits=0.05
            ]
        \addplot plot coordinates { (start, 29) (5min, 60) (30min, 31) (1h, 51) (6h, 55) (12h, 211) (1d, 181) (2d, 110) (1w, 2) (2w, 3) (3w, 3) (4w, 0)  };
        \end{axis}
        \end{tikzpicture}
}\par
\subfloat[tmweb.ru]{
\begin{tikzpicture}
        \begin{axis}[
            minor y tick num = 3,
            ymin=0,
            ybar,
            area style,
            symbolic x coords={start, 5min, 30min, 1h, 6h, 12h, 1d, 2d, 1w, 2w, 3w, 4w},
            enlarge x limits={0.01},
            ymajorgrids = true,
            bar width=0.5em,
            x tick label style={rotate=45},
            xticklabel style={font=\small},
            xtick=data,
            enlarge x limits=0.05
            ]
        \addplot plot coordinates { (start, 7) (5min, 22) (30min, 34) (1h, 88) (6h, 52) (12h, 68) (1d, 73) (2d, 74) (1w, 36) (2w, 7) (3w, 2) (4w, 0)  };
        \end{axis}
        \end{tikzpicture}
}
\subfloat[weebly.com]{
\begin{tikzpicture}
        \begin{axis}[
            minor y tick num = 3,
            ymin=0,
            ybar,
            area style,
            symbolic x coords={start, 5min, 30min, 1h, 6h, 12h, 1d, 2d, 1w, 2w, 3w, 4w},
            enlarge x limits={0.01},
            ymajorgrids = true,
            bar width=0.5em,
            x tick label style={rotate=45},
            xticklabel style={font=\small},
            xtick=data,            
            enlarge x limits=0.05
            ]
        \addplot plot coordinates { (start, 74) (5min, 198) (30min, 193) (1h, 738) (6h, 284) (12h, 222) (1d, 110) (2d, 79) (1w, 21) (2w, 14) (3w, 15) (4w, 0)   };
        \end{axis}
        \end{tikzpicture}
}
    \caption{Uptimes subdomain providers (blogspot.com, yolasite.com, tmweb.ru, eebly.com). \label{fig:uptimessubdomain2}}
\end{figure}
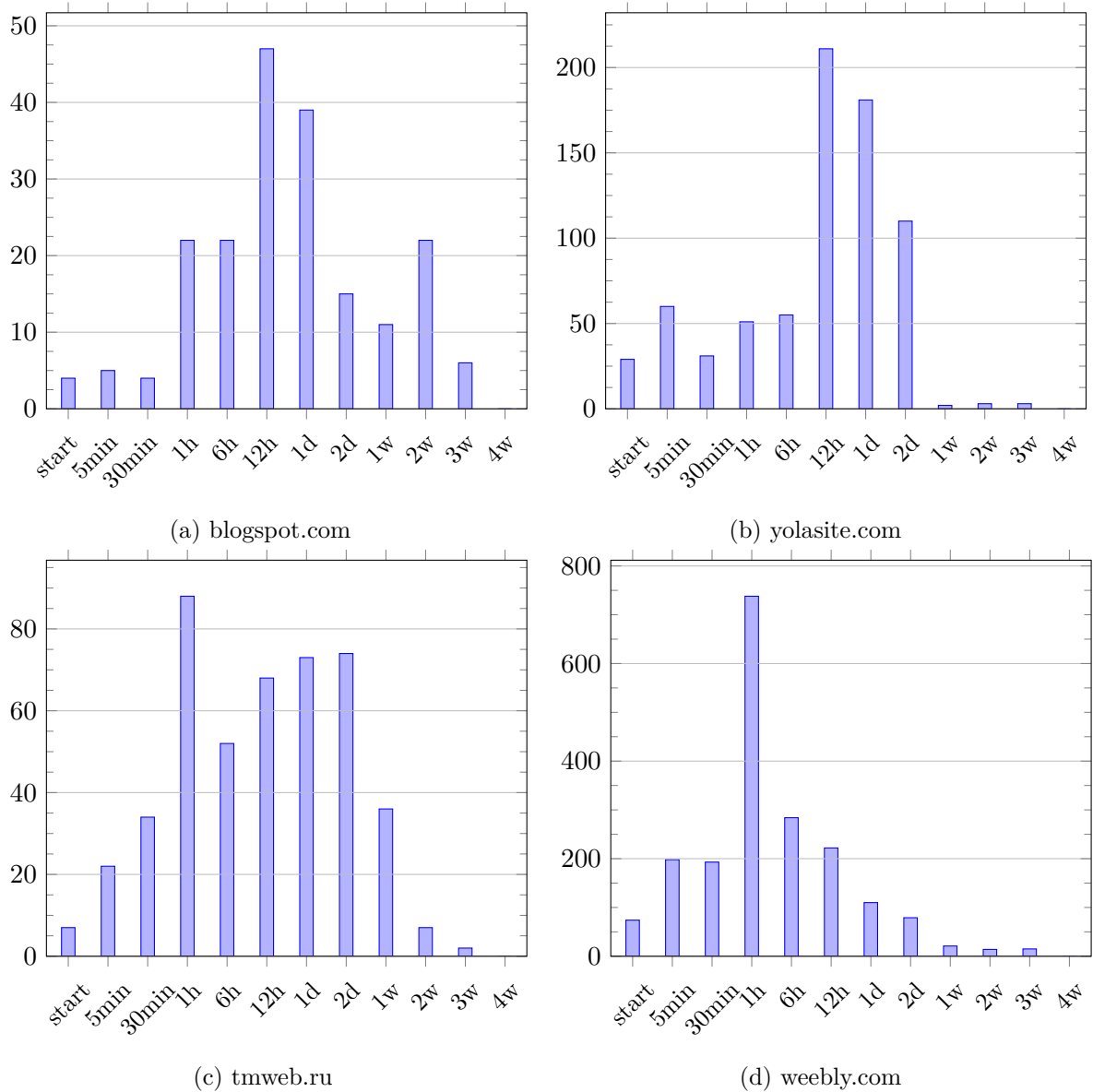

\vspace{+0.5cm}
\begin{tcolorbox}[enhanced,colback=blue!5!white,colframe=blue!75!black,colbacktitle=red!80!black]
  \pagelabel{text:free_service_rec}
  \textbf{Recommendation}: Since free services (e.g., free hosting and subdomains) are commonly exploited in phishing attacks, their operators should employ advanced prevention and remediation solutions to quickly curb abuses of subdomain names and hosting infrastructure. They should proactively detect suspicious domain names containing keywords of the most frequently targeted brands and names and work closely with the most heavily attacked companies and develop trusted notifier programs.
\end{tcolorbox}

  \section{Targeted Brands \label{sec:targeted}}

  \pgfplotstableread[col sep = comma]{data/brands.dat}\datatable
  \begin{figure}[hbt!]
  \resizebox{\textwidth}{!}{
    \begin{tikzpicture}
      \begin{axis}[
        ybar,
        width=\textwidth,
        height=8cm,
        ylabel={Number of attacks},
        xticklabels from table={\datatable}{mapped_brand},
        x tick label style={rotate=45, anchor=north east, inner sep=0mm},
        xtick=data,
        xmin=0,
        xmax=31,
        xticklabel style = {font=\small},
        yticklabel style = {font=\small},
        scaled y ticks=false,
        enlarge x limits={0.01},
        y label style={at={(-0.05,0.5)}},
        ymajorgrids = true,
        bar width=0.5em,
        x=12pt
        ]
        \addplot table [
        x=idx,
        y=count
        ] {\datatable};
      \end{axis}
    \end{tikzpicture}
    }
    \caption{Top 30 most targeted brands.}
    \label{targeted-brands}
      \end{figure}
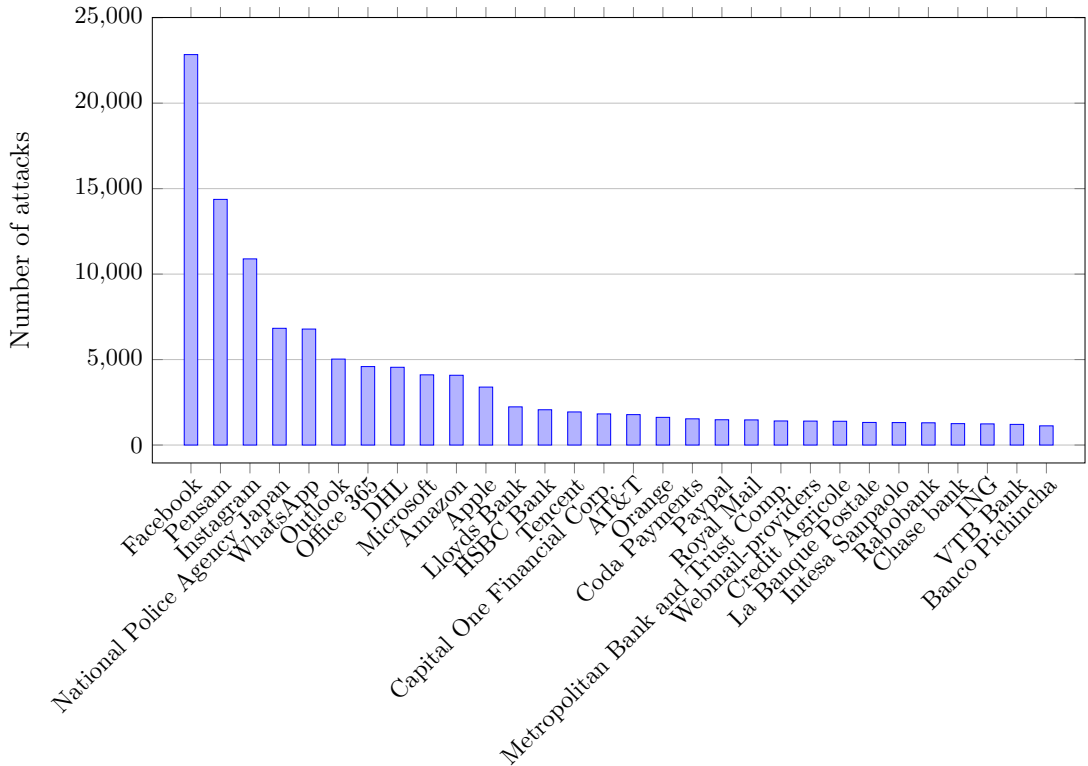

  We also studied 495,085 URLs that APWG, PhishTank, and OpenPhish identified as phishing and containing information about the targeted brands. The brand name in the provided data appears in different variants (e.g. \verb|Facebook| or \verb|Facebook Inc.|) that point to the same entity (\verb|Facebook Inc.|). Therefore, we manually tagged 1,424 variations for 1,076 target brand names. 
  
  Figure~\ref{targeted-brands} shows the 30 most frequent brands in 405,431 URLs (81.89\%) that were reported together with a specific target brand.  As can be seen, Facebook Inc. is the most targeted brand, followed by Pensam (danish  labour market pension fund\footnote{\url{https://www.pensam.dk/in-english}}) and Instagram.
  Among the most targeted brands there are multiple banks such as Lloyds, HSBC, Crédit Agricole, Chase, or Rabobank.
  The results obtained should be considered indicative, as they may be driven (biased) by individual phishing campaigns, or methods of identifying and reporting abuse.

\newpage

\part{DNS Infrastructure Abuse}
\rhead{DNS Infrastructure Abuse}

In this part, we consider the issue of how the technical DNS infrastructure can be abused to perform different types of illegal activities. For example, domain names not deploying Domain Name System Security Extensions (DNSSEC) can be hijacked (cache poisoning \cite{kaminsky} or zone poisoning attacks \cite{zone}) and used in phishing attacks, whereas domains that do not deploy DNS-based email anti-abuse security measures, such as SPF or DMARC, are forged to distribute phishing emails.
Many administrators do not take the precaution of configuring their DNS resolvers to process recursive queries only from internal IP addresses. As a consequence, such open DNS resolvers are increasingly abused to amplify Distributed Denial-of-Service (DDoS) attacks.

In this part, we first discuss the deployment of DNSSEC, followed by DNS resolver vulnerabilities, and the study on SPF and DMARC adoption.

  \section{DNSSEC deployment \label{sec:dnssec}}
  
  \subsection{Motivation}

The domain name system (DNS) was originally designed more than 30 years ago. 
At the time, security was not the primary concern on the Internet. That is why the early DNS standard was found vulnerable to many classes of attacks~\cite{rfc3833}. One of those is \textit{cache poisoning}. When a malicious actor sends a forged reply to a recursive resolver before the genuine reply from an authoritative resolver arrives, it stays in the recursive resolver cache. Such a specifically crafted packet can redirect genuine clients to bogus websites, mail- or name servers.

DNS security extensions (DNSSEC) solve the problem by introducing origin authentication and data integrity~\cite{rfc4033,rfc4035,rfc4034} using a public key infrastructure. DNSSEC is only effective when deployed universally. We analyzed 251 million domain names and found that a small fraction of those attempted to deploy DNSSEC. Even less were correctly signed. We further show that while DNSSEC helps secure certain aspects of DNS, it is also prone to new types of attacks and should be implemented with great caution.

\subsection{Background}

DNSSEC modifies the normal DNS operation by introducing two new concepts - zone signing and response validation. Zone owners generate public/private key pairs. Private keys are used to sign resource record sets (RRsets) and produce \texttt{RRSIG} signatures. The corresponding \texttt{DNSKEY} public keys verify the signatures. Although not required by the DNSSEC standard, there are usually two key pairs - key signing key pair (KSK) and zone signing key pair (ZSK). KSK only signs \texttt{DNSKEY} RRset and its digest is published in the parent zone as \texttt{DS} resource record. ZSK signs the remaining RRsets in the zone.

Zone signing does not protect from manipulation if the keys and signatures are not cryptographically verified. DNSSEC-validating recursive resolvers are pre-configured with one trust anchor, usually the root zone public key (or its digest). The validator follows the chain of \texttt{DS}-\texttt{DNSKEY} resource records from the root zone down the domain name tree to the requested domain name. It ensures that the digests correspond to public keys and public keys verify the signatures. If all the checks are successful, it returns the response with \texttt{NOERROR} status code and \texttt{SERVFAIL} otherwise.

\subsection{DNSSEC Measurements}\label{sec:dnssec_measurements}

We analyze DNSSEC deployment at two different levels. We first show that the majority of TLDs are signed and can be used to publish \texttt{DS} records of their children. We then switch to second-level domain names and observe that DNSSEC suffers low deployment rates.

\textbf{Top-level domains}

Operators of DNSSEC-signed zones assume that validating recursive resolvers will be able to establish a chain of trust from the trust anchor down to the zone. Since 2010, such a universally accepted trust anchor is the root zone KSK~\cite{rootksk}. Once the root zone was signed, TLD operators had an opportunity to sign their zones and upload \texttt{DS} records to the root zone. ICANN publishes a daily report on DNSSEC adoption at the TLD level. As of July 2021, 1,372 out of 1,498 TLDs are signed and publish a key hash at the root~\cite{tld-dnssec}. The last generic top-level domain was signed in December 2020~\cite{gtlds} and all the 126 unsigned TLDs are country-code. 
Note that to implement DNSSEC, the TLD operator must sign the TLD zone. It is the first and most critical step in implementing DNSSEC.
As one of the safeguards proposed by ICANN, all operators of new gTLDs are required to sign the TLD zone \cite{safeguardsAbuse}.

\pagelabel{text:cctld_signed}
\vspace{+0.5cm}
\begin{tcolorbox}[enhanced,colback=blue!5!white,colframe=blue!75!black,colbacktitle=red!80!black]
 \textbf{Recommendation}: Similarly to gTLD registries, the registry operators of ccTLDs should be required to sign TLD zone files with DNS security extensions (DNSSEC) and facilitate its deployment according to good practices. 
\end{tcolorbox}

\textbf{Second-level domains}

DNSSEC-signed zones are different from unsigned ones as they publish additional resource records: \texttt{DS}, \texttt{DNSKEY}, \texttt{RRSIG}, and \texttt{NSEC(3)} that can be queried by recursive resolvers as any other regular resource records, such as \texttt{A}, \texttt{NS} etc. We rely on this fact to enumerate second-level domains that attempted to deploy DNSSEC. We use ZDNS\footnote{https://github.com/zmap/zdns} scanner to send \texttt{DS} and \texttt{DNSKEY}  requests efficiently at scale. We operate it in the nameserver mode so that it forwards all the requests to the recursive resolver of our choice. We then set up a resolver using BIND9.\footnote{https://www.isc.org/bind/} By default, it performs validation of all the received responses. However, we disable this functionality so that we receive the responses even if they are bogus. While scanning for \texttt{DNSKEY}s, we capture all the incoming traffic and extract \texttt{RRSIG} signatures returned along with \texttt{DNSKEY}s. At this stage, we only check for the presence of resource records and not their validity.

\begin{table}[ht!]
\caption{Top 20 TLDs with the highest number of second-level domain in our input list.} 
\vspace{5px}
\label{input} 
\scriptsize
\centering 
\begin{tabular}{lcc}
 \toprule
  TLD & Count & Type\\
   \midrule
\texttt{com} &  145,475,053 & generic \\
\texttt{net} & 12,213,558 &  generic \\ 
\texttt{de} & 9,601,890 & country-code \\ 
\texttt{org} & 9,540,343 &  generic \\ 
\texttt{uk} & 4,263,606 & country-code  \\ 
\texttt{info} & 3,492,481 & generic \\ 
\texttt{ru} & 3,473,332 & country-code  \\ 
\texttt{nl} & 2,741,787 & country-code  \\ 
\texttt{xyz} & 2,516,448 & generic \\ 
\texttt{br} & 2,309,677 & country-code  \\
\texttt{tk} & 2,298,943 & country-code  \\ 
\texttt{ga} & 2,249,643 & country-code  \\ 
\texttt{fr} & 2,098,489 & country-code  \\ 
\texttt{cn} & 1,949,840 & country-code  \\ 
\texttt{it} & 1,758,075 & country-code  \\ 
\texttt{ml} & 1,657,468 & country-code  \\ 
\texttt{eu} & 1,559,517 & country-code  \\ 
\texttt{au} & 1,557,872 & country-code  \\ 
\texttt{cf} & 1,487,356 & country-code  \\ 
\texttt{online} & 1,443,770 &  generic \\ 
 \bottomrule
\end{tabular}
\end{table}

We analyzed DNSSEC deployment of more than 251 million second-level domain names, representing 1,376 TLDs (top 20 TLDs by the number of domains are shown in Table~\ref{input}). Note that for some TLDs for which we have access to their zone files, we evaluate the DNSSEC deployment for all domain names.
However, for most ccTLDs, we assess the deployment based on all enumerated domains rather than all registered domain names (e.g., 9.6 million .de domain names, 3.5 million .ru domain names, or 2.7 .nl domain names). Therefore, the results represent approximate rates of domain names correctly signed with DNSSEC.

Overall, 227 million domain names returned \texttt{NOERROR} responses to our scanner for both queries. We refer to these as \textit{responsive} domains. We exclude from the further analysis the remaining 24 million domains, as we cannot determine whether those are missing some of the resource records or we could not retrieve them for other reasons (temporary network failures etc.).

We first check how many responsive domains contain one or more DNSSEC resource records: \texttt{DNSKEY}, \texttt{RRSIG}, and/or \texttt{DS}. The presence of such records does not necessarily mean that domains are \textit{correctly} signed, but rather signifies that domain owners attempted to do so. 
Only 6.7\%  (15.2 million) of responsive domains publish at least one DNSSEC resource record. Half of those fail to provide all three RRs. Such misconfigurations have different consequences:

\begin{itemize}
    \item \texttt{DNSKEY-RRSIG}, \texttt{DNSKEY}, \texttt{RRSIG}: lack of \texttt{DS} is a common misconfiguration, as this record needs to be manually added to the parent zone (through the registrar control panel). It was previously shown that around 30\% of domains that publish \texttt{DNSKEY} do not have an associated \texttt{DS}~\cite{longitudinal}. The responses from these domains are considered \textit{insecure} by the DNSSEC standard~\cite{rfc4033}. They will not fail the validation check by recursive resolvers, but without a complete chain of trust, we cannot conclude whether the domain is correctly signed. Such DNS zones are referred to as islands of security and can only be used to validate their child zones (if recursive resolvers trust their keys). There are 5.7 million second-level domains from 748 TLDs that fail to provide the \texttt{DS} record while providing the two others (\texttt{DNSKEY} and/or \texttt{RRSIG}).

    \item \texttt{DNSKEY-DS}, \texttt{RRSIG-DS}, \texttt{DS}: domains with \texttt{DS} records at the delegation point have the complete chain of trust and will be verified by validating recursive resolvers. Because of the missing signatures (\texttt{RRSIG}) and/or public keys (\texttt{DNSKEY}), the validation will fail (responses from such domains are called \textit{bogus}) and the end clients will receive \texttt{SERVFAIL} in response to their requests. Such misconfigurations, combined with using validating resolvers, effectively make these domains \textit{unreachable}.
    There are 112,648 second-level domains from 422 TLDs that fail to provide \texttt{DNSKEY} and/or \texttt{RRSIG} while providing \texttt{DS} record. 
  \end{itemize}

These preliminary findings are alarming. The great majority of tested domains do not contain any resource records that would signal the willingness of domain owners to deploy DNSSEC. Only 15.2 million domains contain one or more DNSSEC-related resource records (\texttt{DNSKEY}, \texttt{DS}, \texttt{RRSIG}). However, we see straight away that 37.6\% of them are in the best case, islands of security (because of missing \texttt{DS}) and 0.7\% of them will fail the validation (because of missing public keys and/or signatures).

Note that in addition to TLD registries, registrars also play a key role in the implementation of DNSSEC, as they must add the \texttt{DS} record to the parent zone maintained by the TLD registry. The lack of support from registrars means that all domain names managed by these registrars cannot be signed.
The Danish Ministry of Business has implemented a law requiring the .dk registry to ensure that all registrars who offer domain names in the .dk domain support DNSSEC no later than January 1, 2021 and offer DNSSEC signing to registrants.\footnote{\url{https://www.retsinformation.dk/eli/lta/2020/44} (in danish)}
Some registrars not only facilitate the addition of a \texttt{DS} record to a master zone but provide ``one-click'' DNSSEC deployment as a paid option (e.g., GoDaddy) or even at no cost (e.g., OVH SAS).
The second option is one of the best ways to increase DNSSEC deployment on a massive scale.

\pagelabel{text:dnssec_rec}
\vspace{+0.5cm}
\begin{tcolorbox}[enhanced,colback=blue!5!white,colframe=blue!75!black,colbacktitle=red!80!black]
 \textbf{Recommendation}: To facilitate the implementation of DNSSEC, domain administrators (registrants) should have easy access to DNSSEC signing of domain names within the TLD. TLD registries should require all registrars that offer domain names in the TLD to support DNSSEC signing for registrants.
\end{tcolorbox}

\pagelabel{text:dnssec_correctly_signed}
Domains that do provide all the three resource records (9.4 million) are likely to be correctly signed but need further validation. We switch our BIND9 recursive resolver into validating mode and query these domains for \texttt{SOA} and \texttt{DNSKEY} records. The validating recursive resolver retrieves the requested resource records, performs additional queries to establish the chain of trust and validates the signatures.\pagelabel{text:dnssec_correctly_signed_end}
The results are reassuring: 98.1\% of domains publishing all the three resource records correctly sign both \texttt{DNSKEY} and \texttt{SOA} resource records. Thus, we can conclude that the presence of all the necessary DNSSEC resource records results in a high chance that the zone is \textit{correctly} signed.

Based on our measurements, we categorize all the responsive domains (227 million) into three groups: 
\begin{itemize}
    \item Unsigned (212 million): domains that do not publish any DNSSEC resource records (\texttt{DNSKEY}, \texttt{DS} and \texttt{RRSIG})
    \item Incorrectly signed (6 million): domains that either publish some of DNSSEC resource records or all of them, but the validation fails.
    \item Correctly signed (9 million): domains that publish all the DNSSEC records and when queried by a validating resolver provide correctly signed responses.
\end{itemize}

Tables~\ref{categories_all}, \ref{categories_generic}, \ref{categories_cc}, and \ref{categories_eu} provide information on what TLDs have the highest numbers of second-level domains falling into each category. Table~\ref{categories_all} displays top 20 TLDs of unsigned, incorrectly signed, and correctly signed domains. 
Tables~\ref{categories_generic} and  \ref{categories_cc} show similar ranking among generic and country-code TLDs. 
Table~\ref{categories_eu} shows the number and DNSSEC-deployment rate of European Union TLDs in each category. 

\pagelabel{text:best_dnssec_eucctld}As mentioned earlier, the rates for most ccTLDs were calculated based on a large sample of identified domain names because we do not have access to the zone files and the complete list of domain names. Therefore, the presented rates provide an approximation of the actual adoption.
The DNSSEC adoption rates are not different from the general population and are rather modest---21 out of 34 TLDs consist of more than 90\% of unsigned domains. On the contrary, the \texttt{.cz} TLD exhibits the highest proportion of correctly signed second-level domains.
The \texttt{cz.nic} domain registry achieved it thanks to incentivizing registrars and ISPs economically and supporting them technically~\cite{cz}. Moreover, DNSSEC is a part of the governmental initiative called ``Digital Czech Republic v. 2.0''~\cite{digital}. Swedish country-code TLD comes second with the majority (54.84\%) of correctly signed domains. The \texttt{.se} registry provides guidance on DNSSEC deployment~\cite{sweden} and price incentives. The Dutch TLD \texttt{.nl} has high DNSSEC adoption rates (51.43\%) thanks to the support from both the government and SIDN -- the registry of .nl domains~\cite{sidn-dnssec}. Registrars are charged lower fees for DNSSEC-signed domain names than for unsigned domain names.

The examples of the .cz, .se, .no, or .nl TLDs show that price incentives are the main driving factor behind the implementation of DNSSEC.
All of these registry operators are among those that have used such schemes.\pagelabel{text:best_dnssec_eucctld_end}

\pagelabel{text:dnssec_discounts}
\vspace{+0.5cm}
\begin{tcolorbox}[enhanced,colback=blue!5!white,colframe=blue!75!black,colbacktitle=red!80!black]
 \textbf{Recommendation}: As an incentive to the deployment of DNSSEC, TLD registries might offer discounts for DNSSEC-signed domain names.
 \end{tcolorbox}

\begin{table}[ht!]
\caption{Top 20 TLDs with the highest number of second-level domains falling into each category. The ratio is computed from all the \textit{responsive} domains.} 
\vspace{5px}
\label{categories_all} 
\scriptsize
\centering 
\setlength{\tabcolsep}{4pt}
\begin{tabular}{ccccccccc}
 \toprule
  \multicolumn{3}{c}{Unsigned} & \multicolumn{3}{c}{Incorrectly Signed} & \multicolumn{3}{c}{Correctly Signed} \\
      \cmidrule(lr){1-3}
    \cmidrule(lr){4-6}
    \cmidrule(lr){7-9}
  TLD & Count & Ratio (\%) & TLD & Count & Ratio (\%) & TLD & Count & Ratio (\%) \\
  
 \midrule
\texttt{com} & 122,236,139 & 93.85 & \texttt{com} & 4,905,793 & 3.77 & \texttt{com} & 3,105,826 & 2.38 \\
\texttt{net} & 10,403,214 & 96.30 & \texttt{ru} & 203,715 & 6.43 & \texttt{nl} & 1,367,067 & 51.43 \\
\texttt{de} & 9,230,789 & 97.84 & \texttt{nl} & 162,992 & 6.13 & \texttt{se} & 676,318 & 54.84 \\
\texttt{org} & 8,309,362 & 96.33 & \texttt{net} & 95,869 & 0.89 & \texttt{cz} & 418,299 & 59.23 \\
\texttt{uk} & 3,939,182 & 96.88 & \texttt{org} & 52,810 & 0.61 & \texttt{net} & 303,482 & 2.81 \\
\texttt{info} & 2,970,680 & 97.20 & \texttt{se} & 43,019 & 3.49 & \texttt{fr} & 292,072 & 14.45 \\
\texttt{ru} & 2,959,669 & 93.44 & \texttt{eu} & 41,999 & 2.81 & \texttt{pl} & 279,901 & 21.31 \\
\texttt{ga} & 2,213,920 & 99.95 & \texttt{fr} & 35,661 & 1.76 & \texttt{br} & 265,991 & 12.42 \\
\texttt{tk} & 2,196,953 & 99.88 & \texttt{de} & 31,482 & 0.33 & \texttt{org} & 263,955 & 3.06 \\
\texttt{br} & 1,866,791 & 87.20 & \texttt{cz} & 30,120 & 4.26 & \texttt{eu} & 237,625 & 15.87 \\
\texttt{xyz} & 1,844,580 & 97.92 & \texttt{be} & 28,377 & 3.47 & \texttt{be} & 208,268 & 25.47 \\
\texttt{fr} & 1,693,420 & 83.78 & \texttt{pl} & 24,541 & 1.87 & \texttt{dk} & 200,016 & 29.70 \\
\texttt{it} & 1,685,931 & 99.10 & \texttt{uk} & 21,705 & 0.53 & \texttt{de} & 172,621 & 1.83 \\
\texttt{ml} & 1,644,070 & 99.92 & \texttt{xn--p1ai} & 21,630 & 6.94 & \texttt{no} & 151,435 & 48.82 \\
\texttt{cn} & 1,636,199 & 99.91 & \texttt{co} & 17,798 & 1.60 & \texttt{sk} & 105,044 & 47.74 \\
\texttt{au} & 1,499,598 & 99.56 & \texttt{info} & 15,855 & 0.52 & \texttt{uk} & 104,962 & 2.58 \\
\texttt{cf} & 1,480,148 & 99.95 & \texttt{nu} & 12,690 & 7.94 & \texttt{ch} & 93,150 & 7.42 \\
\texttt{gq} & 1,272,683 & 99.98 & \texttt{no} & 12,614 & 4.07 & \texttt{nu} & 81,041 & 50.68 \\
\texttt{ca} & 1,269,112 & 98.33 & \texttt{hu} & 11,903 & 3.00 & \texttt{hu} & 71,959 & 18.13 \\
\texttt{eu} & 1,217,662 & 81.32 & \texttt{it} & 11,828 & 0.70 & \texttt{info} & 69,699 & 2.28 \\
 \bottomrule
\end{tabular}
\end{table}

\begin{table}[ht!]
\caption{Top 20 generic TLDs with the highest number of second-level domains falling into each category. The ratio is computed for all the \textit{responsive} domains.} 
\vspace{5px}
\label{categories_generic} 
\scriptsize
\centering 
\setlength{\tabcolsep}{3pt}
\begin{tabular}{ccccccccc}
 \toprule
  \multicolumn{3}{c}{Unsigned} & \multicolumn{3}{c}{Incorrectly Signed} & \multicolumn{3}{c}{Correctly Signed} \\
      \cmidrule(lr){1-3}
    \cmidrule(lr){4-6}
    \cmidrule(lr){7-9}
  TLD & Count & Ratio (\%) & TLD & Count & Ratio (\%) & TLD & Count & Ratio (\%) \\
 \midrule
\texttt{com} & 122,236,139 & 93.85 & \texttt{com} & 4,905,793 & 3.77 & \texttt{com} & 3,105,826 & 2.38 \\
\texttt{net} & 10,403,214 & 96.30 & \texttt{net} & 95,869 & 0.89 & \texttt{net} & 303,482 & 2.81 \\
\texttt{org} & 8,309,362 & 96.33 & \texttt{org} & 52,810 & 0.61 & \texttt{org} & 263,955 & 3.06 \\
\texttt{info} & 2,970,680 & 97.20 & \texttt{info} & 15,855 & 0.52 & \texttt{info} & 69,699 & 2.28 \\
\texttt{xyz} & 1,844,580 & 97.92 & \texttt{online} & 10,643 & 0.94 & \texttt{app} & 58,808 & 10.45 \\
\texttt{online} & 1,101,814 & 97.30 & \texttt{xyz} & 8,840 & 0.47 & \texttt{page} & 52,368 & 67.34 \\
\texttt{club} & 748,133 & 98.56 & \texttt{shop} & 5,274 & 1.02 & \texttt{dev} & 48,187 & 24.27 \\
\texttt{vip} & 510,011 & 99.84 & \texttt{site} & 5,131 & 1.06 & \texttt{xyz} & 30,391 & 1.61 \\
\texttt{shop} & 501,711 & 97.50 & \texttt{dev} & 3,719 & 1.87 & \texttt{online} & 19,985 & 1.76 \\
\texttt{app} & 500,579 & 88.94 & \texttt{app} & 3,434 & 0.61 & \texttt{ovh} & 15,413 & 37.29 \\
\texttt{site} & 474,974 & 98.09 & \texttt{store} & 2,860 & 0.91 & \texttt{one} & 12,742 & 25.57 \\
\texttt{top} & 411,033 & 99.45 & \texttt{tech} & 2,568 & 1.28 & \texttt{realty} & 9,272 & 79.63 \\
\texttt{icu} & 398,970 & 99.80 & \texttt{club} & 2,480 & 0.33 & \texttt{club} & 8,453 & 1.11 \\
\texttt{store} & 306,033 & 97.45 & \texttt{cloud} & 1,508 & 1.10 & \texttt{tech} & 8,048 & 4.00 \\
\texttt{live} & 294,922 & 98.35 & \texttt{mobi} & 1,449 & 0.65 & \texttt{shop} & 7,611 & 1.48 \\
\texttt{work} & 266,555 & 99.36 & \texttt{space} & 1,368 & 0.82 & \texttt{cloud} & 6,461 & 4.70 \\
\texttt{mobi} & 220,399 & 98.79 & \texttt{top} & 1,040 & 0.25 & \texttt{store} & 5,139 & 1.64 \\
\texttt{tech} & 190,506 & 94.72 & \texttt{website} & 1,038 & 0.79 & \texttt{studio} & 4,460 & 8.85 \\
\texttt{space} & 161,307 & 97.21 & \texttt{xn--p1acf} & 941 & 6.81 & \texttt{live} & 4,203 & 1.40 \\
\texttt{dev} & 146,628 & 73.86 & \texttt{fun} & 886 & 0.87 & \texttt{site} & 4,112 & 0.85 \\
  \bottomrule
\end{tabular}
\end{table}

\begin{table}[ht!]
\caption{Top 20 country-code TLDs with the highest number of second-level domains falling into each category. The ratio is computed from all the \textit{responsive} domains.} 
\vspace{5px}
\label{categories_cc} 
\scriptsize
\centering 
\begin{tabular}{ccccccccc}
 \toprule
  \multicolumn{3}{c}{Unsigned} & \multicolumn{3}{c}{Incorrectly Signed} & \multicolumn{3}{c}{Correctly Signed} \\
      \cmidrule(lr){1-3}
    \cmidrule(lr){4-6}
    \cmidrule(lr){7-9}
  TLD & Count & Ratio (\%) & TLD & Count & Ratio (\%) & TLD & Count & Ratio (\%) \\
 \midrule
\texttt{de} & 9,230,789 & 97.84 & \texttt{ru} & 203,715 & 6.43 & \texttt{nl} & 1,367,067 & 51.43 \\
\texttt{uk} & 3,939,182 & 96.88 & \texttt{nl} & 162,992 & 6.13 & \texttt{se} & 676,318 & 54.84 \\
\texttt{ru} & 2,959,669 & 93.44 & \texttt{se} & 43,019 & 3.49 & \texttt{cz} & 418,299 & 59.23 \\
\texttt{ga} & 2,213,920 & 99.95 & \texttt{eu} & 41,999 & 2.81 & \texttt{fr} & 292,072 & 14.45 \\
\texttt{tk} & 2,196,953 & 99.88 & \texttt{fr} & 35,661 & 1.76 & \texttt{pl} & 279,901 & 21.31 \\
\texttt{br} & 1,866,791 & 87.20 & \texttt{de} & 31,482 & 0.33 & \texttt{br} & 265,991 & 12.42 \\
\texttt{fr} & 1,693,420 & 83.78 & \texttt{cz} & 30,120 & 4.26 & \texttt{eu} & 237,625 & 15.87 \\
\texttt{it} & 1,685,931 & 99.10 & \texttt{be} & 28,377 & 3.47 & \texttt{be} & 208,268 & 25.47 \\
\texttt{ml} & 1,644,070 & 99.92 & \texttt{pl} & 24,541 & 1.87 & \texttt{dk} & 200,016 & 29.70 \\
\texttt{cn} & 1,636,199 & 99.91 & \texttt{uk} & 21,705 & 0.53 & \texttt{de} & 172,621 & 1.83 \\
\texttt{au} & 1,499,598 & 99.56 & \texttt{xn--p1ai} & 21,630 & 6.94 & \texttt{no} & 151,435 & 48.82 \\
\texttt{cf} & 1,480,148 & 99.95 & \texttt{co} & 17,798 & 1.60 & \texttt{sk} & 105,044 & 47.74 \\
\texttt{gq} & 1,272,683 & 99.98 & \texttt{nu} & 12,690 & 7.94 & \texttt{uk} & 104,962 & 2.58 \\
\texttt{ca} & 1,269,112 & 98.33 & \texttt{no} & 12,614 & 4.07 & \texttt{ch} & 93,150 & 7.42 \\
\texttt{eu} & 1,217,662 & 81.32 & \texttt{hu} & 11,903 & 3.00 & \texttt{nu} & 81,041 & 50.68 \\
\texttt{ch} & 1,158,848 & 92.27 & \texttt{it} & 11,828 & 0.70 & \texttt{hu} & 71,959 & 18.13 \\
\texttt{us} & 1,142,047 & 97.32 & \texttt{dk} & 11,368 & 1.69 & \texttt{co} & 46,556 & 4.20 \\
\texttt{nl} & 1,127,965 & 42.44 & \texttt{us} & 10,369 & 0.88 & \texttt{us} & 21,048 & 1.79 \\
\texttt{co} & 1,044,857 & 94.20 & \texttt{sk} & 8,324 & 3.78 & \texttt{ca} & 16,747 & 1.30 \\
\texttt{pl} & 1,009,203 & 76.82 & \texttt{br} & 8,064 & 0.38 & \texttt{io} & 15,646 & 6.10 \\
   \bottomrule
\end{tabular}
\end{table}

\begin{table}[ht!]
\caption{Country-code TLDs of European Union members with the highest number of second-level domains falling into each category. The ratio is computed for all the \textit{responsive} domains.} 
\vspace{5px}
\label{categories_eu} 
\scriptsize
\centering 
\setlength{\tabcolsep}{3pt}
\begin{tabular}{ccccccccc}
 \toprule
  \multicolumn{3}{c}{Unsigned} & \multicolumn{3}{c}{Incorrectly Signed} & \multicolumn{3}{c}{Correctly Signed} \\
      \cmidrule(lr){1-3}
    \cmidrule(lr){4-6}
    \cmidrule(lr){7-9}
  TLD & Count & Ratio (\%) & TLD & Count & Ratio (\%) & TLD & Count & Ratio (\%) \\
 \midrule
\texttt{xn--qxa6a} & 8 & 100.00 & \texttt{nl} & 162,992 & 6.13 & \texttt{cz} & 418,299 & 59.23 \\
\texttt{xn--qxam} & 321 & 99.69 & \texttt{cz} & 30,120 & 4.26 & \texttt{se} & 676,318 & 54.84 \\
\texttt{ie} & 115,817 & 99.65 & \texttt{xn--e1a4c} & 4 & 3.96 & \texttt{nl} & 1,367,067 & 51.43 \\
\texttt{xn--90ae} & 475 & 99.58 & \texttt{sk} & 8,324 & 3.78 & \texttt{sk} & 105,044 & 47.74 \\
\texttt{hr} & 46,763 & 99.54 & \texttt{se} & 43,019 & 3.49 & \texttt{dk} & 200,016 & 29.70 \\
\texttt{lt} & 94,003 & 99.42 & \texttt{be} & 28,377 & 3.47 & \texttt{be} & 208,268 & 25.47 \\
\texttt{mt} & 4,116 & 99.18 & \texttt{hu} & 11,903 & 3.00 & \texttt{pl} & 279,901 & 21.31 \\
\texttt{si} & 66,691 & 99.15 & \texttt{eu} & 41,999 & 2.81 & \texttt{hu} & 71,959 & 18.13 \\
\texttt{it} & 1,685,931 & 99.10 & \texttt{lv} & 1,392 & 2.67 & \texttt{eu} & 237,625 & 15.87 \\
\texttt{gr} & 211,576 & 99.10 & \texttt{gl} & 44 & 1.90 & \texttt{fr} & 292,072 & 14.45 \\
\texttt{cy} & 5,508 & 98.80 & \texttt{pl} & 24,541 & 1.87 & \texttt{ee} & 8,599 & 13.09 \\
\texttt{bg} & 33,108 & 98.75 & \texttt{fr} & 35,661 & 1.76 & \texttt{xn--e1a4c} & 13 & 12.87 \\
\texttt{at} & 663,527 & 97.95 & \texttt{dk} & 11,368 & 1.69 & \texttt{lv} & 4,425 & 8.49 \\
\texttt{ro} & 301,433 & 97.94 & \texttt{fo} & 30 & 1.59 & \texttt{lu} & 1,620 & 4.31 \\
\texttt{de} & 9,230,789 & 97.84 & \texttt{ro} & 4,753 & 1.54 & \texttt{pt} & 4,692 & 2.91 \\
\texttt{es} & 626,866 & 97.43 & \texttt{pt} & 2,069 & 1.28 & \texttt{fi} & 6,712 & 2.60 \\
\texttt{fo} & 1,831 & 97.14 & \texttt{cy} & 65 & 1.17 & \texttt{gl} & 49 & 2.12 \\
\texttt{fi} & 248,835 & 96.47 & \texttt{fi} & 2,404 & 0.93 & \texttt{es} & 13,302 & 2.07 \\
\texttt{gl} & 2,220 & 95.98 & \texttt{lu} & 345 & 0.92 & \texttt{de} & 172,621 & 1.83 \\
\texttt{pt} & 154,656 & 95.81 & \texttt{mt} & 34 & 0.82 & \texttt{at} & 11,218 & 1.66 \\
\texttt{lu} & 35,633 & 94.77 & \texttt{bg} & 260 & 0.78 & \texttt{fo} & 24 & 1.27 \\
\texttt{lv} & 46,301 & 88.84 & \texttt{it} & 11,828 & 0.70 & \texttt{ro} & 1,572 & 0.51 \\
\texttt{ee} & 56,769 & 86.42 & \texttt{gr} & 1,398 & 0.65 & \texttt{bg} & 158 & 0.47 \\
\texttt{fr} & 1,693,420 & 83.78 & \texttt{si} & 405 & 0.60 & \texttt{xn--90ae} & 2 & 0.42 \\
\texttt{xn--e1a4c} & 84 & 83.17 & \texttt{es} & 3,202 & 0.50 & \texttt{lt} & 242 & 0.26 \\
\texttt{eu} & 1,217,662 & 81.32 & \texttt{ee} & 321 & 0.49 & \texttt{gr} & 527 & 0.25 \\
\texttt{hu} & 313,065 & 78.87 & \texttt{at} & 2,638 & 0.39 & \texttt{si} & 169 & 0.25 \\
\texttt{pl} & 1,009,203 & 76.82 & \texttt{hr} & 159 & 0.34 & \texttt{it} & 3,527 & 0.21 \\
\texttt{be} & 581,024 & 71.06 & \texttt{de} & 31,482 & 0.33 & \texttt{hr} & 58 & 0.12 \\ 
\texttt{dk} & 462,170 & 68.62 & \texttt{lt} & 305 & 0.32 & \texttt{ie} & 128 & 0.11 \\
\texttt{sk} & 106,687 & 48.48 & \texttt{xn--qxam} & 1 & 0.31 & \texttt{cy} & 2 & 0.04 \\
\texttt{nl} & 1,127,965 & 42.44 & \texttt{ie} & 274 & 0.24 & - & - & - \\
\texttt{se} & 513,880 & 41.67 & - & - & - & - & - & -\\
\texttt{cz} & 257,861 & 36.51 & - & - & - & - & - & -\\
   \bottomrule
\end{tabular}
\end{table}

\subsection{Challenges}

DNSSEC has technically solved the problem of forged DNS replies. However, administrators of signed zones face additional maintenance issues, such as key management and signature expiration. We discuss DNSSEC challenges in this section.

\textbf{Amplification of DDoS Attacks}

DNS has long been known as one of the most used protocols to launch reflection and, especially, amplification DDoS attacks~\cite{nxns,exitFromHell,amplificationHell}. DNSSEC introduced a non-negligible overhead to the normal DNS operation because signed responses are larger in size. Van Rijswijk-Deij \textit{et al.}~\cite{ddos} analyzed 2.5 million signed domains and a sample of unsigned domains across 6 TLDs and their amplification factors. While regular queries (\texttt{A}, \texttt{AAAA}, \texttt{DNSKEY}, \texttt{NSEC3}, \texttt{MX}, \texttt{NS}, \texttt{TXT}) do increase the amplification factor compared to normal DNS, it mostly does not exceed the theoretical upper bound. A more serious amplifier is \texttt{ANY} type query, which results in the amplification factor of 47.2 for signed domains versus 5.9 for unsigned. Zone administrators cannot prevent attackers from querying their nameservers. Yet, they can block or provide minimal responses to \texttt{ANY} queries~\cite{rfc8482} and configure the nameservers with response rate/size limiting.

\textbf{Zone Walking}

DNSSEC guarantees that all the responses returned for signed domains are accompanied by signatures. As zones are usually signed once (statically), it is not feasible to precompute and sign all the possible negative responses (for example, when a subdomain does not exist in the zone). Instead of showing that something does not exist in the zone, DNSSEC explicitly lists everything that \textit{does} exist. The new resource records (\texttt{NSEC} and \texttt{NSEC3}) list all the resource record types present for a certain name and the \textit{next} name in the zone in the canonical order. The difference between the two is that \texttt{NSEC} lists names in plaintext while \texttt{NSEC3} hashes them using parameters in \texttt{NSEC3PARAM} resource record. Nevertheless, a determined attacker can attempt to precompute the hashes of all the possible names that could appear in the zone.

Although DNS and DNSSEC were never meant to provide confidentiality, there was no straightforward way to enumerate all the subdomains in the zone. Thus, zone owners could reasonably consider them to be kept private. However, DNSSEC makes it possible to ``walk the zone'' by following the linked list of plaintext \texttt{Next Domain Name} fields in \texttt{NSEC(3)} records. We have scanned all our correctly signed second-level domains and found that 2,4 million of them publish \texttt{NSEC} resource records. It is not a misconfiguration as such, but zone owners should be aware of the information they expose.

\textbf{Signature Validity}

\texttt{RRSIG} signatures introduce the notion of absolute time in DNS. The two fields (\texttt{Signature Inception} and \texttt{Signature Expiration}) are timestamps that specify the time period during which the signature can be used for validation. Validating recursive resolvers use ``their own notion of current time''~\cite{rfc4034} to check that the signature expiration field is greater than or equal to it. We examined 12.8 million signatures across 10.6 million second-level domains and found that 17,376 of them are expired. Responses with such signatures are \textit{bogus}. Zone administrators should make sure that their signatures are always valid. RFC 6781 lists more time-related considerations in DNSSEC~\cite{rfc6781}. For example, signed zones are advised to have TTL values smaller than the signature validity period, which will avoid data being flushed from recursive resolvers caches once signature expiration time is reached. 

\textbf{Key Management}

For DNSSEC to be cryptographically secure, zone administrators should only sign their zones with recommended algorithms, defined in RFC 8624~\cite{rfc8624}. We checked whether domains in our dataset publishing \texttt{DNSKEY} records (15.1 million) adhere to this standard. We found that 25.9\% of all the \texttt{DNSKEY} (25.4 million) implement not recommended algorithms. Very few keys (507) implement algorithms that \textit{must not} be used.

Chung \textit{et al.}~\cite{longitudinal} closely examined some of the common issues when it comes to key management in DNSSEC. Key reuse occurs when one private key is used to sign multiple domains. Although it was found that only 0.5\% of examined keys are shared, one KSK and ZSK were shared among 130,000 domains. If a private key gets compromised, these many domains will be affected at once. Another concern is the key size. The DNSSEC standard does not dictate the key size requirements but the authors refer to NIST recommendations~\cite{nist}. They found that 91.7\% of examined ZSKs were not meeting the minimal size requirements.  

\subsection{Discussion}

DNSSEC remains the most effective way to fight DNS cache poisoning. Yet, it is only effective when deployed universally. 
Surprisingly, many 126 TLDs are still not signed. 
Consequently, their child zones cannot fully deploy DNSSEC because they will not have the complete chain of trust. Out of 
227 million active second-level domain names that we analyzed, a tiny fraction (9.2 million) are correctly signed. 

DNSSEC operation is complex and involves multiple parties: registrants, zone administrators (if different from registants), registrars, TLDs, and operators of recursive DNS resolvers (discussed in the next section). To increase adoption (and validation) of DNSSEC, everyone needs to participate. The remaining unsigned country-code TLDs should adopt DNSSEC to improve their reputation and enable their customers to sign their domains. They should also incentivize registrars to deploy it. Registrars, on their side, can encourage domain owners to deploy DNSSEC by offering them discounts and facilitating the signing process~\cite{registrars}.

\section{DNS Resolvers}\label{sec:dns_resolvers}

In addition to proactive and reactive actions taken by TLD registries, registrars, hosting providers, resellers, and free service providers, 
DNS resolver operators  have also an imperative role in securing the DNS infrastructure. 
Historically, mainly Internet Service Providers (ISPs) were responsible for maintaining DNS resolvers that resolve domain names to their respective IP addresses on behalf of end users.
However, several companies such as Google\footnote{\url{https://developers.google.com/speed/public-dns}}, Cloudflare\footnote{\url{https://www.cloudflare.com/learning/dns/what-is-1.1.1.1/}}, Quad9\footnote{\url{https://www.quad9.net/}}, or OpenDNS\footnote{\url{https://www.opendns.com/setupguide/}} have been offering free and public DNS servers as an alternative way to connect to the Internet in recent years.
One of the main advantages of using public DNS resolvers is to speed up domain name resolution, thereby improving the quality of experience for end users.

One fundamental problem is DNSSEC validation.  In Section \ref{sec:dnssec}, we have described the role of registries and registrars in deploying DNSSEC. However, to protect end-users from cache poisoning attacks, local resolvers must verify the chain of trust to ensure the integrity and authenticity of domain name resolutions.
Even complete deployment of DNSSEC by TLD registries, registrars, and registrants will not protect end users if DNS resolvers do not perform validation.
One of the challenges is to measure whether ISPs perform validation, as it requires performing DNS queries from within the tested networks.
In addition, it is challenging to measure the impact of DNSSEC deployment on global security because detection of cache poisoning attacks can generally be done at the ISP level or using passive DNS data.

\pagelabel{text:isp_dnssec_rec}
\vspace{+0.5cm}
\begin{tcolorbox}[enhanced,colback=blue!5!white,colframe=blue!75!black,colbacktitle=red!80!black]
 \textbf{Recommendation}: Internet Service Providers that operate DNS resolvers should configure DNSSEC validation to protect end users from cache poisoning attacks and ensure the integrity and authenticity of domain name resolutions.
 \end{tcolorbox}

Moreover, regardless of whether DNS resolver service is operated by local ISPs or public resolver operators, they should apply certain measures to improve security of end users.
Service operators may subscribe to blacklists and should not resolve maliciously registered domain names to their IP addresses.
A malicious domain name should resolve with ``NXDOMAIN'' indicating that the domain name does not exist or should be resolved to the DNS service provider own blocking site instead of the IP address of the requested malicious domain.
The Quad9 system uses threat intelligence from more than a dozen leading cybersecurity companies to provide real-time information about which sites contain malware or other threats. 
If the system detects that a site a user wants to visit is infected, it automatically blocks the user from accessing it.\footnote{\url{https://quad9.net/}}
The public resolver operated by Google does not, in principle, perform any blocking.\footnote{\url{https://developers.google.com/speed/public-dns/faq}}
Instead, malicious URLs (and domain names) are blocked by web browsers (e.g., Chrome, Firefox) using Google Safe Browsing.

Another problem is related to open (misconfigured) DNS servers that facilitate amplification reflection Distributed Denial-of-Service (DRDoS) attacks.
Note that open DNS resolvers are not critical for launching DDoS attacks.
For example, in September and October 2016, cybercriminals launched massive DDoS attacks using the Mirai botnet, which did not use the reflection amplification attack vector.\footnote{Scott Hilton. ``Dyn analysis summary of Friday October 21 attack''. In: Dyn Blog, Oct (2016). Available at \url{https://perma.cc/YW5C-MDEV}}
The attack brought down the DNS provider Dyn and several high-profile websites, including Twitter, Guardian, Netflix, Reddit, CNN, and many others in the United States and Europe.
However, reflection and amplification DDoS attacks are mainly carried out using open UDP-based protocols, often DNS resolvers.  Therefore, to increase the barriers to launching modern DDoS attacks, service providers should significantly reduce the number of misconfigured DNS resolvers.
In the following sections, we discuss this problem in detail.

\subsection{Open DNS Resolvers}

Various open services (such as DNS, NTP, SNMP, SSDP, Memcached, etc.) have long been known as efficient DDoS reflectors and powerful amplificators~\cite{amplificationHell,exitFromHell}. 
Among them, DNS resolvers are the most commonly misconfigured and abused.
Open DNS resolvers accept DNS requests from any end-host. These can be misused to either target authoritative nameservers by sending the excessive number of incoming requests or, if combined with IP address spoofing, used to redirect responses to victim end-hosts. In this project, we actively scan for open DNS resolvers in IPv4 and IPv6 address spaces and analyze their distribution across organisations and countries.

\subsection{Methodology}

Scanning for open resolvers requires sending DNS requests to end-hosts and inspecting the received responses. The response codes (\texttt{RCODE}), defined in RFC 1035~\cite{rfc1035}, signal whether the DNS server processes incoming requests. If the query resolution is successful, open resolvers send back the responses to end clients along with \texttt{NOERROR} status code. 

We use the three following datasets to scan for open resolvers: IPv4 BGP prefixes~\cite{routeviews}, IPv6 Hitlist Service~\cite{hitlist}, and IPv6 addresses learned by traversal from IPv4 to resolve IPv6-only domains, as described by Korczy\'{n}ski \textit{et al.}~\cite{korczyski2020closed}. All the three contain globally reachable IP addresses that may be operational recursive resolvers. Each end-host from the list receives an `A' request for the unique domain name under our authority. We developed a software tool that allows us efficiently sending DNS packets at large scale \cite{AXFR}. 

\subsection{Analysis}

\textbf{Scan results}

We performed IPv4 and IPv6 open resolver scans in March 2021. Having tested more than 2.8 billion routable IPv4 addresses and 3.5 million IPv6 addresses, we discovered 3.4 million IPv4 and 18,843 IPv6 open recursive resolvers.

\textbf{Response Integrity}

Although the above mentioned open resolvers returned the \texttt{NOERROR} responses, they are not necessarily \textit{correctly} operating. We closely inspected the answer section of returned packets and found that 18\% IPv4 and 15\% IPv6 open resolvers returned empty responses. More importantly, 8.4\% and 6,6\% of resolvers returned \textit{bogus} replies to our `A' requests. Previous work has shown that this behavior is likely due to censorship, ad redirection and other doubtful activities~\cite{wild}. As the majority of such recursive resolvers return custom responses without contacting authoritative nameservers, their use in DDoS attacks is limited. Thus, we exclude those from further analysis.

\begin{table}[ht!]
\caption{Top 20 IPv4 autonomous systems by the number of open resolvers} 
\vspace{5px}
\label{top_open_v4} 
\scriptsize
\centering 
\begin{tabular}{llc}
 \toprule
  ASN & Organization & IPv4 Resolvers\\
 \midrule
 4134 & China Telecom & 260,649 \\
 4837 & China Unicom & 189,714 \\
 45090 & Tencent-CN & 107,769 \\
 4766 & Korea Telecom & 67,557 \\
 47331 & TTNET A.S. & 58,693 \\
 5617 & Orange Polska & 52,568 \\
 3462 & HiNet & 36,868 \\
 4812 & China Telecom & 33,432 \\
 9318 & SK Broadband & 26,903 \\
 4808 & China Unicom & 26,762 \\
 12389 & Rostelecom  & 24,989 \\
 209 & Centurylink & 24,979 \\
 7713 & Telekomunikasi Indonesia & 21,475 \\
 4538 & China Education and Research Network Center & 18,866 \\
 9808 & China Mobile & 17,838 \\
 58224 & Iran Telecommunication Company & 16,036 \\
 45804 & Meghbela Cable \& Broadband Services & 15,624 \\
 32708 & Root Networks & 15,502 \\
 3269 & Telecom Italia S.p.A. & 12,371 \\
 58659 & Quest Consultancy  & 11,918 \\
 \bottomrule
\end{tabular}
\end{table}

\begin{table}[ht!]
\caption{Top 20 IPv6 autonomous systems by the number of open resolvers.} 
\vspace{5px}
\label{top_open_v6} 
\scriptsize
\centering 
\begin{tabular}{llc}
 \toprule
  ASN & Organization & IPv6 Resolvers\\
 \midrule
 6939 & Hurricane Electric & 4,458 \\
 63949 & Linode & 548 \\
 3462 & HiNet & 415 \\
 4837 & China Unicom & 364 \\
 8966 & Etisalat-AS & 351 \\
 12322 & Free SAS & 332 \\
 4812 & China Telecom & 294 \\
 1241 & Forthnet & 286 \\
 51167 & Contabo & 228 \\
 27839 & Comteco & 184 \\
 16276 & OVH & 179 \\
 7922 & Comcast & 163 \\
 4134 & China Telecom & 159 \\
 37564 & Wirulink Pty Ltd & 153 \\
 8100 & QuadraNet Enterprises LLC & 137 \\
 23910 & China Next Generation Internet CERNET2 & 115 \\
 3303 & Swisscom (Schweiz) AG & 110 \\
 3356 & Level 3 Parent, LLC & 104 \\
 14061 & DigitalOcean, LLC & 102 \\
 8251 & FreeTel, s.r.o. & 102 \\
 \bottomrule
\end{tabular}
\end{table}

\begin{table}[t]
\caption{Top 20 IPv4 autonomous systems by the ratio of open resolvers.} 
\vspace{5px}
\label{top_open_ratio} 
\scriptsize
\centering 
\begin{tabular}{llcc}
 \toprule
  ASN & Organization & AS Size & Ratio \\
 \midrule
  269113 & Uno Telecom LTDA & 1,024 & 99.5 \% \\
  268137 & Net Sini Fiber Home Telecomunicação LTDA & 1,024 & 99.5 \% \\
  136668 & Iana Solutions Digital India & 512 & 99.4 \% \\
  263108 & Opanet Telecomunicacoes LTDA & 2,048 & 99.3 \% \\
  267072 & Veloz Net Serviços e Comunicações LTDA & 768 & 99.2 \% \\
  267007 & Turbo Net Telecom Servicos e Vendas de Equipamento & 1,024 & 99.1 \% \\
  134929 & Orange City Internet Services & 2,048 & 99.0 \% \\
  208070 & TILYTEL B., S.L. & 1,024 & 99.0 \% \\
  270404 & Qualidade Digital Internet e Telecomunicações & 1,024 & 99.0 \% \\
  134924 & Aph Networks & 512 & 99.0 \% \\
  269563 & MAX3 TELECOM LTDA & 1,024 & 98.93 \% \\
  271003 & MARILETE PEREIRA DOS SANTOS & 1,024 & 98.83 \% \\
  270657 & FNET TELECOM & 1,024 & 98.83 \% \\
  34939 & NextDNS & 768 & 98.83 \% \\
  137045 & Athoy Cyber Net & 512 & 98.83 \% \\
  47849 & Global Communication Net Plc & 3,072 & 98.73 \% \\
  269012 & Click Net Link Informatica e Telecomunicações LTDA & 1,024 & 98.73 \% \\
  265276 & SPEED\_MAAX BANDA LARGA LTDA - ME & 1,024 & 98.73 \% \\
  271070 & Ailson Tavares & 1,024 & 98.63 \% \\
  47275 & Torjon Wieslaw Radka & 1,024 & 98.63 \% \\
 \bottomrule
\end{tabular}
\end{table}

\begin{table}[t]
\caption{Top 20 countries/territories by the number of open resolvers.} 
\vspace{5px}
\label{top_countries} 
\scriptsize
\centering 
\begin{tabular}{lc}
 \toprule
  Country & IPv4 Resolvers\\
 \midrule
  China & 758,083 \\
  Brazil & 323,263 \\
  USA & 180,328 \\
  India & 117,363 \\
  Republic of Korea & 116,749 \\
  Russia & 97,287 \\
  Turkey & 78,982 \\
  Indonesia & 75,157 \\
  Poland & 73,189 \\
  Taiwan & 42,577 \\
  Bangladesh & 38,061 \\
  Argentina & 34,858 \\
  France & 31,720 \\
  Italy & 28,916 \\
  Ukraine & 27,348 \\
  Iran & 27,343 \\
  Japan & 24,808 \\
  Thailand & 22,520 \\
  Hong Kong & 20,765 \\
  Bulgaria & 19,992 \\
 \bottomrule
\end{tabular}
\hspace{7px}
\begin{tabular}{lc}
 \toprule
  Country & IPv6 Resolvers\\
 \midrule
  USA & 2,500 \\
  Germany & 1,323 \\
  China & 1,258 \\
  France & 880 \\
  Republic of Korea & 708 \\
  Taiwan & 583 \\
  Russia & 494 \\
  Czech Republic & 409 \\
  Japan & 395 \\
  UK & 376 \\
  Brazil & 367 \\
  United Arab Emirates & 354 \\
  Greece & 342 \\
  Thailand & 310 \\
  Canada & 307 \\
  Iran & 295 \\
  Vietnam & 252 \\
  India & 244 \\
  Switzerland & 242 \\
  South Africa & 239 \\
 \bottomrule
\end{tabular}
\end{table}

\begin{table}[th!]
\caption{Distribution of open resolvers in European Union countries.} 
\vspace{5px}
\label{top_eu} 
\scriptsize
\centering 
\begin{tabular}{lc}
 \toprule
  Country & IPv4 Resolvers\\
 \midrule
  Poland & 73,189 \\
  France & 31,720 \\
  Italy & 28,916 \\
  Bulgaria & 19,992 \\
  Germany & 18,352 \\
  Spain & 12,400 \\
  Hungary & 10,221 \\
  Romania & 7,766 \\
  Czech Republic & 7,508 \\
  Netherlands & 7,165 \\
  Sweden & 5,945 \\
  Greece & 4,962 \\
  Austria & 3,722 \\
  Slovakia & 3,663 \\
  Portugal & 3,646 \\
  Latvia & 3,394 \\
  Croatia & 2,547 \\
  Denmark & 1,877 \\
  Finland & 1,738 \\
  Belgium & 1,734 \\
  Lithuania & 1,178 \\
  Ireland & 1,145 \\
  Cyprus & 694 \\
  Slovenia & 687 \\
  Estonia & 355 \\
  Luxembourg & 313 \\
  Malta & 250 \\
 \bottomrule
\end{tabular}
\hspace{7px}
\begin{tabular}{lc}
 \toprule
  Country & IPv6 Resolvers\\
 \midrule
  Germany & 1,323 \\
  France & 880 \\
  Czech Republic & 409 \\
  Greece & 342 \\
  Netherlands & 181 \\
  Hungary & 119 \\
  Italy & 76 \\
  Romania & 74 \\
  AT & 72 \\
  Lithuania & 64 \\
  SE & 59 \\
  PL & 53 \\
  ES & 51 \\
  BG & 48 \\
  Slovenia & 43 \\
  Finland & 30 \\
  Belgium & 30 \\
  Denmark & 26 \\
  PT & 18 \\
  Ireland & 18 \\
  Croatia & 18 \\
  Cyprus & 17 \\
  LV & 13 \\
  SK & 10 \\
  LU & 7 \\
  EE & 6 \\
 \bottomrule
\end{tabular}
\end{table}

\begin{table}[th!]
\caption{Ratio of open resolvers per region.} 
\vspace{5px}
\label{regions} 
\scriptsize
\centering 
\begin{tabular}{lcc}
 \toprule
  Region & Ratio of IPv4 Resolvers & Ratio of IPv6 Resolvers\\
 \midrule
  Africa & 2.4 \% & 2.1 \% \\
  Asia & 56.4 \% & 34.9 \% \\
  Europe & 6.3 \% & 8.6 \% \\
  European Union & 10.1 \% & 26.9 \% \\
  North America & 7.8 \% & 19.0 \% \\
  Oceania & 0.5 \% & 1.0 \% \\
  South America & 16.5\% & 7.5 \% \\
 \bottomrule
\end{tabular}
\end{table}

\textbf{Autonomous System Distribution}

We map the remaining open resolvers to their autonomous system numbers (ASN) using pyasn\footnote{\url{https://pypi.org/project/pyasn/}} and check the PeeringDB\footnote{\url{https://www.peeringdb.com}} and AS Rank\footnote{\url{https://asrank.caida.org}} for the organization names. Table~\ref{top_open_v4} and Table~\ref{top_open_v6} present the numbers of open DNS resolvers by autonomous systems. Top 20 IPv4 organisations are dominated by Asian telecommunication operators, while IPv6 autonomous systems also include transit and hosting providers. In total, open resolvers are present in 24,087 IPv4 and 1,607 IPv6 autonomous systems (34.2\% and 7,4\% of all those in the BGP routing table as of beginning of March 2021).

The big absolute number of recursive resolvers may not be surprising if belonging to a big autonomous system. Thus, we compute a ratio of open resolvers to the size of the address space announced by the IPv4 autonomous systems. Table~\ref{top_open_ratio}  shows the results. None of the organisations from Table~\ref{top_open_v4} and Table~\ref{top_open_v6} is present in Table~\ref{top_open_ratio}. These small autonomous systems almost entirely consist of open resolvers. In fact, there are 278 IPv4 autonomous systems where more than half of the address space is occupied by open resolvers.

\textbf{Geographic Distribution}

We map all the open resolvers to countries using the MaxMind database.\footnote{https://dev.maxmind.com/geoip/geoip2/geolite2/} Overall, open resolvers are present in 230 countries/territories. Table~\ref{top_countries} shows the top twenty countries by the number of open IPv4 and IPv6 resolvers. Eleven countries dominate both in IPv4 and IPv6 ranking. Importantly, top twenty countries contain the majority of all the open resolvers worldwide: 84.9\% in IPv4 and 80.4\% in IPv6. Table~\ref{top_eu} displays the number of open resolvers in European Union (EU) countries only. 
The top three countries account for more than 50\% IPv4 and 66\% IPv6 open resolvers in EU.

Next, we examine the ratio of open resolvers per regions in Table~\ref{regions}. The majority of IPv4 resolvers are located in Asia. IPv6 resolvers are not dominated by a single region, as more than 60\% of those are shared between Asia and European Union. Africa, Oceania, and Europe (outside the European Union) represent the smallest share of open resolvers. 

\subsection{Discussion}\label{sec:dns_resolvers_discussion}

Open resolvers pose a great security threat---they are prone to misuse by attackers and should only be operated when necessary. We discovered more than \textbf{2.5 million} correctly resolving IPv4 and IPv6 open resolvers worldwide. We have shown that they are distributed both in terms of organisations and geographic territories. Nevertheless, the majority of all the open resolvers originate from very few autonomous systems and countries.

K\"{u}hrer \textit{et al.} fingerprinted 5.4 million open resolvers and concluded that more than 60\% of those were routers, modems, gateways, and embedded devices~\cite{wild}. We hypothesize that telecommunication operators do not configure customer equipment correctly. If this is the case, then some national telecommunication operators could eliminate a significant number of open resolvers in their countries, as it is the case with Orange Polska or Telecom Italia.

Note that the problem is not new. Jared Mauch presented at the NANOG meeting the Open Resolver Project \cite{mauch}.
He uncovered 34 Million DNS servers that responded to UDP/53 probe.
Despite different initiatives to mitigate the problem, such as Computer Emergency Response Team (CERT) alerts \cite{certalert}, research indicating the scale of the problem \cite{amplificationHell,exitFromHell}, and notifications to operators by ShadowServer or locally by the national German CERT \cite{certger}, the issue has still not been resolved.

\pagelabel{text:cert_open_res_rec}
\vspace{+0.5cm}
\begin{tcolorbox}[enhanced,colback=blue!5!white,colframe=blue!75!black,colbacktitle=red!80!black]
 \textbf{Recommendation}: National CERT teams should subscribe to data sources that identify open DNS resolvers. National governments and Computer Emergency Response Team (CERT) teams should intensify notification efforts to reduce the number of open DNS resolvers (and other open services), which are among the root causes of distributed reflective denial-of-service (DRDoS) attacks. 
 \end{tcolorbox}

\section{SPF and DMARC\label{sec:spfdmarc}}

\subsection{Motivation}
Email spoofing is defined as sending emails with a forged sender address in a way that it appears as sent from a legitimate user or on behalf of a company \cite{maroofi2020defensive}. Business Email Compromise (BEC) is one of the most financially damaging online crimes \cite{fbi_bec}, and email spoofing is one of the most common techniques used in BEC. 

The Simple Mail Transfer Protocol (SMTP) does not provide a built-in approach to fight email spoofing. Therefore, the deployment of the email security extensions such as the Sender Policy Framework (SPF) \cite{rfc_spf}, DomainKeys Identified Mail (DKIM) \cite{rfc_dkim}, and Domain-based Message Authentication, Reporting, and  Conformance (DMARC) \cite{rfc_dmarc} is necessary to fight phishing attacks. In this section, we measure the global adoption of email security extensions, namely, SPF and DMARC for all the domain names in our database as described in Section \ref{sec:general_dataset}. We do not measure DKIM since it needs access to DKIM subdomains (also known as the \textit{selector} tag). They are not publicly available and can only be found in the header of the received emails.

\subsection{Methodology}
To measure the deployment of SPF and DMARC, we use the following approach:

\begin{itemize}
    \item Regarding SPF, first, we collect SPF records (as part of DNS TXT resource records) of all 251 Million enumerated domains using the ZDNS tool.
    
    \item Then, for those domains with SPF records, we emulate the \textit{check\_host()} function as described in  RFC 7208 \cite{rfc_spf}, to evaluate the validity and configurations of the records.
    \item The next step is to collect the DMARC rules, which exist in the TXT resource records of the \texttt{\textit{\_dmarc}} subdomains of the registered domains (e.g., \texttt{\_dmarc.example.com}).
    \item Finally, we evaluate DMARC rules to check their strictness in accepting (delivering to the end-users) and/or rejecting the incoming forged emails.
\end{itemize}

The next section shows the results of our scan for SPF and DMARC protocols.

\subsection{Results}\label{sec:spf_dmarc_results}
Table \ref{tbl_spf_result} shows the results of the scan for 247,006,422 domain names that returned \textit{NOERROR}. The second column (status) shows the scan status with ZDNS for SPF rules in the TXT resource record of each domain. \textbf{60.44\%} of the domains have no SPF record, which is a bad practice to protect a domain name from email spoofing since in the absence of the SPF record, for any incoming message, the result of the \textit{check\_host()} function is \textbf{None} (i.e., no strict decision), leaving the decision to the receiver mailing system. Since most of the domains currently do not have SPF records, the receiving email servers make a `softer' decision, generally letting the email be delivered so that the users do not lose any email. This behaviour makes it easier for spammers to send forged emails \cite{hu2018end,maroofi2020defensive, maroofi2021adoption}. 

\vspace{+0.4cm} 
\begin{table}[hbt]
     \centering
     \scalebox{0.8}{
         \begin{tabular}{|c|c|c|c|}
         \hline
         \#& Status & Count & Percentage (\%)  \\ [0.5ex] 
         \hline\hline
         1 & AUTHFAIL & 275,925 &  0.111 \\ 
         \hline
         2& ERROR & 169,679 &  0.068 \\
         \hline
         3& NOERROR & 77,487,889 &  31.370 \\
         \hline
         4& NORECORD & 149,305,756 &  60.446 \\
         \hline
         5& NXDOMAIN & 2,475,409 &  1.002 \\
         \hline
         6& REFUSED & 5,979,033 &  2.420 \\
         \hline
         7& SERVFAIL & 10,616,307 &  4.297 \\
         \hline
         8& TEMPORARY & 348 & 0.0001  \\
         \hline
         9& TIMEOUT & 696,076 &  0.281 \\
         \hline
         \hline
         \multicolumn{2}{c}{Total} & \multicolumn{2}{c}{247,006,422}
                \\
        \hline
        \hline
         \end{tabular}
     }
     \caption{Scan results of the SPF rules.}    
     \label{tbl_spf_result}
 \end{table}

Only 31.37\% of the domains have SPF records. However, having an SPF record does not necessarily guarantee any protection against email spoofing. Table \ref{tbl_spf_emulation} shows the results of the \textit{check\_host()} function emulation for the domains with SPF records (corresponding to the third row of Table \ref{tbl_spf_result} with \textit{NOERROR} status). All the domains with SPF \textit{pass} results are open to email spoofing since they let the sender send emails from any IP address. For other SPF results, the decision is made by the receiver with the help of DMARC rules specified in the TXT resource records of the \textit{\_dmarc} subdomain. The SPF \textit{permerror} result means that there is a problem in either parsing or recursive querying SPF rules, which usually happens because of setting a syntactically wrong SPF rule or defining too much DNS lookups (recursions) in the SPF rule \cite{maroofi2021adoption}. Table \ref{tbl_spf_permerror} shows the most common errors related to the domains with the \textit{permerror} result from the \textit{check\_host()} function emulation.

\begin{table}[hbt]
     \centering
     \scalebox{0.8}{
         \begin{tabular}{|c|c|c|c|}
         \hline
         \#& Status & Count & Percentage (\%)  \\ [0.5ex] 
         \hline\hline
         1 & None & 2,543,870 & 3.28  \\ 
         \hline
         2& Neutral & 5,866,297 & 7.58  \\
         \hline
         3& Pass & 200,362 &  0.26 \\
         \hline
         4& Fail & 29,049,907 & 37.5  \\
         \hline
         5& Softfail & 35,929,956 & 46.37  \\
         \hline
         6& Permerror & 3,207,817 & 4.14  \\
         \hline
         7& Temperror & 689,680 & 0.91  \\
         \hline
         \hline
         \multicolumn{2}{c}{Total} & \multicolumn{2}{c}{77,487,889}
                \\
         \hline
         \hline
         \end{tabular}
     }
     \caption{Results of the check\_host function emulation}    
     \label{tbl_spf_emulation}
 \end{table}

\begin{table}[bht]
    \begin{center}
    \scalebox{0.7}{
        \begin{tabular}{|c|c|c|c|c}
            \hline
            Error type & Example & Correct rule & Frequency \\
            \hline
            Too many DNS lookups & - & SPF rule must generate less than 10 DNS query & 1,638,092 \\
            \hline
            Two or more SPF records found& -  & must set one SPF record for each domain & 691,746 \\
            \hline
            Void lookup limit of 2 exceeded & -  & rules with empty responses must be removed & 64,914\\
            \hline
            More than 10 MX records returned & -  & Total number of lookups must be less than 10 & 27,699 \\
            \hline
            Invalid IP4 address: ip4: & ip4:xxx.xxx.xxx.xx?all & ip4:xxx.xxx.xxx.xx ?all & 16,621\\
            \hline
            \hline
        \end{tabular}
    }
    \end{center}
    \caption{Most common syntactically wrong rules that lead to the \textit{Permerror} result. \label{tbl_spf_permerror}}
\end{table}

\begin{table}[hbt]
     \centering
     \scalebox{0.8}{
         \begin{tabular}{|c|c|c|c|}
         \hline
         \#& Status & Count & Percentage (\%)  \\ [0.5ex] 
         \hline\hline
         1 & AUTHFAIL & 284,939 &  0.115 \\ 
         \hline
         2& ERROR & 166,427 &  0.067\\
         \hline
         3& NOERROR & 8,129,795 &   3.299\\
         \hline
         4& NORECORD & 70,277,952 &  28.518 \\
         \hline
         5& NXDOMAIN & 150,842,488 &  61.212 \\
         \hline
         6& REFUSED & 6,019,716 &  2.442 \\
         \hline
         7& SERVFAIL & 10,037,232 &  4.073 \\
         \hline
         8& TEMPORARY & 10,795 &   0.004\\
         \hline
         9& TIMEOUT & 656,653 &  0.266 \\
         \hline
         \hline
         \multicolumn{2}{c}{Total} & \multicolumn{2}{c}{246,425,997}
                \\
        \hline
        \hline
         \end{tabular}
     }
     \caption{Scan results of the DMARC records}    
     \label{tbl_dmarc_result}
 \end{table}

 \pagelabel{text:spf_dmarc_results_end}Table \ref{tbl_dmarc_result} shows the scan results of the domain names for DMARC records. The status \textit{NXDOMAIN} means that there is no DMARC subdomain for the domain name, and \textit{NOERROR} means that the DMARC record is present, which is only true for \textbf{3\% of the domain names}. However, still having DMARC does not necessarily guarantee any protection. The final decision about the incoming email delivery is up to the \textbf{p}~tag of the DMARC, which specifies the action to do: i) deliver the message, ii) reject the message, or iii) quarantine the message (labeled as spam). Parsing the DMARC record shows that 49.68\% of the domain names with the DMARC record has the \textbf{p=none} rule, which means they specified no strict action against incoming messages sent from unauthorized servers. 11.20\% of the domains have \textbf{p=quarantine} (i.e., labeling the incoming message as spam), and 37.14\% have \textbf{p=reject}, which means rejecting the incoming message with unauthorized sender based on SPF rules.

\subsection{Discussion}\label{sec:spf_dmarc_discussion}  
SPF and DMARC protocols are critical for preventing email spoofing and essential in preventing Business Email Compromise (BEC) fraud, which according to the recent FBI report, caused more than US \$1.8 billion in losses to businesses and individuals in 2020\cite{fbi_bec}. Note that securing domain names with SPF and DMARC does not solve the problem of BEC scams, as criminals can register, e.g., misspelled (e.g., using special characters) or internationalized domain names. However, suppose SPF and DMARC rules are not correctly configured. In that case, a cybercriminal can send emails on behalf of target brand domain names, making recipients unable to distinguish legitimate emails from fraudulent ones.
Correctly implemented and strict SPF and DMARC rules can completely mitigate the problem of domain name spoofing, assuming that recipient mail servers verify and filter emails based on SPF and DMARC~rules.
 
A recent study explored the degree of SPF and DMARC deployment for high-profile domains, including banking domains, and identified misconfigured ones \cite{maroofi2021adoption}. They notified domain owners through local, and national CERT teams, and as many as 23.2\% of the domains were reconfigured.
While this was a one-time notification campaign, such ongoing efforts to measure the deployment and raise awareness of the problem should be promoted by governments and national CERTs, especially in the light of the recent FBI report of financial losses caused by BEC scams.

\pagelabel{text:sec_spf_dmarc_rec}
\vspace{+0.5cm}
\begin{tcolorbox}[enhanced,colback=blue!5!white,colframe=blue!75!black,colbacktitle=red!80!black]
 \textbf{Recommendation}: Security community should intensify efforts to continuously measure the adoption of the SPF and DMARC protocols, especially for high risk domain names and raise awareness of the domain spoofing problem among domain owners and email service providers. Correct and strict SPF and DMARC rules can mitigate email spoofing and provide the first line of defense against Business Email Compromise (BEC) scams.
 \end{tcolorbox}

\clearpage

\label{chapter:emails}

\section{Analysis of RFC-compliant Email Aliases}

In this section, we consider the problem of effective
notifications of domain abuse or vulnerabilities to the domain owners,
administrators, and webmasters.

\subsection{Motivation}

Malicious actors compromise thousands of legitimate domains every day by
exploiting vulnerable content management systems, frameworks, or
libraries used to build  websites. The compromised domains are abused to launch Internet-scale phishing, malware drive-by-download,  
or spam campaigns. 
To prevent vulnerable resources from being exploited and to remediate already
abused domains (URLs), defenders share information about security threats and incidents through collaborative platforms such as Anti-Phishing Working Group
(APWG)\footnote{\url{https://apwg.org}},
PhishTank\footnote{\url{https://www.phishtank.com}}, or URLhaus\footnote{\url{https://urlhaus.abuse.ch}}.

Some types of domain abuse or vulnerabilities should be directly reported to the
domain registrants (owners), administrators, or webmasters (often at scale, so web forms may not be always appropriate). Therefore, the
Internet community needs to maintain the ability of large-scale notification
mechanisms. An alternative approach is to report abuse through intermediaries
such as CERTs.
They validate email notifications 
 and further communicate with the actors responsible for the affected systems.
However, previous work showed that security notifications directly addressed to
the owners of vulnerable resources promote faster remediation than those sent to
national CERTs \cite{Effective}. 
Therefore, several researchers used the contacts of domain administrators and
owners retrieved from the public WHOIS data
\cite{Great,heartbleed,LiGoogle,Effective,hey,atresh,Zeng,CetinJGEM16} and
studied different communications strategies to increase remediation rates.

Retrieving contact information at scale from  public WHOIS 
proved to be highly problematic \cite{Great} and became even more difficult (often impossible) with the introduction of the General Data Protection Regulation (GDPR) on May~25,~2018. 
ICANN adopted the
temporary specification on how to publish
the registration data of individuals \cite{GTLD} that prohibits domain
registrars and registries from storing personal data in the public WHOIS
database, in particular, the email addresses of registrants and
administrators. In the absence of direct contact with the registrant, it is
recommended to contact the relevant registrar who has to provide access to
registrant contact information in ``a~reasonable time''\cite{GTLD1}. However,
this rule may cause delays in patching vulnerabilities and mitigating abuse, and in addition, it does~not~scale.

In this report, based on the previous research \cite{SoussiKMD20}, motivated by the implications of the EU regulation on data
protection, we systematically test available direct contacts of domain owners
and administrators as defined in RFC~2142 \cite{2142}. For a given domain
\texttt{example.com}, RFC~2142 requires to configure valid email aliases such as
\texttt{abuse@example.com} or \texttt{security@example.com} for incident and
vulnerability notifications.  
Rather than quantifying remediation rates of DNS abuse, we test whether five RFC-specific
generic email aliases are correctly configured and whether notifications can be
successfully delivered. 
We also test the reachability of email addresses collected from the DNS Start of
Authority (SOA) resource records (RNAME field).

\subsection{Methodology} \label{sec:eeemethod}

We have developed a scanner to systematically test available direct contacts of
domain owners and administrators. We first scan for the DNS MX records of the
domain and select a mail server with the highest priority.
Afterwards, we establish different connections using the Simple Mail Transfer
Protocol (SMTP) \cite{SMTP} to the selected mail server. We do not send emails, but we verify the existence of an email address using the RCPT TO query
followed by the destination email address.
If the mail server replies with code 250, the recipient address is considered as valid.
In a single SMTP session, we only test one email address to avoid triggering
mechanisms preventing email address enumeration \cite{SMTP}, which may close the
SMTP connection and blacklist the IP address of the sender.
 
Another countermeasure used by mail servers is to accept any given recipient,
even a non-existent one, and return code 250.  
This procedure is called CATCH ALL (or wildcard email address) \cite{postfix}. 
Our scanner is designed to detect if a mail server uses the CATCH ALL mechanism 
by checking for the existence of a randomly generated email. If such a contact
is validated, then the mail server is most likely validating all, even
non-existent, addresses. 

For each sampled domain name, we generate email aliases using the names defined
in RFC 2142~\cite{2142}: for the domain \texttt{example.com}, we test the
validity of the following email aliases: \texttt{hostmaster@example.com},
\texttt{webmaster@example.com} (for DNS and HTTP issues),
\texttt{abuse@example.com} (for generic abuse and vulnerability notifications),
\texttt{noc@example.com}, and \texttt{security@example.com} (for network
security).
    
We scan for DNS SOA records, extract the hostmaster contact stored 
in the RNAME field as defined in RFC 1035 \cite{1035}, and check whether the
syntax of the email address is correct. 
Note that the domain name of an email gathered from the RNAME field may be
different from the tested domain itself implying that we also need to lookup the
MX record of the hostmaster domain.

To compare the results of the most popular websites with less known ones, we
leverage top 1~M domains from Tranco---a domain ranking list oriented toward
research~\cite{tranco}.
   
To measure the reachability rates of the different TLDs, we categorize the domains based on their TLD and we sample domains from each~group. We calculate the size of each sample using binomial approximation
\cite{binomial1} as explained in \cite{SoussiKMD20}. 
We perform the
experiment on generated samples of \texttt{.com} (633 K domain names) and \texttt{.net} (772 K domains) gTLD names, new gTLD names (232 K), and ccTLD names (967 K). 
Finally, we remove duplicates present in more than one sample: e.g., \texttt{google.com} is present
in the Tranco list and has been randomly drawn for the \texttt{.com} sample.

\subsection{Results of Email Validation Scans} \label{sec:eeeres}

The scan results are shown in Table \ref{tab:results}. 
We consider a domain as reachable if at least one RFC-specific generic email alias has been validated. 
It is then labeled as CONTACT FOUND.
We find that in all the studied categories, domains are more reachable using SOA contacts.
For instance, 39.74\% of Tranco top 1~M domains are reachable using an email
leveraged from the SOA RNAME field while only 24.16\% are reachable using
RFC-specific contacts. 
We also find significantly more missing MX records for RFC-specific contacts
than for mail servers of SOA contacts because DNS SOA records are often
maintained by DNS service operators and less frequently by the domain owners who
often do not have enough expertise in configuring DNS servers.

 \begin{table}[ht!]
\caption{Results of email validation scan on the selected TLDs and popular domains.\label{tab:results}} 
 \begin{tabular}{|cc|cccc|c|}
        \hline
        \textbf{} & \textbf{} & \multicolumn{5}{c|}{\textbf{Domains (\%)}}\\
                \cline{3-7}
        \textbf{} & \textbf{Category} & \multicolumn{4}{c|}{\textbf{Selected TLDs}} & 
        \multicolumn{1}{c|}{\textbf{Popular}}
        \\
                \textbf{} & \textbf{} & \textbf{.com} & \textbf{.net} & \textbf{ccTLD} & \textbf{ngTLD} & \multicolumn{1}{c|}{\textbf{Tranco 1~M}} \\
        \hline
        \multirow{6}{*}{\rotatebox[origin=c]{90}{\textbf{~~~RFC emails}}} &
        NO MX RECORD& 35.48	& 35.59 & 18.43 & 61.63 & \multicolumn{1}{c|}{21.46} \\
        
        &CONN. ERROR& 25.58 & 27.32 & 19,11 & 9.68 & \multicolumn{1}{c|}{13.00} \\

        &CATCH ALL& 12.33 & 11.42 & 18.86 & 7.68 &  \multicolumn{1}{c|}{20.36} \\
         
        &NO CONTACT FOUND& 17.63 &	17.80 & 30.91 & 17.03 & \multicolumn{1}{c|}{21.02} \\
         
        &CONTACT FOUND& 8.97 & 7.87 & 12.68 & 3.97 & \multicolumn{1}{c|}{24.16} \\
        
        \hline
        \multirow{6}{*}{\rotatebox[origin=c]{90}{\textbf{SOA contact}}} &
        NO SOA RECORD& 4.49 & 4.66 & 3.63 & 11.01 & \multicolumn{1}{c|}{2.84} \\
        
        &NO MX FOR SOA& 9.01 & 7.70 & 8.31 & 10.47 & \multicolumn{1}{c|}{9.04} \\
        
        &CONN. ERROR& 34.68 & 35.83 & 15.30 & 15.86 & \multicolumn{1}{c|}{18,60} \\
        
        &CATCH ALL& 18.09 & 17.54 & 22.36 & 12.12  & \multicolumn{1}{c|}{19.38} \\
       
        &NO CONTACT FOUND& 13.53 & 13.28 & 14.99 & 29.17 & \multicolumn{1}{c|}{10.40} \\
 
        &CONTACT FOUND& 20.20 & 20.97 & 35.40 & 21.35  & \multicolumn{1}{c|}{39.74} \\
        
        \hline
 \end{tabular}
\end{table}

The ccTLD names seem to be the most reachable for both SOA (35.4\%) and RFC-specific (12.68\%) contacts 
when compared to
\texttt{.com} (20.20\% and 8.97\%) and especially  
to new gTLDs (21.35\% and only 3.97\%). New gTLDs names are 
far less reachable 
in the sample with 61.63\%  of 
domains without MX records, three times more than ccTLDs and almost twice more than \texttt{.com} domains.

 \begin{figure}[ht!]
    \vspace{-0.4cm}
        \includegraphics[width=1.0\linewidth] {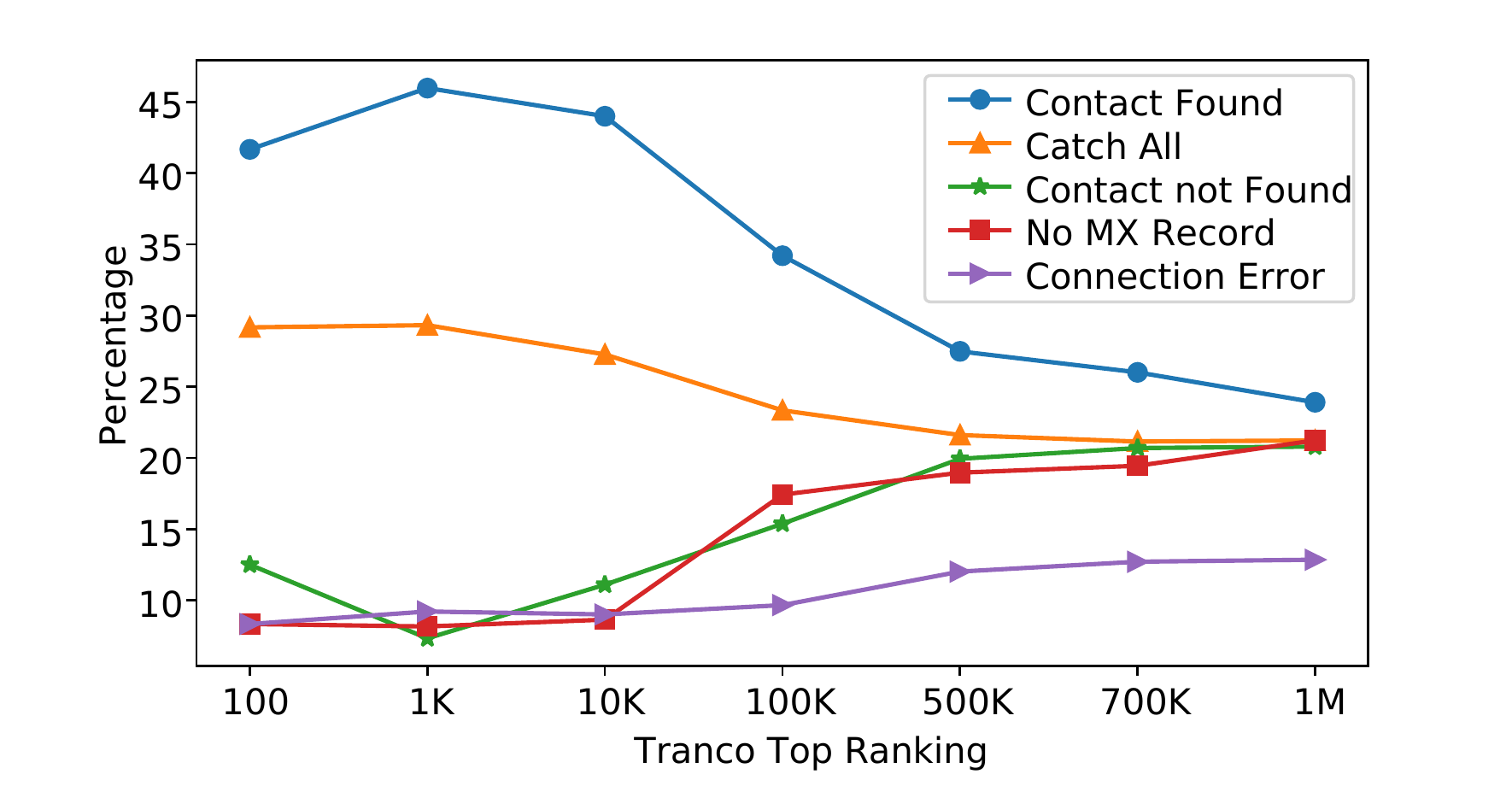}
        \vspace{-0.6cm}
                \caption{Results of validation scans of RFC-specific email aliases for
          the Tranco top 100 to Tranco top 1~M lists.} 
        \label{fig:a-prog}
        \vspace{-0.2cm}
 \end{figure}

The results for the Tranco top 100 to 1~M lists (Figure~\ref{fig:a-prog}) show
that the better the rank of the domain is, the more likely the domain is
reachable and complies with RFC 2142. 
For the top 1 K domains, at least 43.9\% are reachable and the proportion
gradually decreases to 22.6\% for Tranco top 1~M. 
Similarly, the rate of CATCH ALL domains decreases from 28\% (top 100) to 19\% (top 1~M). 
The results can be explained by the fact that the operators of more popular
websites put more emphasis on preventing email address enumeration of their
clients. 

We also observe an increasing rate of domains
with invalid contacts: from 7\% (top 1~K) to 19.7\% (top 1~M) and
domains without MX records: from 8\% (top 1~K) to 20\% (top 1~M). The connection error rate has also slightly increased from 7\% (top 100) to 11.5\% (top 1~M).
   
\begin{figure}
            \centering
           \vspace{-0.5cm} \includegraphics[width=\linewidth] {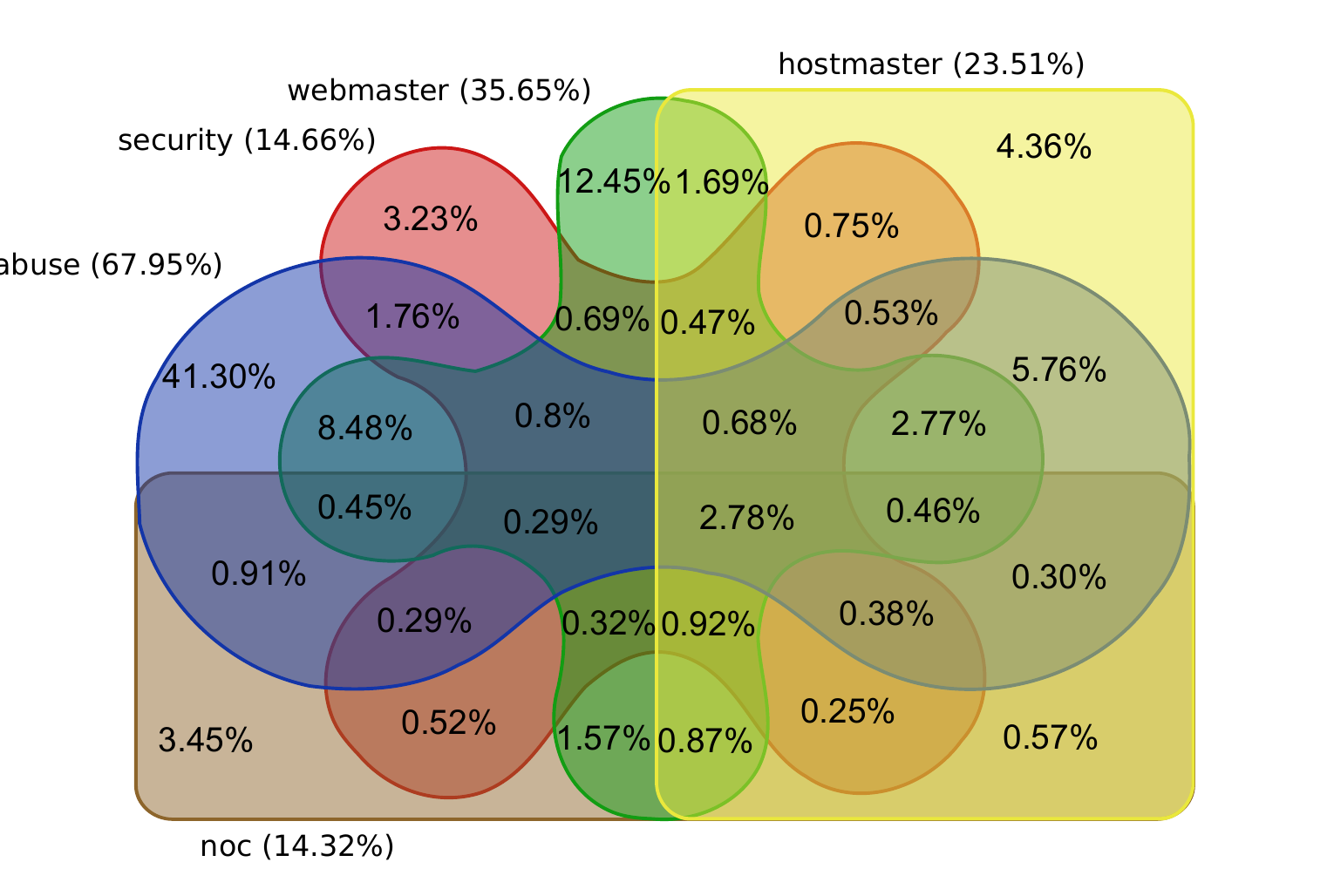}
            \caption{Venn diagram of the used RFC-specific email aliases}
            \label{fig:venn}
            \vspace{-0.2cm}
\end{figure}

We next briefly analyze the most common RFC-specific email aliases. 
Figure \ref{fig:venn} presents the results in the Venn diagram.
We observe that the most frequently used contact is \textit{abuse} (67,95\%). As
many as 64.79\% of domains only use one valid contact: 41.3\% of domains use the
\textit{abuse} contact, while 23.46\% use one of the other 4 aliases (half of
them use \textit{webmaster}). 35.21\% of domains use multiple RFC-specific
contacts.

\subsection{Discussion}\label{sec:eee_disc}

The Internet community needs a large-scale notification system that covers the largest part of Internet domains for direct reporting of threats or abuses.

\pagelabel{text:whois_rfc_email}
\vspace{+0.5cm}
\begin{tcolorbox}[enhanced,colback=blue!5!white,colframe=blue!75!black,colbacktitle=red!80!black]
 \textbf{Recommendation}: With no direct contact with domain name registrants and administrators via the public WHOIS database, domain name administrators should also maintain specific email aliases for given domain names (e.g., abuse, hostmaster, webmaster) so that they can be contacted directly in the event of vulnerabilities and domain name abuse.
 \end{tcolorbox}

 \pagelabel{text:whois_anonym_rec}
\vspace{+0.5cm}
\begin{tcolorbox}[enhanced,colback=blue!5!white,colframe=blue!75!black,colbacktitle=red!80!black]
 \textbf{Recommendation}: The email addresses of registrants and domain name administrators that are not visible in the public WHOIS could be displayed as anonymized email addresses to ensure the ability to contact domain owners and administrators directly to notify security vulnerabilities and abuses.
 \end{tcolorbox}

\label{chapter:inbound}

\section{Inbound Source Address Validation}\label{sec:isav}

The Internet relies on IP packets to enable communication between hosts with the
destination and source addresses specified in packet headers. However, there is
no packet-level authentication mechanism to ensure that the source address has
not been altered~\cite{Beverly:2009:UED:1644893.1644936}. The modification of a
source IP address is referred to as ``IP spoofing''. It results in the anonymity
of the sender and prevents a packet from being traced to its
origin. Reflection-based Distributed Denial-of-Service (DRDoS) attacks
leverage this mechanism and become even more effective using
amplification~\cite{hell, Kuhrer:2014:EHR:2671225.2671233, unchained}. As
it is not possible in general to prevent packet header modification, concerted
efforts have been undertaken to prevent spoofed packets from reaching potential
victims. Filtering packets at the network edge formalized in RFC~2827 and called
\textit{Source Address Validation} (SAV)~\cite{Ferguson:2000:NIF:RFC2827,sav-encyclopedia} can
achieve this goal. 

Given the prevalent role of IP spoofing in cyberattacks, there is a need to
estimate the level of SAV deployment by network providers.  Projects such as
Spoofer~\cite{Spoofer} already enumerate networks that do not implement packet
filtering. However, a great majority of this existing work concentrates on
\textit{outbound} SAV \cite{bb-spoofer-sruti,Beverly:2009:UED:1644893.1644936,Spoofer,Lichtblau,marketplaces,loops,spoofer_new,saving} it can prevent DRDoS attacks near
their origin~\cite{Kuhrer:2014:EHR:2671225.2671233}. While less obvious, the
lack of \textit{inbound} filtering enables an external attacker to masquerade as
an internal host of a network, which may reveal valuable information about the
network infrastructure usually not seen from the outside. Inbound IP spoofing
can serve as a vector for DNS zone poisoning
attacks~\cite{zone} that may lead to domain
hijacking or cache poisoning attacks~\cite{kaminsky} even~if~the DNS resolver is correctly configured as a closed~resolver. A closed
resolver only accepts DNS queries from known clients and does so by matching the
source IP address of a query against a list of allowed addresses. 

The lack of SAV for inbound traffic can also have devastating consequences when
combined with the DNS Unchained \cite{unchained} or the NXDOMAIN attack (also
known as the Water Torture Attack)~\cite{LuoWXCYT18}, or the recently discovered
NXNSAttack~\cite{NXNSAttack}. These attacks result in Denial-of-Service against
both recursive resolvers and authoritative servers with a maximum packet
amplification factor of 1,620 for the NXNSAttack~\cite{NXNSAttack}. IP spoofing
is not required for this attack to succeed because any client can attack a
resolver if it is allowed to query it.  However, IP spoofing can greatly
increase the number of affected resolvers by allowing an external attacker to
target closed DNS resolvers: the attacker simply needs to masquerade as a
legitimate client by spoofing its source IP address. Deploying inbound SAV at
the edge of a network is an effective way of protecting closed DNS resolvers
from this type of external attacks as well as possible zero-day  vulnerabilities  that reside  within  the  DNS  server  software.

The goal of the recently launched Closed Resolver Project \cite{korczyski2020dont, closed, korczyski2020closed, korczyski2020anrw} is to enumerate networks vulnerable to inbound spoofing across the Internet as a first step in estimating the scale of the problem.
We have presented a novel method to infer the deployment of inbound SAV for the IPv4 and IPv6 address spaces. We have measured the filtering
policies of 52\% of routable IPv4 autonomous systems (26\% for IPv6) and 28\% of
all the IPv4 BGP prefixes (almost 9\% for IPv6).
We showed that the vast majority of the networks for which we obtained measurements are (consistently or partially) vulnerable to inbound spoofing and that the vulnerability is not limited to any geographic territory and is spread worldwide.

\pagelabel{text:sav_rec}
\vspace{+0.5cm}
\begin{tcolorbox}[enhanced,colback=blue!5!white,colframe=blue!75!black,colbacktitle=red!80!black]
 \textbf{Recommendation}: Network operators should deploy IP Source Address Validation (SAV) not only for outgoing but also for \textit{incoming} traffic at the edge of a network. It provides an effective way of protecting closed DNS resolvers from different external attacks against DNS infrastructure, including possible zero-day vulnerabilities within the DNS server software.
 \end{tcolorbox}

\newpage
\pagestyle{plain}
\section*{Acknowledgements}
This study was commissioned by the European Commission (EC reference VIGIE 2020/0653).
We would like to thank EU and international institutions and agencies, law enforcement authorities, brand owners, trade and industry associations, TLD registries, registrars, hosting providers, other intermediaries, and security experts for their constructive and valuable comments.
We would like to thank Spamhaus, SURBL, Anti-Phishing Working Group, Abuse.ch, Phishtank, and OpenPhish for providing access to their blacklist feeds. The authors also thank  Roman Huessy for providing the uptime data for the URLhaus feed, the CENTR community for sharing the sizes of ccTLDs, and Sourena Maroofi for providing valuable comments and discussions on the paper.

\newpage
\setstretch{1.0}
\bibliographystyle{ieeetrans}
\bibliography{references}

\begin{thebibliography}{10}
\providecommand{\url}[1]{#1}
\csname url@samestyle\endcsname
\providecommand{\newblock}{\relax}
\providecommand{\bibinfo}[2]{#2}
\providecommand{\BIBentrySTDinterwordspacing}{\spaceskip=0pt\relax}
\providecommand{\BIBentryALTinterwordstretchfactor}{4}
\providecommand{\BIBentryALTinterwordspacing}{\spaceskip=\fontdimen2\font plus
\BIBentryALTinterwordstretchfactor\fontdimen3\font minus
  \fontdimen4\font\relax}
\providecommand{\BIBforeignlanguage}[2]{{%
\expandafter\ifx\csname l@#1\endcsname\relax
\typeout{** WARNING: IEEEtran.bst: No hyphenation pattern has been}%
\typeout{** loaded for the language `#1'. Using the pattern for}%
\typeout{** the default language instead.}%
\else
\language=\csname l@#1\endcsname
\fi
#2}}
\providecommand{\BIBdecl}{\relax}
\BIBdecl

\bibitem{comar}
S.~Maroofi, M.~Korczy\'nski, C.~Hesselman, B.~Ampeau, and A.~Duda, ``{COMAR:
  Classification of Compromised versus Maliciously Registered Domains},'' in
  \emph{2020 IEEE European Symposium on Security and Privacy (EuroS\&P)}, 2020.

\bibitem{zone}
M.~Korczy\'{n}ski, M.~Kr\'{o}l, and M.~van Eeten, ``{Zone Poisoning: The How
  and Where of Non-Secure DNS Dynamic Updates},'' in \emph{Proceedings of the
  2016 ACM on Internet Measurement Conference}, ser. IMC '16.\hskip 1em plus
  0.5em minus 0.4em\relax ACM, 2016, pp. 271--278.

\bibitem{spamhaus}
``{The Spamhaus Project},'' \url{www.spamhaus.org}.

\bibitem{surbl}
``{SURBL - URI reputation data},'' \url{http://www.surbl.org}.

\bibitem{abuse.ch}
``{ABUSE CH},'' \url{https://abuse.ch}.

\bibitem{apwg}
``{APWG: Cross-industry Global Group Supporting Tackling the Phishing
  Menace},'' \url{http://antiphishing.org}.

\bibitem{phishtank}
``{PhishTank: A Nonprofit Anti-phishing Organization},''
  \url{http://www.phishtank.com}.

\bibitem{openphish}
``{OpenPhish},'' \url{https://openphish.com}.

\bibitem{dbl}
``{The Domain Block List},'' \url{https://www.spamhaus.org/dbl}.

\bibitem{surbllists}
``{SURBL Lists},'' \url{http://www.surbl.org/lists}.

\bibitem{godaddy_merge}
P.~Nicks, ``{Uniregistry’s registrar and marketplace joining GoDaddy to bring
  domain investing to the next level},''
  \url{https://www.godaddy.com/garage/uniregistrys-registrar-and-marketplace-joining-godaddy},
  February 2020.

\bibitem{maroofi2020defensive}
S.~Maroofi, M.~Korczynski, and A.~Duda, ``{From Defensive Registration to
  Subdomain Protection: Evaluation of Email Anti-Spoofing Schemes for
  High-Profile Domains},'' in \emph{Proc. Network Traffic Measurement and
  Analysis Conference (TMA)}, 2020.

\bibitem{amplificationHell}
C.~Rossow, ``{Amplification Hell: Revisiting Network Protocols for DDoS
  Abuse},'' in \emph{Network and Distributed System Security Symposium}, 2014.

\bibitem{exitFromHell}
M.~K\"{u}hrer, T.~Hupperich, C.~Rossow, and T.~Holz, ``{Exit from Hell?
  Reducing the Impact of Amplification DDoS Attacks},'' in \emph{USENIX
  Conference on Security Symposium}, 2014.

\bibitem{rfc1035}
\BIBentryALTinterwordspacing
``{Domain names - implementation and specification},'' RFC 1035, Nov. 1987.
  [Online]. Available: \url{https://rfc-editor.org/rfc/rfc1035.txt}
\BIBentrySTDinterwordspacing

\bibitem{routeviews}
``{University of Oregon Route Views Project},''
  \url{http://www.routeviews.org/routeviews/}.

\bibitem{hitlist}
O.~Gasser, Q.~Scheitle, P.~Foremski, Q.~Lone, M.~Korczynski, S.~D. Strowes,
  L.~Hendriks, and G.~Carle, ``Clusters in the expanse: Understanding and
  unbiasing ipv6 hitlists,'' in \emph{Proceedings of the 2018 Internet
  Measurement Conference}.\hskip 1em plus 0.5em minus 0.4em\relax New York, NY,
  USA: ACM, 2018.

\bibitem{korczyski2020closed}
M.~Korczy\'{n}ski, Y.~Nosyk, Q.~Lone, M.~Skwarek, B.~Jonglez, and A.~Duda,
  ``{The Closed Resolver Project: Measuring the Deployment of Source Address
  Validation of Inbound Traffic},'' in \emph{arXiv, 2006.05277}, 2020.

\bibitem{AXFR}
M.~Skwarek, M.~Korczy\'{n}ski, W.~Mazurczyk, and A.~Duda, ``{Characterizing
  Vulnerability of DNS AXFR Transfers with Global-Scale Scanning},'' in
  \emph{IEEE Security and Privacy Workshops (SPW)}, 2019.

\bibitem{reputation-encyclopedia}
\BIBentryALTinterwordspacing
M.~Korczy{\'{n}}ski and A.~Noroozian, ``Security reputation metrics,'' in
  \emph{Encyclopedia of Cryptography, Security and Privacy}.\hskip 1em plus
  0.5em minus 0.4em\relax Springer Berlin Heidelberg, 2021. [Online].
  Available: \url{https://doi.org/10.1007/978-3-642-27739-9_1625-1}
\BIBentrySTDinterwordspacing

\bibitem{korczynski2018cybercrime}
M.~Korczy\'nski, M.~Wullink, S.~Tajalizadehkhoob, G.~Moura, A.~Noroozian,
  D.~Bagley, and C.~Hesselman, ``Cybercrime after the sunrise: A statistical
  analysis of dns abuse in new gtlds,'' in \emph{Proceedings of the 2018 on
  Asia Conference on Computer and Communications Security}.\hskip 1em plus
  0.5em minus 0.4em\relax ACM, 2018, pp. 609--623.

\bibitem{tld_maciej}
M.~Korczy\'nski, S.~Tajalizadehkhoob, A.~Noroozian, M.~Wullink, C.~Hesselman,
  and M.~van Eeten, ``Reputation metrics design to improve intermediary
  incentives for security of tlds,'' in \emph{2017 IEEE European Symposium on
  Security and Privacy (Euro SP)}, April 2017.

\bibitem{noroozian2015}
A.~Noroozian, M.~Korczy\'nski, S.~Tajalizadehkhoob, and M.~van Eeten,
  ``Developing security reputation metrics for hosting providers,'' in
  \emph{8th Usenix Workshop on Cyber Security Experimentation and Test (CSET
  15)}, 2015.

\bibitem{sunrise}
M.~Korczy\'nski, M.~Wullink, S.~Tajalizadehkhoob, G.~C. Moura, A.~Noroozian,
  D.~Bagley, and C.~Hesselman, ``{Cybercrime After the Sunrise: A Statistical
  Analysis of DNS Abuse in New gTLDs},'' in \emph{ACM AsiaCCS}, 2018.

\bibitem{sadag-cct}
``{Competition, Consumer Trust, and Consumer Choice Review Team - Final
  Report},''
  \url{https://www.icann.org/en/system/files/files/cct-final-08sep18-en.pdf },
  2018.

\bibitem{sadag-ssr2}
``{Second Security, Stability, and Resiliency (SSR2) Review Team Final
  Report},''
  \url{https://www.icann.org/en/system/files/files/ssr2-review-team-final-report-25jan21-en.pdf
  }, 2021.

\bibitem{APWG2016}
\BIBentryALTinterwordspacing
(2016) Global phishing survey: Trends and domain name use in 2016. [Online].
  Available:
  \url{https://docs.apwg.org/reports/APWG_Global_Phishing_Report_2015-2016.pdf}
\BIBentrySTDinterwordspacing

\bibitem{sadag}
\BIBentryALTinterwordspacing
M.~Korczy\'nski, M.~Wullink, S.~Tajalizadehkhoob, G.~C. Moura, and
  C.~Hesselman, ``{Statistical Analysis of DNS Abuse in gTLDs Final Report},''
  Tech. Rep., 2017. [Online]. Available:
  \url{https://www.icann.org/en/system/files/files/sadag-final-09aug17-en.pdf}
\BIBentrySTDinterwordspacing

\bibitem{landscape}
\BIBentryALTinterwordspacing
G.~Aaron, L.~C.~D. Piscitello, and C.~Strutt, ``{Phishing Landscape 2020 A
  Study of the Scope and Distribution of Phishing},'' Tech. Rep., 2020.
  [Online]. Available: \url{interisle.net/PhishingLandscape2020.pdf}
\BIBentrySTDinterwordspacing

\bibitem{Halvorson}
T.~Halvorson, M.~F. Der, I.~Foster, S.~Savage, L.~K. Saul, and G.~M. Voelker,
  ``{From .Academy to .Zone: An Analysis of the New TLD Land Rush},'' in
  \emph{Proc. of ACM IMC}, 2015, pp. 381--394.

\bibitem{arman}
A.~Noroozian, M.~Korczy\'nski, S.~Tajalizadehkhoob, and M.~van Eeten,
  ``{Developing Security Reputation Metrics for Hosting Providers},'' in
  \emph{Proc. of the 8th USENIX CSET}, 2015, pp. 1--8.

\bibitem{human}
S.~Maroofi, M.~Korczy\'nski, and A.~Duda, ``{Are You Human?: Resilience of
  Phishing Detection to Evasion Techniques Based on Human Verification},'' in
  \emph{{ACM} Internet Measurement Conference, Virtual Event}.\hskip 1em plus
  0.5em minus 0.4em\relax {ACM}, 2020, pp. 78--86.

\bibitem{kaminsky}
D.~Kaminsky, ``{It's the End of the Cache as We Know It},''
  \url{https://www.slideshare.net/dakami/dmk-bo2-k8}.

\bibitem{rfc3833}
\BIBentryALTinterwordspacing
D.~Atkins and R.~Austein, ``{Threat Analysis of the Domain Name System
  (DNS)},'' RFC 3833, Aug. 2004. [Online]. Available:
  \url{https://rfc-editor.org/rfc/rfc3833.txt}
\BIBentrySTDinterwordspacing

\bibitem{rfc4033}
\BIBentryALTinterwordspacing
R.~Arends, R.~Austein, M.~Larson, D.~Massey, and S.~Rose, ``{DNS Security
  Introduction and Requirements},'' Internet Requests for Comments, RFC 4033,
  March 2005. [Online]. Available:
  \url{http://www.rfc-editor.org/rfc/rfc4033.txt}
\BIBentrySTDinterwordspacing

\bibitem{rfc4035}
\BIBentryALTinterwordspacing
S.~Rose, M.~Larson, D.~Massey, R.~Austein, and R.~Arends, ``{Protocol
  Modifications for the DNS Security Extensions},'' RFC 4035, Mar. 2005.
  [Online]. Available: \url{https://rfc-editor.org/rfc/rfc4035.txt}
\BIBentrySTDinterwordspacing

\bibitem{rfc4034}
\BIBentryALTinterwordspacing
S.~Rose, M.~Larson, D.~Massey, R.~Austein, and R.~Arends, ``{Resource Records
  for the DNS Security Extensions},'' RFC 4034, Mar. 2005. [Online]. Available:
  \url{https://rfc-editor.org/rfc/rfc4034.txt}
\BIBentrySTDinterwordspacing

\bibitem{rootksk}
IANA, ``{Root KSK Ceremony 1},'' 2010,
  \url{https://www.iana.org/dnssec/ceremonies/1}.

\bibitem{tld-dnssec}
I.~Research, ``{TLD DNSSEC Report (2021-05-18 00:05:03)},'' 2021,
  \url{http://stats.research.icann.org/dns/tld_report/archive/20210518.000101.html}.

\bibitem{gtlds}
ICANN, ``{Domain Name System Security Extensions Now Deployed in all Generic
  Top-Level Domains},'' 12 2020,
  \url{https://www.icann.org/en/announcements/details/domain-name-system-security-extensions-now-deployed-in-all-generic-top-level-domains-23-12-2020-en}.

\bibitem{safeguardsAbuse}
ICANN, ``{New gTLD Program Safeguards Against DNS Abuse},''
  \url{https://newgtlds.icann.org/en/reviews/dns-abuse/safeguards-against-dns-abuse-18jul16-en.pdf},
  July 2016.

\bibitem{longitudinal}
\BIBentryALTinterwordspacing
T.~Chung, R.~van Rijswijk-Deij, B.~Chandrasekaran, D.~Choffnes, D.~Levin, B.~M.
  Maggs, A.~Mislove, and C.~Wilson, ``A longitudinal, end-to-end view of the
  {DNSSEC} ecosystem,'' in \emph{26th {USENIX} Security Symposium ({USENIX}
  Security 17)}.\hskip 1em plus 0.5em minus 0.4em\relax Vancouver, BC: {USENIX}
  Association, Aug. 2017, pp. 1307--1322. [Online]. Available:
  \url{https://www.usenix.org/conference/usenixsecurity17/technical-sessions/presentation/chung}
\BIBentrySTDinterwordspacing

\bibitem{cz}
O.~Filip, ``{DNSSEC.CZ},'' 10 2012,
  \url{https://archive.icann.org/en/meetings/toronto2012/bitcache/DNSSEC.CZ-vid=41901&disposition=attachment&op=download.pdf}.

\bibitem{digital}
``{Digital Czech Republic v. 2.0 - The Way to the Digital Economy},'' 04 2014,
  \url{https://www.mpo.cz/dokument149132.html}.

\bibitem{sweden}
{Internetstiftelsen}, ``{Recommendations for DNSSEC deployment},''
  \url{https://internetstiftelsen.se/en/domains/tech-tools/recommendations-for-dnssec-deployment/}.

\bibitem{sidn-dnssec}
{SIDN}, ``{Frequently Asked Questions},''
  \url{https://www.sidn.nl/en/faq/dnssec}.

\bibitem{nxns}
\BIBentryALTinterwordspacing
Y.~Afek, A.~Bremler{-}Barr, and L.~Shafir, ``Nxnsattack: Recursive {DNS}
  inefficiencies and vulnerabilities,'' in \emph{29th {USENIX} Security
  Symposium, {USENIX} Security 2020, August 12-14, 2020}, S.~Capkun and
  F.~Roesner, Eds.\hskip 1em plus 0.5em minus 0.4em\relax {USENIX} Association,
  2020, pp. 631--648. [Online]. Available:
  \url{https://www.usenix.org/conference/usenixsecurity20/presentation/afek}
\BIBentrySTDinterwordspacing

\bibitem{ddos}
\BIBentryALTinterwordspacing
R.~van Rijswijk-Deij, A.~Sperotto, and A.~Pras, ``Dnssec and its potential for
  ddos attacks: A comprehensive measurement study,'' in \emph{Proceedings of
  the 2014 Conference on Internet Measurement Conference}, ser. IMC '14.\hskip
  1em plus 0.5em minus 0.4em\relax New York, NY, USA: Association for Computing
  Machinery, 2014, p. 449–460. [Online]. Available:
  \url{https://doi.org/10.1145/2663716.2663731}
\BIBentrySTDinterwordspacing

\bibitem{rfc8482}
\BIBentryALTinterwordspacing
J.~Abley, Ólafur Guðmundsson, M.~Majkowski, and E.~Hunt, ``{Providing
  Minimal-Sized Responses to DNS Queries That Have QTYPE=ANY},'' RFC 8482, Jan.
  2019. [Online]. Available: \url{https://rfc-editor.org/rfc/rfc8482.txt}
\BIBentrySTDinterwordspacing

\bibitem{rfc6781}
\BIBentryALTinterwordspacing
O.~Kolkman, M.~Mekking, and R.~M. Gieben, ``{DNSSEC Operational Practices,
  Version 2},'' RFC 6781, Dec. 2012. [Online]. Available:
  \url{https://rfc-editor.org/rfc/rfc6781.txt}
\BIBentrySTDinterwordspacing

\bibitem{rfc8624}
\BIBentryALTinterwordspacing
P.~Wouters and O.~Surý, ``{Algorithm Implementation Requirements and Usage
  Guidance for DNSSEC},'' RFC 8624, Jun. 2019. [Online]. Available:
  \url{https://rfc-editor.org/rfc/rfc8624.txt}
\BIBentrySTDinterwordspacing

\bibitem{nist}
E.~Barker and A.~Roginsky, ``\BIBforeignlanguage{en}{{Transitions:
  Recommendation for Transitioning the Use of Cryptographic Algorithms and Key
  Lengths}},'' in \emph{\BIBforeignlanguage{en}{Special Publication (NIST SP),
  National Institute of Standards and Technology, Gaithersburg, MD}}, 01 2011.

\bibitem{registrars}
\BIBentryALTinterwordspacing
T.~Chung, R.~van Rijswijk-Deij, D.~Choffnes, D.~Levin, B.~M. Maggs, A.~Mislove,
  and C.~Wilson, ``Understanding the role of registrars in dnssec deployment,''
  in \emph{Proceedings of the 2017 Internet Measurement Conference}, ser. IMC
  '17.\hskip 1em plus 0.5em minus 0.4em\relax New York, NY, USA: Association
  for Computing Machinery, 2017, p. 369–383. [Online]. Available:
  \url{https://doi.org/10.1145/3131365.3131373}
\BIBentrySTDinterwordspacing

\bibitem{wild}
M.~K\"{u}hrer, T.~Hupperich, J.~Bushart, C.~Rossow, and T.~Holz, ``{Going Wild:
  Large-Scale Classification of Open DNS Resolvers},'' in \emph{Internet
  Measurement Conference}.\hskip 1em plus 0.5em minus 0.4em\relax ACM, 2015.

\bibitem{mauch}
J.~Mauch, ``{Spoofing ASNs},'' \url{http://seclists.org/nanog/2013/Aug/132}.

\bibitem{certalert}
``{Alert (TA14-017A) UDP-Based Amplification Attacks},''
  \url{https://us-cert.cisa.gov/ncas/alerts/TA14-017A/}.

\bibitem{certger}
``{Reports on openly accessible services},''
  \url{https://www.bsi.bund.de/EN/Topics/IT-Crisis-Management/CERT-Bund/CERT-Reports/Reports/openly-accessible-services/openly-accessible-services_node.html}.

\bibitem{fbi_bec}
``{Business Email Compromise},''
  \url{https://www.ic3.gov/Media/PDF/AnnualReport/2020_IC3Report.pdf}, 2020.

\bibitem{rfc_spf}
\BIBentryALTinterwordspacing
S.~Kitterman, ``{RFC 7208: Sender Policy Framework (SPF) for Authorizing Use of
  Domains in Email},'' Internet Requests for Comments, 2014. [Online].
  Available: \url{http://tools.ietf.org/html/rfc7208}
\BIBentrySTDinterwordspacing

\bibitem{rfc_dkim}
\BIBentryALTinterwordspacing
D.~Crocker, T.~Hansen, and M.~Kucherawy, ``{RFC 6376: DomainKeys Identified
  Mail (DKIM) Signatures},'' Internet Requests for Comments, 2011. [Online].
  Available: \url{http://tools.ietf.org/html/rfc6376}
\BIBentrySTDinterwordspacing

\bibitem{rfc_dmarc}
\BIBentryALTinterwordspacing
E.~Kucherawy, M.~Zwicky, and E.~Zwicky, ``{RFC 7489: Domain-Based Message
  Authentication, Reporting, and Conformance (DMARC)},'' Internet Requests for
  Comments, 2015. [Online]. Available: \url{http://tools.ietf.org/html/rfc7489}
\BIBentrySTDinterwordspacing

\bibitem{hu2018end}
H.~Hu and G.~Wang, ``End-to-end measurements of email spoofing attacks,'' in
  \emph{27th USENIX Security Symposium}, 2018, pp. 1095--1112.

\bibitem{maroofi2021adoption}
S.~Maroofi, M.~Korczy{\'n}ski, A.~H{\"o}lzel, and A.~Duda, ``{Adoption of Email
  Anti-Spoofing Schemes: A Large Scale Analysis},'' \emph{IEEE Transactions on
  Network and Service Management}, 2021.

\bibitem{Effective}
F.~Li, Z.~Durumeric, J.~Czyz, M.~Karami, M.~Bailey, D.~McCoy, S.~Savage, and
  V.~Paxson, ``{You've Got Vulnerability: Exploring Effective Vulnerability
  Notifications},'' in \emph{{USENIX Security}}, 2016.

\bibitem{Great}
O.~Cetin, C.~Ga{\~{n}}{\'{a}}n, M.~Korczy\'nski, and M.~van Eeten, ``{Make
  Notifications Great Again: Learning How to Notify in the Age of Large-Scale
  Vulnerability Scanning},'' in \emph{WEIS}, 2017.

\bibitem{heartbleed}
Z.~Durumeric, J.~Kasten, D.~Adrian, J.~A. Halderman, M.~Bailey, F.~Li,
  N.~Weaver, J.~Amann, J.~Beekman, M.~Payer, and V.~Paxson, ``{The Matter of
  Heartbleed},'' in \emph{Proc. {IMC}}, 2014.

\bibitem{LiGoogle}
F.~Li, G.~Ho, E.~Kuan, Y.~Niu, L.~Ballard, K.~Thomas, E.~Bursztein, and
  V.~Paxson, ``{Remedying Web Hijacking: Notification Effectiveness and
  Webmaster Comprehension},'' in \emph{WWW}, 2016.

\bibitem{hey}
B.~Stock, G.~Pellegrino, C.~Rossow, M.~Johns, and M.~Backes, ``{Hey, You Have a
  Problem: On the Feasibility of Large-scale Web Vulnerability Notification},''
  in \emph{USENIX Security}, 2016.

\bibitem{atresh}
B.~Stock, G.~Pellegrino, F.~Li, M.~Backes, and C.~Rossow, ``{Didn’t You Hear
  Me?—Towards More Successful Web Vulnerability Notifications},'' in
  \emph{NDSS}, 2018.

\bibitem{Zeng}
E.~Zeng, F.~Li, E.~Stark, A.~P. Felt, and P.~Tabriz, ``{Fixing HTTPS
  Misconfigurations at Scale: An Experiment with Security Notifications},'' in
  \emph{WEIS}, 2019.

\bibitem{CetinJGEM16}
O.~{\c{C}}etin, M.~H. Jhaveri, C.~Ga{\~{n}}{\'{a}}n, M.~van Eeten, and
  T.~Moore, ``{Understanding the Role of Sender Reputation in Abuse Reporting
  and Cleanup},'' \emph{J. Cybersecur.}, vol.~2, no.~1, pp. 83--98, 2016.

\bibitem{GTLD}
``{Temporary Specification for gTLD Registration Data},''
  \url{https://www.icann.org/resources/pages/gtld-registration-data-specs-en}.

\bibitem{GTLD1}
``{Advisory Statement: Temporary Specification for gTLD Registration Data},''
  \url{https://www.icann.org/en/system/files/files/advisory-statement-gtld-registration-data-specs-17may18-en.pdf}.

\bibitem{SoussiKMD20}
W.~Soussi, M.~Korczynski, S.~Maroofi, and A.~Duda, ``{Feasibility of
  Large-Scale Vulnerability Notifications after {GDPR}},'' in \emph{{IEEE}
  European Symposium on Security and Privacy Workshops, EuroS{\&}P Workshops
  2020}.\hskip 1em plus 0.5em minus 0.4em\relax {IEEE}, 2020, pp. 532--537.

\bibitem{2142}
D.~Crocker, ``Mailbox names for common services, roles and functions,'' RFC
  2142, 1997.

\bibitem{SMTP}
J.~Klensin, ``{Simple Mail Transfer Protocol},'' RFC 5321, 2008.

\bibitem{postfix}
\BIBentryALTinterwordspacing
{Postfix Virtual Domain Hosting Howto}. [Online]. Available:
  \url{http://www.postfix.org/VIRTUAL_README.html}
\BIBentrySTDinterwordspacing

\bibitem{1035}
P.~Mockapetris, ``{Domain Names - Implementation and Specification},'' RFC
  1035, 1987.

\bibitem{tranco}
V.~L. Pochat, T.~Van~Goethem, S.~Tajalizadehkhoob, M.~Korczy{\'n}ski, and
  W.~Joosen, ``{Tranco: A Research-Oriented Top Sites Ranking Hardened Against
  Manipulation},'' in \emph{NDSS}, 2019.

\bibitem{binomial1}
{S. B. Hulley and S. R. Cummings and W. S. Browner and D. Grady and T. B.
  Newman}, \emph{{Designing Clinical Research: an Epidemiologic Approach. 4th
  ed.}}\hskip 1em plus 0.5em minus 0.4em\relax Philadelphia, PA: Lippincott
  Williams \& Wilkins, July 2013.

\bibitem{Beverly:2009:UED:1644893.1644936}
R.~Beverly, A.~Berger, Y.~Hyun, and k.~claffy, ``{Understanding the Efficacy of
  Deployed Internet Source Address Validation Filtering},'' in \emph{Internet
  Measurement Conference}.\hskip 1em plus 0.5em minus 0.4em\relax ACM, 2009.

\bibitem{hell}
C.~Rossow, ``{Amplification Hell: Revisiting Network Protocols for DDoS
  Abuse},'' in \emph{Network and Distributed System Security Symposium}, 2014.

\bibitem{Kuhrer:2014:EHR:2671225.2671233}
M.~K\"{u}hrer, T.~Hupperich, C.~Rossow, and T.~Holz, ``{Exit from Hell?
  Reducing the Impact of Amplification DDoS Attacks},'' in \emph{USENIX
  Conference on Security Symposium}, 2014.

\bibitem{unchained}
J.~Bushart and C.~Rossow, ``{DNS Unchained: Amplified Application-Layer DoS
  Attacks Against DNS Authoritatives},'' in \emph{Research in Attacks,
  Intrusions, and Defenses}, M.~Bailey, T.~Holz, M.~Stamatogiannakis, and
  S.~Ioannidis, Eds.\hskip 1em plus 0.5em minus 0.4em\relax Cham: Springer
  International Publishing, 2018, pp. 139--160.

\bibitem{Ferguson:2000:NIF:RFC2827}
\BIBentryALTinterwordspacing
D.~Senie and P.~Ferguson, ``{Network Ingress Filtering: Defeating Denial of
  Service Attacks which Employ IP Source Address Spoofing},'' RFC 2827, May
  2000. [Online]. Available: \url{https://rfc-editor.org/rfc/rfc2827.txt}
\BIBentrySTDinterwordspacing

\bibitem{sav-encyclopedia}
\BIBentryALTinterwordspacing
M.~Korczy{\'{n}}ski and Y.~Nosyk, ``{Source Address Validation},'' in
  \emph{Encyclopedia of Cryptography, Security and Privacy}.\hskip 1em plus
  0.5em minus 0.4em\relax Springer Berlin Heidelberg, 2021. [Online].
  Available: \url{https://doi.org/10.1007/978-3-642-27739-9_1626-1}
\BIBentrySTDinterwordspacing

\bibitem{Spoofer}
CAIDA, ``{The Spoofer Project},''
  \url{https://www.caida.org/projects/spoofer/}.

\bibitem{bb-spoofer-sruti}
R.~Beverly and S.~Bauer, ``{The Spoofer Project: Inferring the Extent of Source
  Address Filtering on the Internet},'' in \emph{{USENIX} Steps to Reducing
  Unwanted Traffic on the Internet Workshop}, Jul. 2005.

\bibitem{Lichtblau}
F.~Lichtblau, F.~Streibelt, T.~Kr\"{u}ger, P.~Richter, and A.~Feldmann,
  ``{Detection, Classification, and Analysis of Inter-domain Traffic with
  Spoofed Source IP Addresses},'' in \emph{Internet Measurement
  Conference}.\hskip 1em plus 0.5em minus 0.4em\relax ACM, 2017.

\bibitem{marketplaces}
Q.~Lone, M.~Luckie, M.~Korczy\'{n}ski, H.~Asghari, M.~Javed, and M.~van Eeten,
  ``{Using Crowdsourcing Marketplaces for Network Measurements: The Case of
  Spoofer},'' in \emph{Traffic Monitoring and Analysis Conference}, 2018.

\bibitem{loops}
Q.~Lone, M.~Luckie, M.~Korczy\'{n}ski, and M.~van Eeten, ``{Using Loops
  Observed in Traceroute to Infer the Ability to Spoof},'' in \emph{Passive and
  Active Measurement Conference}.\hskip 1em plus 0.5em minus 0.4em\relax
  Springer International Publishing, 2017.

\bibitem{spoofer_new}
M.~Luckie, R.~Beverly, R.~Koga, K.~Keys, J.~Kroll, and k.~claffy, ``{Network
  Hygiene, Incentives, and Regulation: Deployment of Source Address Validation
  in the Internet},'' in \emph{Computer and Communications Security
  Conference}.\hskip 1em plus 0.5em minus 0.4em\relax ACM, 2019.

\bibitem{saving}
Q.~Lone, M.~Korczy\'{n}ski, C.~Ga\~n\'an, and M.~van Eeten, ``{SAVing the
  Internet: Explaining the Adoption of Source Address Validation by Internet
  Service Providers},'' in \emph{Workshop on the Economics of Information
  Security}, 2020.

\bibitem{LuoWXCYT18}
X.~Luo, L.~Wang, Z.~Xu, K.~Chen, J.~Yang, and T.~Tian, ``{A Large Scale
  Analysis of {DNS} Water Torture Attack},'' in \emph{{Conference on Computer
  Science and Artificial Intelligence}}, 2018.

\bibitem{NXNSAttack}
L.~Shafir, Y.~Afek, and A.~Bremler-Barr, ``{NXNSAttack: Recursive DNS
  Inefficiencies and Vulnerabilities},'' in \emph{USENIX Security Symposium},
  2020.

\bibitem{korczyski2020dont}
M.~Korczyński, Y.~Nosyk, Q.~Lone, M.~Skwarek, B.~Jonglez, and A.~Duda,
  ``{Don't Forget to Lock the Front Door! Inferring the Deployment of Source
  Address Validation of Inbound Traffic},'' in \emph{{Passive and Active
  Measurement}}.\hskip 1em plus 0.5em minus 0.4em\relax Springer International
  Publishing, 2020.

\bibitem{closed}
``{The Closed Resolver Project},'' \url{https://closedresolver.com}.

\bibitem{korczyski2020anrw}
M.~Korczy\'{n}ski, Y.~Nosyk, Q.~Lone, M.~Skwarek, B.~Jonglez, and A.~Duda,
  ``Inferring the deployment of inbound source address validation using dns
  resolvers,'' in \emph{Proceedings of the Applied Networking Research
  Workshop}, ser. ANRW'20.\hskip 1em plus 0.5em minus 0.4em\relax Association
  for Computing Machinery, 2020, p. 9–11.

\end{thebibliography}

\glsaddallunused
\end{document}